\pdfoutput=1
\documentclass{article}
\usepackage[utf8]{inputenc}
\usepackage[utf8]{inputenc}
\usepackage{graphicx, fancyhdr}
\usepackage{mathrsfs}
\usepackage{amsmath,amssymb,verbatim,amsthm,xfrac,bbm,mathabx}
\usepackage{apacite}
\usepackage{textcomp} 
\usepackage{stmaryrd}
\usepackage{setspace}
\usepackage{mathtools,mathabx}
\usepackage[round]{natbib}
\usepackage[colorlinks,allcolors=blue]{hyperref}
\usepackage{multirow}
\usepackage{booktabs}
\usepackage[shortlabels]{enumitem}
\usepackage[scr=esstix]{mathalfa} 
\usepackage{tikz}
\usepackage{multirow}
\usepackage[top=1in, bottom=1in, right=1in, left=1in]{geometry}

\newtheorem{theorem}{Theorem}[section]
\newtheorem{lemma}{Lemma}[section]

\theoremstyle{definition}

\theoremstyle{plain}

\newtheorem*{prop*}{Proposition}
\newtheorem*{theorem*}{Theorem}
\newtheorem*{lemma*}{Lemma}
\newtheorem*{assumption*}{Assumption}
\newtheorem*{defin*}{Definition}
\newtheorem*{corollary*}{Corollary}
\theoremstyle{definition}
\newtheorem*{example*}{Example}
\newtheorem*{remark*}{Remark}
\theoremstyle{plain}
\newtheorem*{exercise*}{Exercise}

\newcommand{\Ex}{\mathbb{E}}

\newcommand{\Narg}[1]{\mathbb{N}_{#1}}
\newcommand{\R}{\mathbb{R}}

\newcommand{\tharg}[1]{${#1}^{\mathrm{th}}$\:}

\newcommand{\phe}{\phantom{{}={}}}

\newcommand{\smallcirc}{~\raisebox{1pt}{\tikz \draw[line width=0.6pt] circle(1.1pt);}~}

\newcommand{\Eta}{\mathrm{H}}

\newcommand{\Tau}{\mathrm{T}}

\newcommand{\bzero}{\pmb{0}}
\newcommand{\bone}{\pmb{1}}

\newcommand{\calF}{\mathcal{F}}

\newcommand{\calH}{\mathcal{H}}
\newcommand{\calI}{\mathcal{I}}

\newcommand{\calL}{\mathcal{L}}

\newcommand{\calN}{\mathcal{N}}

\newcommand{\calQ}{\mathcal{Q}}

\newcommand{\calS}{\mathcal{S}}

\newcommand{\calY}{\mathcal{Y}}

\newcommand{\ba}{\pmb{a}}
\newcommand{\bb}{\pmb{b}}

\newcommand{\bu}{\pmb{u}}

\newcommand{\bz}{\pmb{z}}

\newcommand{\bA}{\pmb{A}}
\newcommand{\bB}{\pmb{B}}
\newcommand{\bC}{\pmb{C}}
\newcommand{\bD}{\pmb{D}}

\newcommand{\bF}{\pmb{F}}
\newcommand{\bG}{\pmb{G}}

\newcommand{\bI}{\pmb{I}}

\newcommand{\bL}{\pmb{L}}
\newcommand{\bM}{\pmb{M}}
\newcommand{\bN}{\pmb{N}}

\newcommand{\bQ}{\pmb{Q}}
\newcommand{\bR}{\pmb{R}}

\newcommand{\bV}{\pmb{V}}
\newcommand{\bW}{\pmb{W}}
\newcommand{\bX}{\pmb{X}}
\newcommand{\bY}{\pmb{Y}}

\newcommand{\bbeta}{\pmb{\beta}}
\newcommand{\bgamma}{\pmb{\gamma}}

\newcommand{\boldeta}{\pmb{\eta}}

\newcommand{\blambda}{\pmb{\lambda}}

\newcommand{\bEta}{\pmb{\Eta}}

\newcommand{\bTau}{\pmb{\Tau}}

\newcommand{\tillambda}{\widetilde{\lambda}}

\newcommand{\btilE}{\widetilde{\pmb{E}}}

\newcommand{\btilX}{\widetilde{\pmb{X}}}

\newcommand{\btillambda}{\widetilde{\pmb{\lambda}}}

\newcommand{\hath}{\widehat{h}}

\newcommand{\hatq}{\widehat{q}}

\newcommand{\haty}{\widehat{y}}

\newcommand{\hatP}{\widehat{P}}

\newcommand{\hatS}{\widehat{S}}

\newcommand{\hatalpha}{\widehat{\alpha}}

\newcommand{\hatlambda}{\widehat{\lambda}}

\newcommand{\hatPi}{\widehat{\Pi}}

\newcommand{\bhatE}{\widehat{\pmb{E}}}

\newcommand{\bhatQ}{\widehat{\pmb{Q}}}

\newcommand{\bhatY}{\widehat{\pmb{Y}}}

\newcommand{\bhatbeta}{\widehat{\pmb{\beta}}}
\newcommand{\bhatgamma}{\widehat{\pmb{\gamma}}}

\newcommand{\bhatlambda}{\widehat{\pmb{\lambda}}}

\newcommand{\bhatEta}{\widehat{\pmb{\Eta}}}

\newcommand{\bbarY}{\widebar{\pmb{Y}}}

\newcommand{\tr}{{\rm tr}}

\newcommand{\proj}[1]{\mathbb{P}_{#1}}
\newcommand{\diag}{\mathrm{diag}}

\newcommand{\ind}{\mathbbm{1}}

\DeclarePairedDelimiter\inner{\langle}{\rangle}%
\DeclarePairedDelimiter\abs{\lvert}{\rvert}%
\DeclarePairedDelimiter\norm{\lVert}{\rVert}%
\DeclarePairedDelimiter\floor{\lfloor}{\rfloor}

\DeclareMathOperator*{\argmin}{arg\,min}

\makeatletter
\let\oldabs\abs
\def\abs{\@ifstar{\oldabs}{\oldabs*}}
\let\oldnorm\norm
\def\norm{\@ifstar{\oldnorm}{\oldnorm*}}
\makeatother

\let\oldpropto\propto
\newcommand{\garbtemp}{\:\oldpropto\:} 
\let\propto\garbtemp

\usepackage[linesnumbered,ruled,vlined]{algorithm2e}
\SetKwInput{KwInput}{Input}
\SetKwInput{KwOutput}{Output}
\defcitealias{CDC:2023}{CDC, 2023}

\begin{document}

\onehalfspacing

\begingroup
\centering
\Large \textbf{Fast variable selection for distributional regression\\with application to continuous glucose monitoring data}\\
\normalsize
\begin{table}[h!]
    \centering
    \begin{tabular}{l|l}
    \textbf{Alexander Coulter} & \textit{Department of Statistics, Texas A\&M University}\\
    \textbf{Rashmi N. Aurora} & \textit{Grossman School of Medicine, New York University}\\
    \textbf{Naresh M. Punjabi} & \textit{Miller School of Medicine, University of Miami}\\
    \textbf{Irina Gaynanova} & \textit{Department of Biostatistics, University of Michigan}
\end{tabular}
\end{table}
\today\par
\endgroup

\begin{abstract}

With the growing prevalence of diabetes and the associated public health burden, it is crucial to identify modifiable factors that could improve patients’ glycemic control. In this work, we seek to examine associations between medication usage, concurrent comorbidities, and glycemic control, utilizing data from continuous glucose monitors (CGMs). CGMs provide high-frequency interstitial glucose measurements, but reducing data to simple statistical summaries is common in clinical studies, resulting in substantial information loss. Recent advancements in the Fréchet regression framework allow to utilize more information by treating the full distributional representation of CGM data as the response, while sparsity regularization enables variable selection. However, the methodology does not scale to large datasets. Crucially, variable selection inference using subsampling is computationally infeasible. We develop a new algorithm for sparse distributional regression by deriving a new explicit characterization of the gradient and Hessian of the underlying objective function, while also utilizing rotations on the sphere to perform feasible updates. The updated method is up to 10000+ fold faster than the original approach, opening the door for applying sparse distributional regression to large-scale datasets and enabling previously unattainable subsampling-based inference. Applying our method to CGM data from patients with type 2 diabetes and obstructive sleep apnea, we found a significant association between sulfonylurea medication and glucose variability without evidence of association with glucose mean. We also found that overnight oxygen desaturation variability showed a stronger association with glucose regulation than overall oxygen desaturation levels.

\end{abstract}

\section{Introduction}

Diabetes mellitus is a chronic illness characterized by elevated glucose levels. Type 1 diabetes results from an autoimmune destruction of the pancreas' beta cells which are responsible for the production of insulin. In contrast, type 2 diabetes is characterized by a state of insulin resistance which is accompanied by a decrease in insulin production.  Diabetes is a leading cause of death among U.S. adults \citep{Ahmad:2023} and is linked to development of various complications \citep{Sobrin:2011, Resnick:2002,Moxey:2011, Kodl:2008}.  While diabetes-related mortality has fallen in recent decades \citep{Callaghan:2020} due to improvements in treatment and interventions \citep{Jonas:2021, Rosenquist:2018}, the growing diabetes prevalence \citepalias{CDC:2023} and the associated public health burden make it crucial to identify
modifiable factors that could improve patients’ glycemic control.

In this work, we seek to examine associations between medication usage, obstructive sleep apnea (OSA) and glycemic control. OSA is a sleep-related disorder characterized by cessation of breathing during sleep, and has a high estimated prevalence ($54\%$-$86\%$) among patients with type 2 diabetes \citep{Singh:2021, Lam:2010, Foster:2009}. Our major motivating application is ``Hyperglycemic Profiles in Obstructive Sleep Apnea (HYPNOS)" study  \citep{Rooney:2021}, which aimed to characterize the effects of positive airway pressure (PAP) therapy in patients with OSA and concurrent type 2 diabetes. For each patient, interstitial glucose levels were recorded via Dexcom G4 continuous glucose monitor (CGM), which has a measurement frequency of 5 minutes with an average of 10 days of data per patient.  Herein, our goal is to quantify the association between patient characteristics and glycemic control based on CGM data at baseline (before the institution of PAP therapy). The effects of patients' medication usage and degree of OSA severity are of particular interest, considering the range of available medication types and the possibility of treating OSA with PAP therapy. Given a large number of available medication types, demographic characteristics and measures of OSA severity, we are interested in performing data-driven variable selection and inference with CGM data serving as response.

In clinical practice, it's common to reduce CGM data to crude summaries (e.g., mean, time-in-range), and subsequently evaluate the effect of specific covariates on those summaries \citep{ battelinoContinuousGlucoseMonitoring2022}.
However, a large number of available summaries (40$+$) makes it difficult to discern which are most appropriate in a given context \citep{gaynanovaDigitalBiomarkersGlucose2022}. Utilization of all summary measures leads to the loss of statistical power due to the necessity of multiple comparisons adjustment.  Functional data analysis (FDA) \citep{wangFunctionalDataAnalysis2016a} provides an alternative approach by treating the entire CGM profile as a functional response. However, due to misalignment of CGM profiles across patients (due to differences in time the sensor is worn, meal intake, physical activity, and sleep patterns), FDA applications are either restricted to time periods less confounded by environmental
factors (e.g., sleep time in \citet{Gaynanova:2022, Sergazinov:2023}), or are based on aggregated data over 24h domain \citep{Law:2015, Scott:2020}. As a consequence, despite numerous studies identifying associations between OSA and glucose metabolism \citep{Punjabi:2002, Lindberg:2012}, the effects of OSA severity on glycemic control are not well elucidated, remaining an open area of study \citep{Reutrakul:2017}. In the context of HYPNOS study, initial analysis using CGM summaries post randomization failed to elucidate the effects of PAP therapy on glycemic metrics \citep{auroraGlucoseProfilesObstructive2022}.  Without advancements in algorithms that provide better characterization of temporal features of CGM data, the adverse effects of OSA on glucose metabolism will likely remain underappreciated.

\begin{figure}[!t]
    \centering
    \includegraphics[width = \textwidth, keepaspectratio]{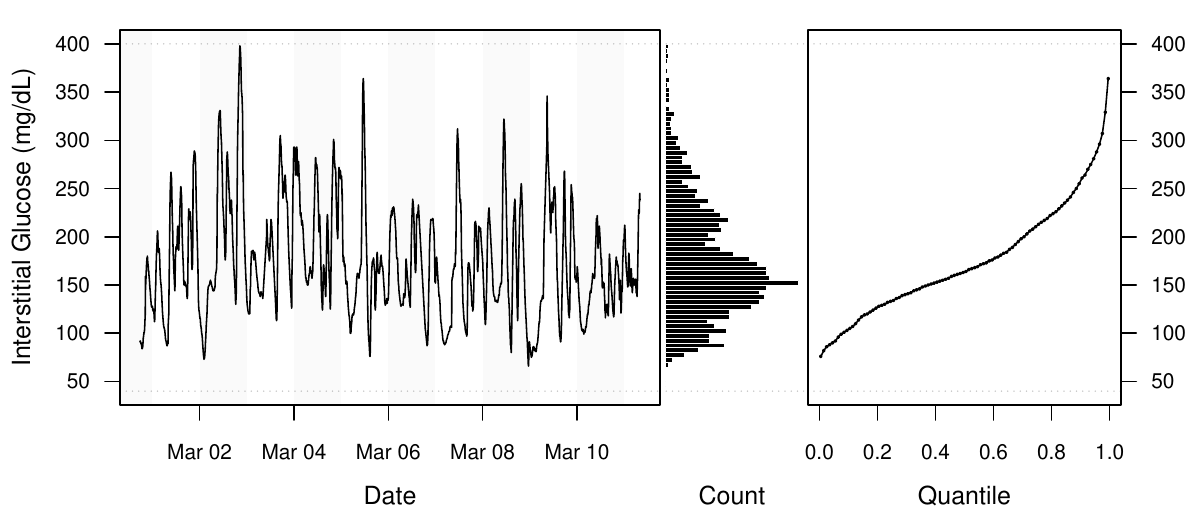}
    \caption{CGM profile (\textit{left}) of a selected subject from HYPNOS study, represented via histogram (\textit{center}) and empirical quantile function (\textit{right}).}
    \label{fig:CGMEQF}
\end{figure}

Distributional analysis provides an alternative framework for CGM data as it advances on traditional summaries by using the whole distribution function of glucose levels \citep{Matabuena:2021, petersenModelingProbabilityDensity2022}, while avoiding time alignment issues in FDA methods. Figure~\ref{fig:CGMEQF} illustrates the full CGM profile of one selected study participant, with corresponding distributional representation via histogram and empirical quantile function. Recent advancements in the Fréchet regression framework  \citep{Petersen:2019} allow to treat the full distributional representation of CGM data as the response, while sparsity regularization as in \citet{Tucker:2023} enables variable selection as is the goal in our study. However, the original methodology does not scale to large datasets. In our application with sample size $n=207$, covariate dimension $p=34$ and a sequence of 20 sparsity tuning parameters, the original method took approximately 1.5 hours on a standard laptop. While this model fitting step is still feasible on our data, albeit slow, it is not feasible for data with hundreds of covariates or data from population-scale cohorts. For example, distributional approaches are also advantageous for data from actigraphy devices \citep{ghosalDistributionalDataAnalysis2021, Matabuena:2023}, while large-scale repositories, such as UK Biobank \citep{doherty:2017}, contain both actigraphy data and large number of individual's characteristics. Currently, sparse distributional regression cannot be applied to such large-scale studies. More crucially, the associated computing costs of model fitting make it infeasible to perform variable selection inference using subsampling-based techniques, such as stability selection \citep{Meinshausen:2010}, even on our data. 

In light of our scientific goal to perform variable selection with inference on HYPNOS data, our main innovation is in developing a new algorithm to enable the application of the sparse distributional regression methodology on a large scale. From the statistical perspective, variable selection with Fréchet regression \citep{Tucker:2023} leverages an assumption that only a subset of covariates is relevant for the prediction of distributional response. From the optimization perspective, this assumption translates into estimating a vector $\bhatlambda \in \R^p$ with a simplex constraint, where either $\hatlambda_k > 0$ (the $k^{\text{th}}$ covariate is selected) or $\hatlambda_k = 0$ (the $k^{\text{th}}$ covariate is not selected), leading to highly non-trivial embedded optimization problem with constraints. The existing approach is based on updating one element of $\blambda$ at a time, making it challenging to scale the approach to settings with many covariates. Furthermore, the existing algorithm does not take advantage of the gradient information, which we believe is due to associated difficulties in characterizing the gradient. Our proposed algorithm is methodologically new in several ways: (i) we obtain closed-form characterization of the gradient and Hessian for distributional responses by taking advantage of the optimality conditions of the embedded problem, thus enabling the use of gradient-based methods; (ii) we replace $\bhatlambda$'s simplex constraint with a spherical constraint, thus enabling feasible updates of the whole vector $\blambda$ by performing rotation steps in the directions against the gradient. The resulting geodesic gradient descent is 10,000+ folds faster than the original approach. Crucially, it allows us to perform variable selection inference for our scientific study via stability selection \citep{Meinshausen:2010}; such inference was previously computationally infeasible.

In summary, our contributions are the following:
\begin{enumerate}[nolistsep]
\item We develop a new algorithm for sparse distributional regression that opens the door to applying the methodology to large-scale datasets;
\item We combine our algorithm with stability selection to enable previously unattainable variable selection inference within sparse distributional regression framework; 
\item In the analysis of HYPNOS data, we find a significant association between sulfonylurea medication and glucose variability without evidence of association with glucose mean. We also find that overnight oxygen desaturation variability shows a stronger association with glucose regulation than overall oxygen desaturation levels. 
\end{enumerate}

The rest of the paper is organized as follows. Section~\ref{sec:data} describes the HYPNOS data and relevant scientific questions. Section~\ref{sec:frechet} reviews Fréchet regression with variable selection.  Section~\ref{sec:proposed_methods} provides development of the proposed algorithm.  Section~\ref{sec:stability_selection} describes stability selection for inference. Section~\ref{sec:simulation} evaluates performance on synthetic data. Section~\ref{sec:dataAnalysis} describes the application to HYPNOS data. Section~\ref{sec:discussion} concludes with discussion.

\textbf{Notation.} For a matrix $\bA = (a_{i,j}) \in \R^{n\times m}$, let $\bA_j = (\begin{matrix} a_{1,j} & \hdots & a_{n,j} \end{matrix})^{\top}$ be its $j^{\mathrm{th}}$ column, and $\mathbf{a}_i = (\begin{matrix} a_{i,1} & \hdots & a_{i,m} \end{matrix})^{\top}$ its $i^{\mathrm{th}}$ row; let $\norm{\bA}_F^2 = \sum_{i,j} a_{i,j}^2$; let $\bA^-$ be a generalized inverse of $\bA$; let $\bA_+$ be the positive part of $\bA$, i.e. $(a_{i,j})_+ = \max\{a_{i,j}, 0\}$; and let $\proj{\bA}$ be the orthogonal projector onto the column space of $\bA$. For two matrices $\bA$ and $\bB$ of the same size, we write $\bA \leq \bB$ to denote $a_{i,j} \leq b_{i,j}$ for all $i,j$; and write $\bA \circ \bB$ to denote $(\bA \circ \bB)_{i,j} = a_{i,j} b_{i,j}$.  For $\bu \in \R^n$, let $\underline{\bu} := \bu/\norm{\bu}_2$; the standard unit basis vectors are denoted $\underline{\textbf{e}}_i \in \R^n$, where the $i^{\mathrm{th}}$ entry is one and the others zero.  We write diagonal matrices as $\bD_{\ba}$, where $(\bD_{\ba})_{i,i} = a_i$; we write the identity matrix as $\bI_n \in \R^{n\times n}$.  We define the $\tau$-simplex $\bTau := \{ \blambda \in \R^p : \blambda^{\top}\bone_p = \tau, \blambda \geq \bzero\}$, and the $\sqrt{\tau}$-radius sphere surface manifold $\calS_{\sqrt{\tau}} := \{\bgamma \in \R^p : \norm{\bgamma}_2 = \sqrt{\tau} \}$.  For consistency, we index samples with $i \in \{1, \hdots, n\}$, quantiles with $j \in \{1, \hdots, m\}$, covariates with $k \in \{1, \hdots, p\}$, and constraints with $c \in \{1, \hdots, m+1\}$.  Finally, we denote a sequence of integers starting at one by $[n] := \{1, \hdots, n\}$.

\section{Data description}\label{sec:data}

\subsection{Data collection}

The data used in this paper were collected as part of the hyperglycemic profiles in obstructive sleep apnea (HYPNOS) randomized clinical trial. The study population consisted of adults between 21 and 75 years old with type 2 diabetes recruited from the community. The primary objective of the trial was to determine whether treatment with positive airway pressure (PAP) therapy is associated with improvements in glycemic measures. Study participants were screened based on the point-of-care HbA1c measured with a DC Vantage Analyzer (Siemens, Malvern, PA) and a home sleep apnea test using the Apnealink monitor (Resmed, San Diego, CA). Participants with HbA1c $\geq$ 6.5\% and obstructive sleep apnea were invited to enroll in the study. Exclusion criteria included pregnancy, any prior therapy for OSA, use of insulin, change in glycemic medications in the previous 6 weeks, current oral steroid use, other sleep disorders, habitual sleep duration of $<$ 6 h/night, and any unstable medical conditions. Study participants used Dexcom G4 Platinum CGM sensor, which produces one measurement every 5 min. The CGM was placed 6 cm lateral to the umbilicus. Participants were instructed to wear the monitor for at least 7 days and provide calibration glucose data for the sensor twice a day according to the manufacturer's instructions. To investigate the effects of OSA severity and other variables on the glucose control of patients with type 2 diabetes, we consider the data at the baseline visit (prior to the randomization of study participants into control and PAP therapy groups). The research protocol was approved by the Institutional Review Board on human research (Number: NA\_00093188). A detailed description of the trial protocol and implementation can be found in \citet{Rooney:2021}. 

\subsection{Data processing}

For each patient, we calculate an empirical quantile function of their CGM measurements corresponding to a baseline visit using a grid of $m = 100$ equally spaced points in $(0, 1)$, leading to response vector $\mathbf{y}_i\in \R^m$ corresponding to patient $i \in [n]$ with $n=207$ patients. On average, there are 2871 glucose values for each patient, ranging from 742 to 5234 (given sensor's 5 min frequency, this corresponds to 2.6 days and 18.2 days, respectively).  All but 10 subjects have $\geq$ 5 days of data. For each patient, we form a vector of baseline covariates $\mathbf{x}_i\in \R^p$ that contains demographic information (i.e., age, sex), clinical information (i.e., BMI, HbA1c), class of medications used, and variables related to OSA severity. We remove any medication that was used by four or fewer patients from the analysis, leading to $p=34$ covariates total.  Table~\ref{tab:description} summarizes select characteristics of the study sample based on 
demographics, as well as covariates related to diabetes and OSA severity (Supplement~\ref{sup:hypnos} describes the full list of covariates). HbA1c values were obtained from point-of-care measurements taken before placement of the CGM sensor. Biguanides and sulfonylureas are families of oral drugs commonly prescribed for type 2 diabetes to reduce glucose levels. ODI$_4$ is the oxygen desaturation index corresponding to the rate of oxygen desaturation events of 4\% or more,  mean saturation is the overall average of the oxygen saturation during the recording period during sleep, and TST90\% is the total sleep time with an oxygen saturation below 90\% (represented as percentage).

\begin{table}[!t]
    \renewcommand{\arraystretch}{1.2}
    \centering
    \begin{tabular}{l|l|c|c}
        \textbf{Demographic} & Age & 61 & (28 - 75) \\
        & Male Sex & 113 & 54.3\% \\
        \hline
        \textbf{Diabetes-related} & HbA1c & 7.2\% & (6.5\% - 11.2\%) \\
        & Biguanide use & 177 & 85.1\% \\
        & Sulfonylurea use & 70 & 33.7\% \\
        \hline\textbf{OSA-related} & ODI$_4$ & 14.9 & (2.0 - 128.6) \\
        & Mean Saturation & 92.7\% & (74.2\% - 96.8\%) \\
        & TST90\% & 8.4\% & (0.2\% - 98.9\%) \\ 
    \end{tabular}
    \caption{Selected descriptive statistics of the $n=207$ patients used for analysis. Values are the medians (min-max) for continuous covariates or size (\%) for binary covariates.}\label{tab:description}
\end{table}

\subsection{Research questions and statistical challenges}

Our primary scientific goal is to investigate the effects of measures of OSA and hyperglycemic medications on glucose distributions after accounting for other covariates. Fréchet regression with variable selection \citep{Tucker:2023} provides a methodological framework for selecting covariates that affect empirical quantile functions of glucose; however, the dimensions of HYPNOS data present computational challenges for the existing algorithm. Consequently, characterizing the uncertainty in variable selection within this framework is infeasible. Our methodological goals are to develop a fast algorithm for Fréchet regression with variable selection, and characterize variable selection uncertainty by evaluating the stability of selected covariates across subsamples \citep{Meinshausen:2010}.

\section{Distributional regression via Fréchet regression} \label{sec:frechet}

\subsection{Fréchet regression} \label{sec:globalfrechet}

\citet{Petersen:2019} propose Fréchet regression by reformulating the conditional mean in linear regression to make underlying Euclidean geometry explicit. Given a random $Y \in \R$ and a realization $\mathbf{x}_*$ of $X \in \R^p$, the conditional mean model in linear regression is
$
    \Ex(Y|X = \mathbf{x}_*) = \beta_0 + \mathbf{x}^{\top}_*\bbeta,
$
where $\bbeta\in \R^p$ is a vector of coefficients. Consider an iid sample $\{(\mathbf{x}_i , y_i) \in \R^p \times \R : i \in [n] \}$, where $\mathbf{y} := ( y_1 \ \hdots \ y_n)^{\top}\in \R^n$, $\bX := ( \mathbf{x}_1 \ \hdots \ \mathbf{x}_n )^{\top}\in \R^{n\times p}$ and we assume $\bX$ is column-centered. \citet{Petersen:2019} show that least squares estimation from this sample leads to an estimator $\haty_*$ of the conditional mean which satisfies
\begin{equation}\label{eq:linear}
 \haty_* :=\argmin_{y \in \R} \sum_{i=1}^n s_i(\mathbf{x}_*) \cdot (y - y_i)^2, \qquad s_i(\mathbf{x}_*) = \frac{1}{n} + \mathbf{x}_*^{\top}( \bX^{\top} \bX)^- \mathbf{x}_i.
\end{equation}

\citet{Petersen:2019} propose to replace the Euclidean space $(\R, d_E^2)$ in~\eqref{eq:linear}—that is, response $y \in \R$ with distance $d_E^2(y, y_i) = (y- y_i)^2$—with a general metric space $(\Omega, d^2)$, leading to the Fréchet regression
\begin{equation}\label{eq:Frechet}
 \haty_* :=\argmin_{y \in \Omega} \sum_{i=1}^n s_i(\mathbf{x}_*) \cdot d^2(y, y_i), \qquad s_i(\mathbf{x}_*) = \frac{1}{n} + \mathbf{x}_*^{\top}( \bX^{\top} \bX)^{-1} \mathbf{x}_i.
\end{equation}
We consider the case of Fréchet regression where $\Omega$ is the space of univariate quantile functions $\mathbf{Q} = \{\mathbf{q} : [0, 1] \mapsto \R\}$, equipped with the 2-Wasserstein distance $d_W^2$, leading to
\begin{align}\label{eq:2wass}
    \widehat{\mathbf{q}}_* = \argmin_{\mathbf{q} \in \mathbf{Q}} \sum_{i=1}^n s_i(\mathbf{x}_*) \cdot d^2_W(\mathbf{q}, \mathbf{y}_i), \qquad d^2_W(\mathbf{q}, \: \mathbf{p}) = \int_0^1 \left\{ \mathbf{q}(u) - \mathbf{p}(u)\right\}^2 du \quad \forall \mathbf{q}, \mathbf{p}\in \mathbf{Q}.
\end{align}
In practice, we consider responses that are empirical quantile functions evaluated on the same uniformly spaced $m$-sequence in $(0, 1)$, i.e. $\bY = (\mathbf{y}_1 \ \hdots \ \mathbf{y}_n)^{\top} \in \R^{n\times m}$ and $y_{i,j} \leq y_{i,j'}$ for $j \leq j'$,
leading to the formulation
\begin{align}\label{eq:qfEstimationOne}
 \widehat{\mathbf{q}}_* = \argmin_{\mathbf{q} \in \R^m} \sum_{i=1}^n s_i(\mathbf{x}_*) \cdot \oldnorm{\mathbf{q} - \mathbf{y}_i}_2^2,\quad\mbox{subject to}\quad \mathbf{b} - \bA^{\top}\mathbf{q} \leq \bzero.
\end{align}
The constraints $\mathbf{b} - \bA^{\top}\mathbf{q} \leq \bzero$ enforce monotonicity of $\widehat{\mathbf{q}}$, as well as adherence to specified distribution support via box constraints $b_L \leq q_j \leq b_U$. The latter is necessary in our context as Dexcom CGM only records values in a $[40, 400]$ mg/dL range.

It is often necessary to evaluate predicted quantile functions $\widehat{\mathbf{q}}_*$ at multiple $\mathbf{x}_*$, e.g., evaluate predictions on the training data. Below we show these predictions can be evaluated concurrently in a quadratic optimization problem; the proof is in Supplement~\ref{sup:proofs}.

\begin{theorem}\label{prop:FrobeniusNormMinimization}
Let $\bX \in \R^{n\times p}$ be a column-centered covariate matrix, let $\bX_* \in \R^{d\times p}$ consist of $d$ covariate vectors $\mathbf{x}_* \in \R^p$ centered the same as $\bX$, and let $\bY \in \R^{n\times m}$ row-wise consist of $n$ empirical quantile functions $\{\mathbf{y}_i\}$ evaluated on a shared, uniformly dispersed $m$-grid over $(0, 1)$.  Then for each row $\mathbf{x}_*$ in $\bX_*$, the corresponding  $\widehat{\mathbf{q}}_*$ in \eqref{eq:qfEstimationOne} satisfies
\begin{align*}
    \widehat{\mathbf{q}}_* = \argmin_{\mathbf{q} \in \R^m} \frac{1}{2}\oldnorm{\mathbf{q} - \widehat{\mathbf{y}}_*}_2^2,\quad\mbox{subject to}\quad \mathbf{b} - \bA^{\top}\mathbf{q} \leq \bzero,
\end{align*}
where $\widehat{\mathbf{y}}_*^{\top} = \{n^{-1}\bone_n^{\top} + \mathbf{x}_*^{\top}(\bX^{\top} \bX)^- \bX^{\top}\}\bY$. Furthermore, the matrix $\bhatQ_* \equiv \bhatQ(\bX_*) \in \R^{d\times m}$, which row-wise consists of $\widehat{\mathbf{q}}_*$'s, satisfies
\begin{align}\label{eq:bhatQ_noLambda}
    \bhatQ_* = \argmin_{\bQ \in \R^{d\times n}} \frac{1}{2}\oldnorm{\bQ - \bhatY_*}_F^2,\quad\mbox{subject to}\quad \bB - \bQ\bA \leq \bzero,
\end{align}
where the rows of $\bB$ are each $\mathbf{b} $, and $\bhatY_* := \{n^{-1}\bone_{d\times n} + \bX_*(\bX^{\top} \bX)^- \bX^{\top}\}\bY$.
\end{theorem}

\subsection{Variable Selection in Fréchet Regression} \label{sec:frechetvarselect}

\citet{Tucker:2023} extend Fréchet regression \citep{Petersen:2019} to allow for variable selection.  They introduce (what we call) an ``allowance vector" $\blambda \in \bTau := \{ \blambda \in \R^p : \blambda^{\top}\bone_p = \tau,\: \blambda \geq \bzero\}$ whose non-negative entries add to ``total allowance" $\tau > 0$, with zero entries corresponding to eliminated variables. \citet{Tucker:2023} replace $\bhatY$ in Proposition~\ref{prop:FrobeniusNormMinimization} with $\bhatY(\blambda) = n^{-1}\{ \bone_{n\times n} + \bX( n^{-1}\bX^{\top} \bX + \bD_{\blambda^{-1}})^{-1} \bX^{\top} \} \bY$, which involves a generalized ridge penalty.  This is an idea based on \citet{Wu:2021}, who shows that generalized ridge regression can perform variable selection via tuning parameters inverse-proportional to elements of $\blambda$. To prevent division by zero, $\bhatY(\blambda)$ can be rewritten
\begin{align}\label{eq:hatY}
    \bhatY(\blambda) &= n^{-1}\bone_{n\times n}\bY + \bY - (\btilX \bD_{\blambda} \btilX^{\top} + \bI_n )^{-1} \bY,
\end{align}
and the best $\bhatlambda(\tau)$ is found as
\begin{align}
    \bhatlambda(\tau) &:= \argmin_{\blambda \in \bTau} \frac{1}{2}\oldnorm{\bhatQ(\blambda) - \bY}_F^2, \label{eq:friso}\\
    \bhatQ(\blambda) &:= \argmin_{\bQ \in \R^{n\times m}} \frac{1}{2}\oldnorm{\bQ - \bhatY(\blambda)}_F^2, \quad\mbox{subject to}\quad \bB - \bQ\bA \leq \bzero. \label{eq:embeddedProblem}
\end{align}
We refer to~\eqref{eq:embeddedProblem} as the embedded problem and to~\eqref{eq:friso} as the sparsity problem.  A collection of solution paths $\{ \hatlambda_k(\tau) : \tau \in (0, \tau_{\mathrm{max}}] \}_{k=1}^p$ can be obtained by varying $\tau$ over a user-specified range.

\subsection{Review of existing algorithms} \label{sec:existing_algorithm}

The existing algorithm to solve the embedded problem~\eqref{eq:embeddedProblem} is the \verb+quadprog::solve.QP+ \textsf{R} function \citep{Turlach:2019}, a wrapper for a Fortran implementation of the dual-active set method of \citet{Goldfarb:1983} for convex quadratic programming problems. However, the algorithm can only be applied to a single vector $\widehat{\mathbf{q}}_i$ at a time, requiring an outer loop structure which slows the computations. 

The existing algorithm to solve the sparsity problem~\eqref{eq:friso} is a modified coordinate descent algorithm implemented in \textsf{R} language. The complete algorithm is described in Supplement~\ref{sup:algorithms}. A single \tharg{t} iteration comprises a sequential loop through all coordinates $k \in [p]$. Inside the loop, the current $\blambda^{(t)}$ is moved along the simplex-spanning line segment connecting it to the $k^{\text{th}}$ corner of the simplex $\bTau$.   The optimal step size is determined using \verb+stats::optimize+ function in \textsf{R}, which implements a golden-section line search method. The latter only requires evaluating the objective function value of~\eqref{eq:friso} at a given $\blambda$, making the algorithm sufficiently flexible for Fréchet regression on general metric spaces $(\Omega, d^2)$. However, the overall algorithm is very slow, as only one coordinate of $\blambda$ is updated at a time, and the update does not take advantage of the gradient of the objective function.

\section{Proposed method}
\label{sec:proposed_methods}

In this section, we develop a new fast algorithm for~\eqref{eq:friso}.  First, we develop a dedicated algorithm for the embedded problem~\eqref{eq:embeddedProblem} that allows us to avoid matrix multiplications (Section~\ref{sec:dedicated_solver_embedded}).  Second, we derive the gradient and Hessian of the objective function for the sparsity problem (Section~\ref{sec:gradients}). Finally, we utilize the gradient information to develop an unconstrained geodesic descent algorithm (Section~\ref{sec:GSD}).

\subsection{Dedicated solver for embedded problem}\label{sec:dedicated_solver_embedded}

In this section, we outline our proposed algorithm for solving the embedded problem \eqref{eq:embeddedProblem}. Specifically, we propose to perform projected gradient ascent on the Lagrange multipliers for the associated dual problem. The advantage of this approach is that it allows us to directly solve for the whole matrix $\bQ\in \R^{n\times m}$ (rather than work with one $\mathbf{q}_i\in \R^{m}$ at a time), and take advantage of the special structure of matrix $\bA$ in constraints to make efficient matrix multiplications.

Let $\bhatY = \bhatY(\blambda)$, and consider $\calL(\bQ, \bEta) = \frac{1}{2}\oldnorm{\bQ - \bhatY}_F^2 + \tr\left\{ (\bB - \bQ\bA)\bEta^{\top}\right\}$, the Lagrangian for~\eqref{eq:embeddedProblem}, where $\bEta \geq \bzero_{n\times (m + 1)}$ is the matrix of associated Lagrange multipliers. Given the step size $\alpha > 0$, the update has the form
$
\bEta^{(t+1)} = \{ \bEta^{(t)} +\alpha \nabla h(\bEta^{(t)}) \}_+,
$
where $\nabla h(\bEta^{(t)})$ is the gradient of the dual function $h(\bEta) = \inf_{\bQ}\calL(\bQ, \bEta)$ evaluated at $\bEta^{(t)}$. By taking the gradients of the Lagrangian, we derive a closed-form update
\begin{align*}
    \bEta^{(t+1)} &= \left[ \alpha \{ \bEta^{(t)}(\alpha^{-1}\bI_m - \bA^{\top}\bA) + (\bB - \bhatY\bA) \} \right]_+ ,
\end{align*}
with final $\bQ^{(t+1)} = \bhatY +  \bEta^{(t+1)}\bA^{\top}$. We use $\alpha = 1/2$ in our implementation; thus each update requires evaluation of the matrix product $\bEta^{(t)}(2\bI_m - \bA^{\top}\bA)$. We show that this operation can be reduced to an addition operation by taking advantage of the specific structure of $\bA$ matrix. See Supplement~\ref{sup:fastEtaGradient} for full derivations.

\subsection{Gradient and Hessian}\label{sec:gradients}

In this section, we present the gradient and Hessian of the objective function in \eqref{eq:friso}.  We utilize the KKT conditions of the embedded optimizer $\bhatQ$ to derive the gradient, from which the Hessian easily follows.  Derivations are in Supplement~\ref{sup:proofs}.

\begin{theorem}\label{prop:gradient_Hessian_lambda} Let $\bhatQ=\bhatQ(\blambda)$ from~\eqref{eq:embeddedProblem}. For each $i \in [n]$, let $\bA = [\begin{matrix} \bA_{0,i}&\bA_{-,i}\end{matrix}]$ be the block decomposition of $\bA$ into those columns corresponding to the active constraints on $\widehat{\mathbf{q}}_i$ (i.e. $\bA_{0,i}$) and those columns corresponding to inactive constraints (i.e. $\bA_{-,i}$), with $\proj{\bA_{0,i}}$ the orthogonal projector onto $\bA_{0,i}$.  If no constraints are active, let $\proj{\bA_{0,i}} = \bzero$.  Let $f(\blambda) := \oldnorm{\bhatQ - \bY}_F^2/2$ be the objective function in~\eqref{eq:friso}, and let $\bG := (\btilX \bD_{\blambda}\btilX^{\top} + \bI_n)^{-1}$ with $\btilX = \bX/\sqrt{n}$. Then
the gradient and Hessian of $f(\blambda)$ are respectively
\begin{align}\label{eq:gradient_and_N}
    \nabla f(\blambda) &= \diag(\bN),\ \mbox{where}\quad\bN := \btilX^{\top}\bG \Big\{ \sum_{i=1}^n \underline{\mathbf{e}}_i\underline{\mathbf{e}}_i^{\top}(\bhatY - \bY)(\bI_m - \proj{\bA_{0,i}}) \Big\} \bY^{\top}\bG\btilX; \\
    \begin{split}
        \nabla^2 f(\blambda) &= (\btilX^{\top}\bG\bY\bY^{\top}\bG\btilX) \circ (\btilX^{\top}\bG^2 \btilX) - (\btilX^{\top}\bG\btilX) \circ (\bN + \bN^{\top}) \: - \\
        &\phe \quad\sum_{i=1}^n (\btilX^{\top} \bG \bY \proj{\bA_{0,i}} \bY^{\top} \bG \bX) \circ (\btilX^{\top}\bG \underline{\mathbf{e}}_i\underline{\mathbf{e}}_i^{\top} \bG \btilX).
    \end{split}\label{eq:Hessian_lambda}
\end{align}
\end{theorem}

\subsection{Geodesic descent algorithm}\label{sec:GSD}

We propose to utilize the obtained gradient and Hessian (Proposition~\ref{prop:gradient_Hessian_lambda}) in creating a faster algorithm for solving sparsity problem~\eqref{eq:friso}. Specifically, we develop a second-order geodesic descent algorithm by (i) changing the constraint set in~\eqref{eq:friso} from simplex to sphere via a change of variables (Section~\ref{sec:simplex_to_sphere}); (ii) deriving constraint-preserving rotation steps along geodesics on the sphere via rotations (Section~\ref{sec:gradients_geodesics}); (iii) choosing the optimal rotation angle at each step using the local Hessian of the reparameterized function (Section~\ref{sec:GSD_angle_selection}).

\subsubsection{Simplex to spherical constraint}\label{sec:simplex_to_sphere}

We propose to reparameterize the simplex constraints in \eqref{eq:friso} into spherical constraints, allowing unconstrained updates (Section~\ref{sec:gradients_geodesics}). Let $\calF : \calS_{\sqrt{\tau}} \mapsto \bTau$ map the $\sqrt{\tau}$-radius sphere to the simplex by $\calF(\bgamma) = \bgamma \circ \bgamma$. Then for $\blambda \in \bTau$, $\calF^{\leftarrow}(\blambda) = \{\bgamma \in \calS_{\sqrt{\tau}} : \bgamma \circ \bgamma = \blambda\}$ defines an equivalence class in $\calS_{\sqrt{\tau}}$ leading to reparameterized sparsity problem \eqref{eq:friso} in terms of $\bgamma$
\begin{align}\label{eq:friso_gamma}
    \bhatlambda = \argmin_{\blambda \in \bTau} f (\blambda)\quad \to \quad \bhatgamma = \argmin_{\bgamma \in \calS_{\sqrt{\tau}}} (f \smallcirc \calF)(\bgamma),\quad \bhatgamma \in \calF^{\leftarrow}(\bhatlambda).
\end{align}
We propose performing gradient descent on \eqref{eq:friso_gamma} instead of \eqref{eq:friso}. We next derive the gradient and Hessian of the objective function with respect to $\bgamma$, $g(\bgamma) = (f \smallcirc \calF)(\bgamma)$, by taking advantage of Proposition~\ref{prop:gradient_Hessian_lambda} and the chain rule. The full derivations are in Supplement~\ref{sup:proofs}.

\begin{theorem}\label{prop:gradient_Hessian_gamma}
    Let $\blambda = \calF(\bgamma) = \bgamma \circ \bgamma$, and define composite function $g(\bgamma) := (f \smallcirc \calF)(\bgamma)$ as in~\eqref{eq:friso_gamma}.  The gradient and Hessian of $g(\bgamma)$ are respectively
    \begin{align}
        \nabla g(\bgamma) &= 2\bgamma \circ (\nabla f \smallcirc \calF)(\bgamma); \label{eq:gradient_gamma}\\
        \nabla^2 g(\bgamma) &= 2 \bD_{(\nabla f \smallcirc \calF)(\bgamma)} + 4(\bgamma\bgamma^{\top}) \circ (\nabla^2 f \smallcirc \calF)(\bgamma),\label{eq:Hessian_gamma}
    \end{align}
    where $(\nabla f \smallcirc \calF)(\bgamma) \equiv \nabla f(\blambda)$ is from \eqref{eq:gradient_and_N} and $(\nabla^2 f \smallcirc \calF)(\bgamma) \equiv \nabla^2 f(\blambda)$ is \eqref{eq:Hessian_lambda}.
\end{theorem}

\subsubsection{Geodesic updates}\label{sec:gradients_geodesics}

Given the starting $\bgamma^{(0)}\in \calS_{\sqrt{\tau}}$, we propose to utilize constraint-preserving optimization updates by conducting steps along geodesics in the sphere in the direction against the gradient. Specifically, the updates take the form
\begin{align}\label{eq:rotation}
    \bgamma^{(t+1)} = \bR^{(t)}\bgamma^{(t)}
\end{align}
where rotations $\bR^{(t)}$ on $\calS_{\sqrt{\tau}}$ are geodesic steps, in contrast to linear steps in $\R^p$ that may not be constraint-preserving. To rotate against the gradient, we decompose the gradient at the current $\bgamma^{(t)}$ as the sum of tangent and normal components to $\calS_{\sqrt{\tau}}$ at $\bgamma^{(t)}$:
\begin{align*}
    \nabla g(\bgamma^{(t)}) = (\bI_p - \proj{\bgamma^{(t)}})\nabla g(\bgamma^{(t)}) + \proj{\bgamma^{(t)}} \nabla g(\bgamma^{(t)}).
\end{align*}
For brevity, we define
\begin{align}\label{eq:v}
    \mathbf{v}^{(t)} = -(\bI_p - \proj{\bgamma^{(t)}})\nabla g(\bgamma^{(t)}),
\end{align}
which is orthogonal to $\bgamma^{(t)}$.  The rotation between orthonormal vectors $\underline{\bgamma}^{(t)} \rightarrow \underline{\mathbf{v}}^{(t)}$ by angle $\theta^{(t)}$ while fixing the orthogonal complement of $\text{span}(\bgamma^{(t)}, \mathbf{v}^{(t)})$ is
\begin{align}\label{eq:gamma_update}
    \bgamma^{(t+1)} = \bR^{(t)}\bgamma^{(t)} = \cos(\theta^{(t)})\bgamma^{(t)} + \sqrt{\tau} \sin(\theta^{(t)})\underline{\mathbf{v}}^{(t)}
\end{align}
(proof is given in Supplement~\ref{sup:lemmas}).  By construction, $\bgamma^{(t+1)}$ is guaranteed to be feasible for~\eqref{eq:friso_gamma}. The remaining part in executing~\eqref{eq:gamma_update} is choosing a step size, that is, choosing the rotation angle $\theta^{(t)} > 0$.

\subsubsection{Selecting the rotation angle}\label{sec:GSD_angle_selection}

To choose the rotation angle $\theta^{(t)}$ in~\eqref{eq:gamma_update}, we modify the optimized second-order Taylor expansion of $g(\bgamma)$ around the circle. Specifically, we let  
\begin{align}\label{eq:theta}
    \theta^{(t)} = \min\left\{ \abs{ \frac{\frac{d}{d\theta}g_*^{(t)}(0)}{\frac{d^2}{d\theta^2}g_*^{(t)}(0)} }, \: \theta_{\text{max}} \right\},
\end{align}
where $\theta_{\text{max}} = \pi / 4$, and $g_*^{(t)}(\theta) := (g \smallcirc \bR^{(t)})(\bgamma^{(t)})$.
Further, we derive the closed form of the first and second total derivatives of the composite function $g_*(\theta)$ with respect to $\theta$, evaluated at $\theta = 0$.  Derivations are given in Supplements~\ref{sup:proofs}; see also Supplement~\ref{sup:Implementation}.

\begin{theorem}\label{prop:theta_derivatives} Let $\bgamma \in \calS_{\sqrt{\tau}}$ and assume $\mathbf{v} \neq \bzero$. The first and second total derivatives of $g_*(\theta) := (g \smallcirc \bR_{\theta})(\bgamma)$, evaluated at $\theta = 0$, are respectively
\begin{align}
  g'_*(0) &= -\sqrt{\tau} \cdot \norm{\mathbf{v}}_2, \label{eq:theta_first_derivative} \\
  g''_*(0) &= \tau \cdot \underline{\mathbf{v}}^{\top}[\nabla^2 g(\bgamma)]\underline{\mathbf{v}} - \bgamma^{\top}\nabla g(\bgamma). \label{eq:theta_second_derivative}
\end{align}
\end{theorem}

\subsubsection{Summary}\label{sec:summary}
In summary, we solve~\eqref{eq:friso} by recasting the problem as~\eqref{eq:friso_gamma}.  We use iterative rotations of $\bgamma^{(t)}$ via~\eqref{eq:gamma_update} with the angle $\theta^{(t)}$ set as in~\eqref{eq:theta}.
We perform these rotations until a norm convergence condition $\oldnorm{\bgamma^{(t+1)}-\bgamma^{(t)}}_{\infty} \leq \varepsilon$ is met. We use $\bhatgamma$ at convergence to set $\bhatlambda = \bhatgamma \circ \bhatgamma$, and select variables based on corresponding nonzero elements of $\bhatlambda$. We implemented our algorithm in \textsf{C++} and made it available in \textsf{R} using \textsf{Rcpp} interface \citep{eddelbuettelRcppSeamlessIntegration2011}. We use algebraically equivalent implementations to avoid large square matrices in e.g. the $p \gg n$ or $n \gg p$ cases; see Supplement~\ref{sup:Implementation} for details.

\section{Stability selection}\label{sec:stability_selection}
The selected model for~\eqref{eq:friso} depends on $\tau$. \citet{Tucker:2023} propose $K$-fold cross-validation with refitting to select $\tau$ which minimizes out-of-sample error.  This method has limitations: the results of cross-validation may be split-dependent; the final set of selected variables must correspond to one of the models on the solution path of~\eqref{eq:friso}, which may not contain the true model; and cross-validation for variable selection comes with no general theoretical guarantees. In sparse linear regression, stability selection is an alternative variable selection approach that overcomes these limitations of cross-validation \citep{Meinshausen:2010}. At a high level, stability selection consists of taking repeated subsamples of the full data, performing variable selection for each, and then measuring selection frequency across subsamples. The final model is determined by selecting variables with the highest selection frequency, with theoretical guarantees provided on the number of false discoveries as a function of the selection threshold. Our new proposed geodesic descent algorithm for \eqref{eq:friso} is fast, in contrast to the original algorithm, making computationally demanding subsampling-based methods for variable selection feasible. We adopt the complementary pairs stability selection \citep{Shah:2013}, which leads to tighter error bounds compared to \citet{Meinshausen:2010}.

 At each subsampling step $b \in [B]$, let $\calI_b, \calI_b^c \subset [n]$ be randomly drawn complementary index sets of size $\floor{n/2}$ (half of the samples) i.e. $\calI_b \cap \calI_b^c = \emptyset$.
Let $(\bX^{\calI_b}, \bY^{\calI_b})$ and $(\bX^{\calI_b^c}, \bY^{\calI_b^c})$ denote the corresponding subsets of the full $(\bX, \bY)$. Given $\tau > 0$, the selected variables for each subset are determined based on non-zero elements of $\bhatlambda$ from~\eqref{eq:friso}, that is
$$
\hatS_b(\tau) := \{k : \hatlambda_k(\tau; \bX^{\calI_b}, \bY^{\calI_b}) > 0 \}, \qquad \hatS_b^c(\tau) := \{k : \hatlambda_k(\tau; \bX^{\calI_b^c}, \bY^{\calI_b^c}) > 0 \}.
$$
The stability measure of the \tharg{k} variable is defined as its empirical selection probability across all subsamples and complementary subsets
\begin{align}\label{eq:stability}
    \hatPi_B(k; \tau) := \frac1{2B}\sum_{b=1}^B \left[ \ind\{k \in \hatS_b(\tau) \} + \ind\{k \in \hatS_b^c(\tau) \} \right].
\end{align}
Higher $\hatPi_B(k; \tau)$ means more frequent selection of the \tharg{k} variable, giving more evidence for its importance. A selected variable set is defined with only those variables that have high selection frequency, i.e. $\hatS(\tau) := \{k : \hatPi_B(k; \tau) \geq \pi_{\mathrm{thr}}\}$ for some threshold $0 < \pi_{\mathrm{thr}} \leq 1$. The value of $\pi_{\mathrm{thr}}$ can be chosen to control the variable selection error, as we describe below.

\citet{Shah:2013} derive several bounds on variable selection error with complementary pairs stability selection, which rely on individual selection probabilities of the underlying base procedure (in our case, selection based on non-zero elements in $\bhatlambda$ from~\eqref{eq:friso}). The bounds control the expected number of variables in $\hatS(\tau)$ that have low selection probability under the base procedure without making any assumptions on the base procedure itself. These bound can be interpreted in terms of falsely selected variables under additional assumptions that the base procedure is no worse than random guessing, and that the selection of noise variables is exchangeable. Here we adopt the tightest bound, which is obtained under mild shape restrictions on the distribution of the proportion of times a variable is selected ($r$-concavity, we refer to \citet{Shah:2013} for details).

\begin{theorem}[Bound (8) from \citet{Shah:2013}]\label{prop:shah} Let $\phi\in (0, 1)$, and let $L_{\phi} \subset [p]$ be the set of variables with low selection probability (below $\phi$) under base procedure~\eqref{eq:friso}. Let $\hatS(\tau)$ be the set of variables selected with complementary pairs stability selection using $B$ subsamples and threshold $\pi_{\mathrm{thr}}$. Then, under the assumption of $r$-concavity,
\begin{align}\label{eq:shah_bound}
    E_L(\tau) = \Ex\oldabs{\hatS(\tau) \cap L_{\phi}} \leq d(B, \pi_{\mathrm{thr}}, \phi)|L_{\phi}|.
\end{align}
\end{theorem}
In~\eqref{eq:shah_bound}, $d(B, \pi_{\mathrm{thr}}, \phi)$ is a known function that depends on $B$, $\pi_{\mathrm{thr}}$ and $\phi$; and $|L_{\phi}|$ is the cardinality of $L_{\phi}$. \citet{Shah:2013} recommend choosing $B=50$ to satisfy assumption of $r$-concavity, and setting $\phi = q/p$, where $q \equiv q(\tau)$ is the expected model size from~\eqref{eq:friso} with a given $\tau$ and sample size $\floor{n/2}$. The latter choice implies that the variables with low selection probability have a selection frequency at most as high as random guessing, which makes them noise variables when the base procedure is no worse than random guessing. Since the number of noise variables is unknown, in practice bound~\eqref{eq:shah_bound} is applied with $|L_{\phi}|$ replaced by $p$. To estimate $q$, we sum $\hatPi_B(k; \tau)$ across all variables $k$.
Finally, the value of  $\pi_{\mathrm{thr}}(\tau)$ is selected to control the bound at a fixed level, i.e.,  such that $E_L(\tau) \leq K$ for a given $K$. The higher is $\pi_{\mathrm{thr}}(\tau)$, the lower is the bound.

In simulations (Section~\ref{sec:variable_selection_performance}), we evaluate how tight is the bound using $K=1$ and $K=2$. Since $\hatPi_B(k; \tau)$—and consequently, $q$ and selected $\pi_{\mathrm{thr}}$—depend on $\tau$, we consider how $\hatPi_B(k; \tau)$ change across a range of $\tau$ values (stability paths), and choose those variables whose stability paths exceed $\pi_{\mathrm{thr}}(\tau)$ at least once over the path. Since large values of $\tau$ in~\eqref{eq:friso} lead to all variables being selected, in practice we restrict the range of $\tau$ values so that the selected model size is at most $2/3$ of the total variables, that is $q(\tau)/p\leq 2/3$.

\section{Simulation Studies} \label{sec:simulation}

\subsection{Algorithmic performance}\label{sec:sim_algo}
In this section, we evaluate the performance of our proposed geodesic second-order descent (GSD) algorithm and compare it to existing modified coordinate descent (MCD) algorithm of \citet{Tucker:2023}. For MCD, we use the default implementation in \textsf{R} provided as supplement to \citet{Tucker:2023}. For GSD, we use our implementation with $\varepsilon = 0.0075$ (Section~\ref{sec:summary}), chosen to ensure comparable or favorable convergence accuracy against default MCD implementation. In each experiment, we evaluate the total wall clock time (in seconds), and objective function values of~\eqref{eq:friso} at convergence. We consider two settings for synthetic data generation, which we refer to Experiments A and B, described below. 

\textbf{Experiment A.} We replicate Example 5.2.2 from \citet{Tucker:2023}. The random quantile function is generated conditionally on $\mathbf{x}\in \R^p$ as $\mathbf{q}(\:\cdot\:) = \mu_{\mathbf{x}} + \sigma_{\mathbf{x}} \Phi^{-1}(\:\cdot\:)$, with $\mu_{\mathbf{x}} \sim \calN(\mu_0 + \beta(x_2 + x_3), \nu_1)$ and $\sigma_{\mathbf{x}} \sim \mathrm{Gamma}((\sigma_0 + \kappa x_1)^2/\nu_2, \: \nu_2/(\sigma_0 + \kappa x_1))$. Hyper-parameters are as used in \citet{Tucker:2023}. Only the first three covariates affect $\mathbf{q}$.

\textbf{Experiment B.} We consider a less favorable setting for GSD, where the constraints in~\eqref{eq:embeddedProblem} are more likely to be active, requiring more computationally demanding evaluations of $\proj{\bA_{0,i}}$ for~\eqref{sec:gradients}. We use the zero-inflated negative binomial distribution with pmf $p(z) = \alpha_{\mathbf{x}}\ind\{z=0\} + (1 - \alpha_{\mathbf{x}})p_{Z|\mathbf{x}}(z)$, where $Z|{\mathbf{x}} \sim \text{nbinom}(r_{\mathbf{x}}, \pi_{\mathbf{x}})$. The random quantile function is generated conditionally on $\mathbf{x}\in \R^p$ as $\mathbf{q}(\:\cdot\:) = F_{Z|\mathbf{x}}^{-1}\{ ( \:\cdot\:-\alpha_{\mathbf{x}})/(1 - \alpha_{\mathbf{x}})\}$. We set
$\text{logit}(\alpha_{\mathbf{x}}) \sim \calN(\mu_{\alpha} + \beta_{\alpha} x_4, \nu_{\alpha})$, $\text{logit}(\pi_{\mathbf{x}}) \sim \calN(\mu_{\pi} + \beta_{\pi} x_3, \nu_{\pi})$, $
\log(r_{\mathbf{x}}) \sim \calN(\mu_r + \beta_r (x_1 + x_2), \nu_r).$
We set $\mu_{\alpha} = \text{logit}(0.2)$, $\beta_{\alpha} = 0.4$, and $\nu_{\alpha} = 0.1^2$; $\mu_{\pi} = \text{logit}(0.5)$, $\beta_{\pi} = 0.1$, and $\nu_{\pi} = 0.15^2$; and $\mu_r = \log(10)$, $\beta_r = 0.2$, and $\nu_r = 0.15^2$.  Only the first four covariates affect $\mathbf{q}$.

\begin{figure}[!t]
    \centering
    \includegraphics[width=5.5in]{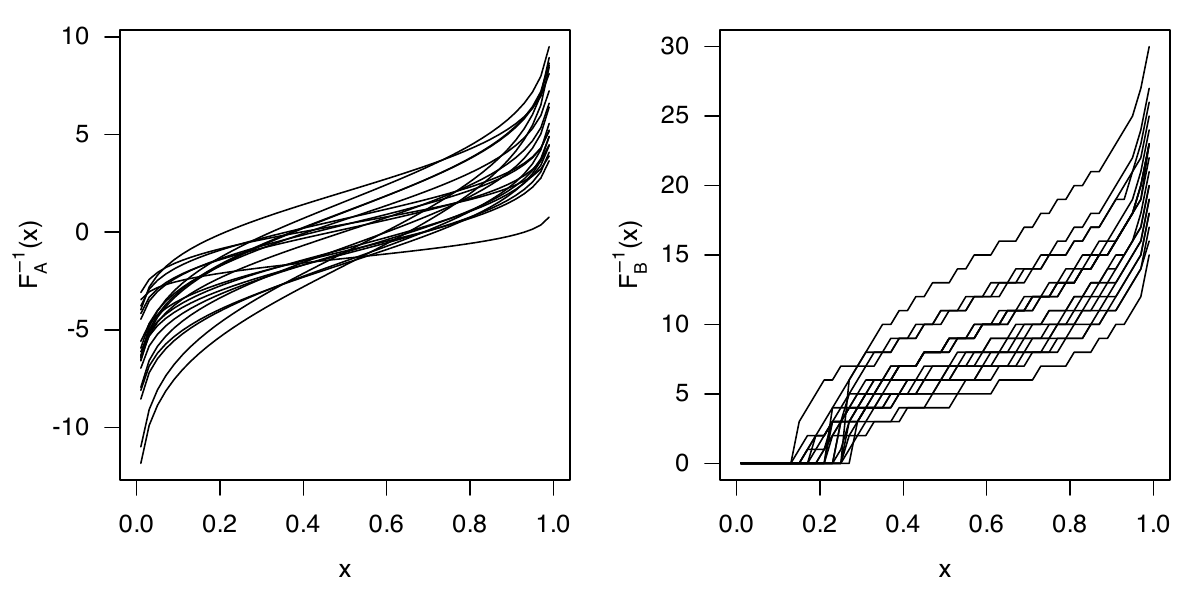}
    \caption{Example empirical quantile functions from Experiment A (\textit{left}) and Experiment B (\textit{right}) using $m=50$ discretization.}
    \label{fig:example_EQFs}
\end{figure}

Figure \ref{fig:example_EQFs} illustrates 20 example discretized quantile functions from each experiment.

\begin{figure}[!t]
    \centering
    \includegraphics[width=5.5in, keepaspectratio]{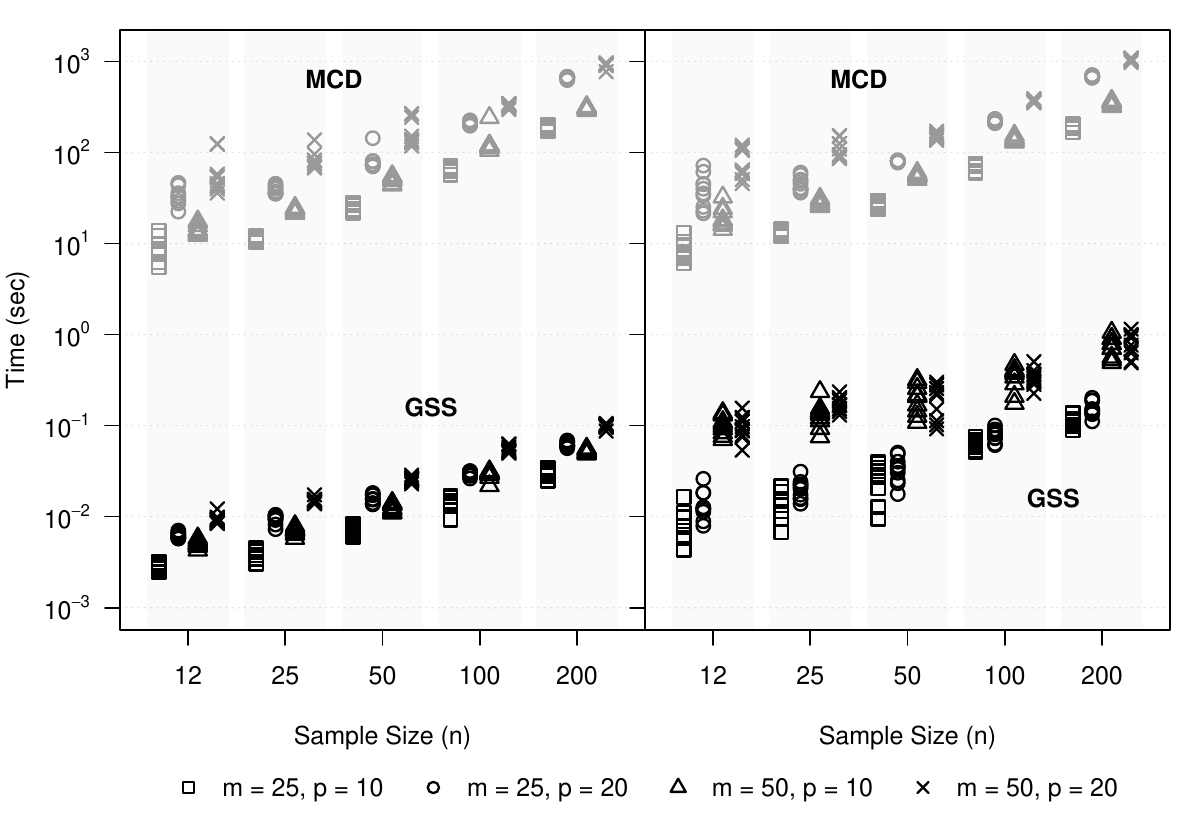}
    \caption{Sparsity algorithm wall clock times on $\tau \in [20]$, Experiment A (\textit{left}) and Experiment B (\textit{right}).  MCD times in grey, GSD times in black.}
    \label{fig:speed}
\end{figure}

\textbf{Run Times.} Figure~\ref{fig:speed} compares wall clock times for MCD and GSD to solve the sparsity problem~\eqref{eq:friso}.  For each combination of parameter settings $n \in \{12, 25, 50, 100, 200\}$, $p \in \{10, 20\}$, and $m \in \{25, 50\}$, we generated ten data sets from Experiment A and Experiment B each.  Solution paths $\blambda(\tau \in [20])$ were generated independently with MCD and GSD with initial allowance vector $\blambda^{(0)}(\tau = 20) = (\tau/p)\bone_p$.  Wall clock times were estimated for one replication each with \textsf{R}'s \verb+microbenchmark+ package \citep{Mersmann:2019} on a Mac OS system with Apple M1 Max chip. While our implementation of GSD has an advantage over MCD due to the underlying code being in C++ rather than R, the conclusions from timing comparisons still hold true with our earlier version of algorithm in R. As expected, the higher $n$, $p$, and $m$, the more difficult the problem.  In Experiment B, larger $m$ generally means more instances of active lower box constraints on $\bhatQ$, lending to more frequent and expensive evaluations of $\proj{\bA_{0,i}}$ in~\eqref{eq:gradient_and_N}.  Such a problem does not occur for MCD, which does not evaluate the gradient.  Despite this, the GSD algorithm is several orders of magnitude faster, and has more favorable scaling with $n$ and $p$ than MCD.

\begin{figure}[!t]
    \centering
    \includegraphics[width=5.5in, keepaspectratio]{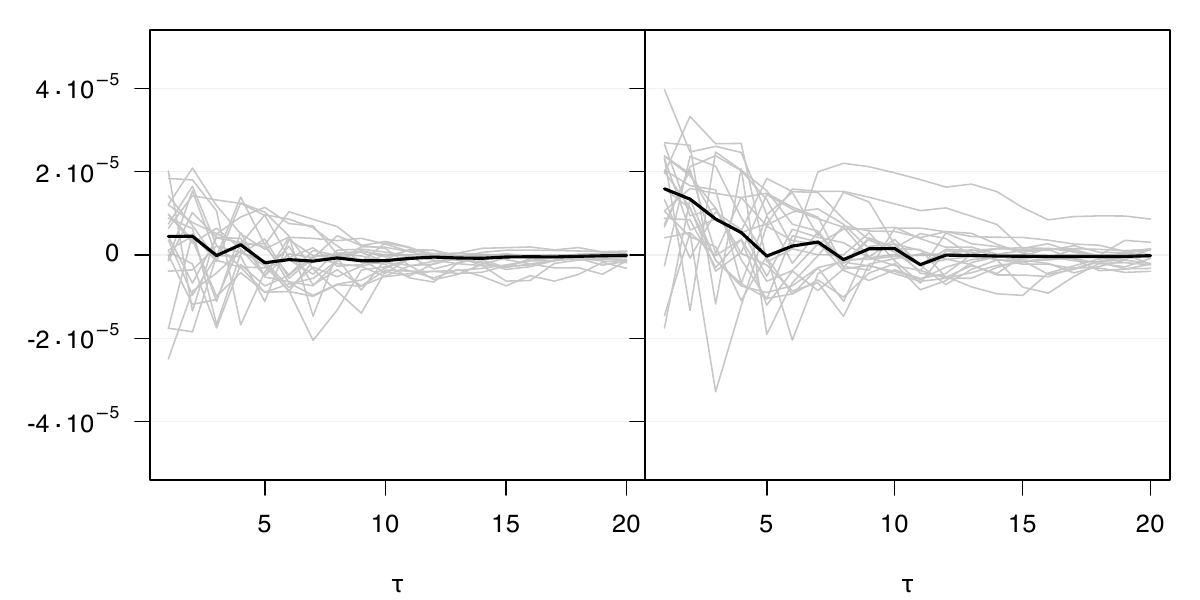}
    \caption{Sparsity algorithm optimization accuracy comparison on 21 synthetic data sets from Experiment A (\textit{left}) and Experiment B (\textit{right}).  Gray lines are $\log_{10} \{f(\blambda^{(t)}_{\text{MCD}})/f(\blambda^{(t)}_{\text{GSD}})\}$, evaluated on $\tau \in \{1, \hdots, 20\}$ for each data set; the median values per $\tau$ across the respective 21 data sets are given by thick black lines.}
    \label{fig:Accuracy}
\end{figure}

\textbf{Accuracy.} Figure \ref{fig:Accuracy} compares optimization accuracy between MCD and GSD across $\tau \in [20]$, fixing $n = 50$, $p = 10$, and $m = 50$.  Within each experiment setting, the two algorithms were pairwise compared using 21 replications. For each replication, we plot $\log_{10} \frac{f(\blambda^{(t)}_{\text{MCD}})}{f(\blambda^{(t)}_{\text{GSD}})}$ against $\tau$, so that positive values indicate GSD out-performing MCD.  The two algorithms have comparable accuracy in both simulation settings.

\subsection{Variable selection performance}\label{sec:variable_selection_performance}

\begin{table}[!t]
    \renewcommand{\arraystretch}{1.2}
    \centering
    \begin{tabular}{l l | c c c| c c c}
        \multicolumn{2}{c}{\multirow{2}{*}{}} & \multicolumn{3}{c}{$\hatP[\text{true selection}]$} & \multicolumn{3}{c}{$\widehat{\Ex}\oldabs{\text{false selections}}$} \\
        \multicolumn{2}{c}{} & $\text{CV}_{K=10}$ & $\text{SS}_{E\leq 1\:}$ & $\text{SS}_{E\leq 2\:}$ & $\text{CV}_{K=10}$ & $\text{SS}_{E\leq 1\:}$ & $\text{SS}_{E\leq 2\:}$ \\ \cline{1-8}
        $p = 10$ & $n = 50$  & $0.83$ & $0.75$ & $0.82$ & $0.92$ & $0.09$ & $0.17$ \\
                 & $n = 100$ & $0.98$ & $0.98$ & $0.98$ & $0.49$ & $0.08$ & $0.13$ \\
                 & $n = 200$ & $1.00$ & $1.00$ & $1.00$ & $0.30$ & $0.07$ & $0.09$ \\
        $p = 30$ & $n = 50$  & $0.78$ & $0.74$ & $0.82$ & $1.34$ & $0.35$ & $0.91$ \\
                 & $n = 100$ & $0.93$ & $0.96$ & $0.98$ & $0.45$ & $0.37$ & $0.83$ \\
                 & $n = 200$ & $1.00$ & $1.00$ & $1.00$ & $0.24$ & $0.35$ & $0.82$ \\
    \end{tabular}
    \caption{Variable selection performance comparison between 10-fold cross validation ($\text{CV}_{K=10}$), and complementary pairs stability selection with bound~\eqref{eq:shah_bound} controlled at  $\{1, 2\}$ ($\text{SS}_{E\leq 1\:}$, $\text{SS}_{E\leq 2\:}$).  $\hatP[\text{true selection}]$ is estimated power (probability of true variable being selected), and $\widehat{\Ex}\oldabs{\text{false selections}}$ is the average number of falsely selected variables. All results are averaged over $100$ replications of Experiment A.}
    \label{tab:cv_cpss_comparison}
\end{table}

In this section, we compare the variable selection performance of $10$-fold cross-validation against complementary pairs stability selection as described in Section~\ref{sec:stability_selection}. We consider Experiment A with $n \in \{50, 100, 200\}$, $p \in \{10, 30\}$, and $m = 50$, with $100$ replicates for each setting.  We use a fine grid $\tau_{\mathrm{CV}} \in \{10, 9.9, \hdots, 0.1\}$ for cross-validation; and coarse grid $\tau_{\mathrm{SS}} \in \{10, 9.5, \hdots, 0.5\}$ for stability selection. To avoid edge cases with stability selection (all variables being selected with sufficiently large $\tau$), we only consider stability paths over the range of $\tau$ that lead to model size at most $2/3$ of the total $p$, that is $q(\tau)/p\leq 2/3$. For stability selection, we calculate minimal $\pi_{\mathrm{thr}}(\tau_{\mathrm{SS}})$ for each $\tau_{\mathrm{SS}}$ to control~\eqref{eq:shah_bound} as $E_L(\tau_{\mathrm{SS}}) \leq 1$ and $E_L(\tau_{\mathrm{SS}}) \leq 2$, respectively. 

We evaluate $\hatP[\text{true selection}]$, the average proportion of the true variable being selected (power, averaged across the first three true variables), and $\widehat{\Ex}\oldabs{\text{false selections}}$, average number of falsely selected variables. Table~\ref{tab:cv_cpss_comparison} summarizes the results for each method. Stability selection provides comparable or improved power compared to cross-validation, while false discovery comparison depends on $p$. As expected, the increase in bound for stability selection from $E_L(\tau) \leq 1$ to $E_L(\tau) \leq 2$ increases power when $n=50$. The number of actual false discoveries is well-within the bound in both cases, confirming the observation in~\citet{Shah:2013} that the bound is conservative and not tight, especially when $p$ is relatively small; even with the more lenient $E_L(\tau) \leq 2$ bound, stability selection has vlow number of expected false discoveries ($< 1$).

\section{Analysis of HYPNOS data}\label{sec:dataAnalysis}

In this section, we utilize Fréchet regression to analyze the HYPNOS data described in Section~\ref{sec:data}. We apply the proposed geodesic descent algorithm to solve~\eqref{eq:friso}, and combine it with complementary pairs stability selection described in Section~\ref{sec:stability_selection} to perform variable selection inference. The entire data have empirical quantile functions of $n=207$ patients stored at $m=100$ quantile discretization, with $p=34$ covariates for each patient. As Dexcom G4 CGM measures glucose values in $[40, 400]$ mg/dL range, the embedded problem~\eqref{eq:embeddedProblem} is solved with box constraints (in addition to monotonicity constraints).
The overarching goal of the analysis is to elucidate which patients' characteristics affect their glucose distribution and quantify in which way, with a specific focus on the effect of potentially modifiable factors (medications and measures of OSA severity).

\textbf{Covariate correlation for variables related to oxygen saturation:} Before applying stability selection, we calculate pairwise correlations across all $p = 34$ covariates. High correlations (positive or negative) may present issues for stability selection as has been demonstrated by \citet{Faletto:2022} in the context of sparse linear regression with LASSO. In the presence of high correlations, variable selection becomes unstable, with stability selection ``splitting the vote" across the variables in the correlated group, leading to low individual stability measures and, consequently, missed selections of true variables. In the HYPNOS data, we found that most covariates have low correlations, except for the five variables associated with measurements of oxygen saturation overnight, which form a correlated cluster: (1) ODI$_4$, (2) mean saturation, (3) minimum saturation, (4) standard deviation of oxygen saturation, and (5) TST90\%.  To circumvent the correlation issue, we perform principal component analysis on these five variables, and use the extracted principal components as covariates instead of original variables, still leading to $p=34$.  Supplement~\ref{sup:hypnos} provides additional details.

\textbf{Computation time:} We calculated stability measures~\eqref{eq:stability} for each covariate using $B = 50$ complementary pairs on each of $\tau \in \{0.5, 1, \hdots, 20\}$ leading to a total of $2000$ evaluations of~\eqref{eq:friso_gamma}. Computationally, using convergence tolerance $\varepsilon = 10^{-5}$ (Section~\ref{sec:summary}), this analysis took 4 minutes total using the proposed geodesic descent algorithm.  Based on comparisons in simulations (Figure~\ref{fig:speed}), such analysis would not be feasible with the original algorithm of \citet{Tucker:2023}.

\begin{figure}[!t]
    \centering
    \includegraphics[width=5.5in, keepaspectratio]{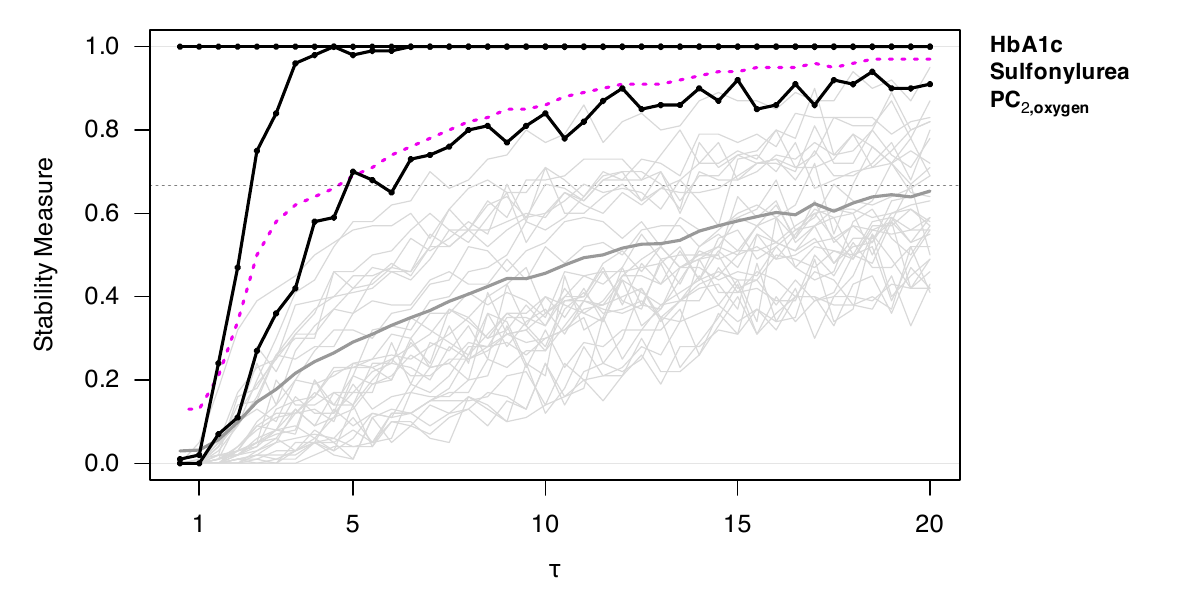}
    \caption{HYPNOS data set stability paths~\eqref{eq:stability} from $B = 50$ complementary pairs subsets at each $\tau \in \{0.5, 1, \hdots, 20\}$. Highlighted are the stability threshold $\pi_{\mathrm{thr}}(\tau)$ guaranteeing $E_L(\tau) \leq 2$ (bold magenta line); variables selected by the ``any vote" method (bold black lines), i.e. paths which exceed $\pi_{\mathrm{thr}}(\tau)$ for some $\tau$; and average relative model size $\hatq(\tau)/p$ (bold gray line).  A horizontal dotted line at $2/3$ is plotted for reference.}
    \label{fig:HYPNOS_stability}
\end{figure}

\textbf{Stability selection results:} Figure~\ref{fig:HYPNOS_stability} displays stability paths for each covariate, with the $\pi_{\mathrm{thr}}(\tau)$ path guaranteeing $E_L(\tau) \leq 2$ (see Section~\ref{sec:stability_selection}).  We used this bound based on simulation results in Section~\ref{sec:simulation}, to maximize power while maintaining acceptable error bounds. For comparison, we also include estimated relative model size $\hatq(\tau)/p$, which peaks just below $2/3$ at $\tau = 20$.  The final selected variable set includes three covariates: HbA1c, sulfonylurea medication, and the $2^{\mathrm{nd}}$ principal component for oxygen saturation variables ($\text{PC}_{2,\text{oxygen}}$). The solution paths of three other covariates come within $0.1$ of the $\pi_{\mathrm{thr}}(\tau)$ path, all indicators for use of a medication: non-steroidal anti-inflammatory drug (NSAID), H1 antagonist, and angiotensin-II receptor blocker (A2RB).

\begin{figure}[!t]
    \centering
    \includegraphics[width=\textwidth, keepaspectratio]{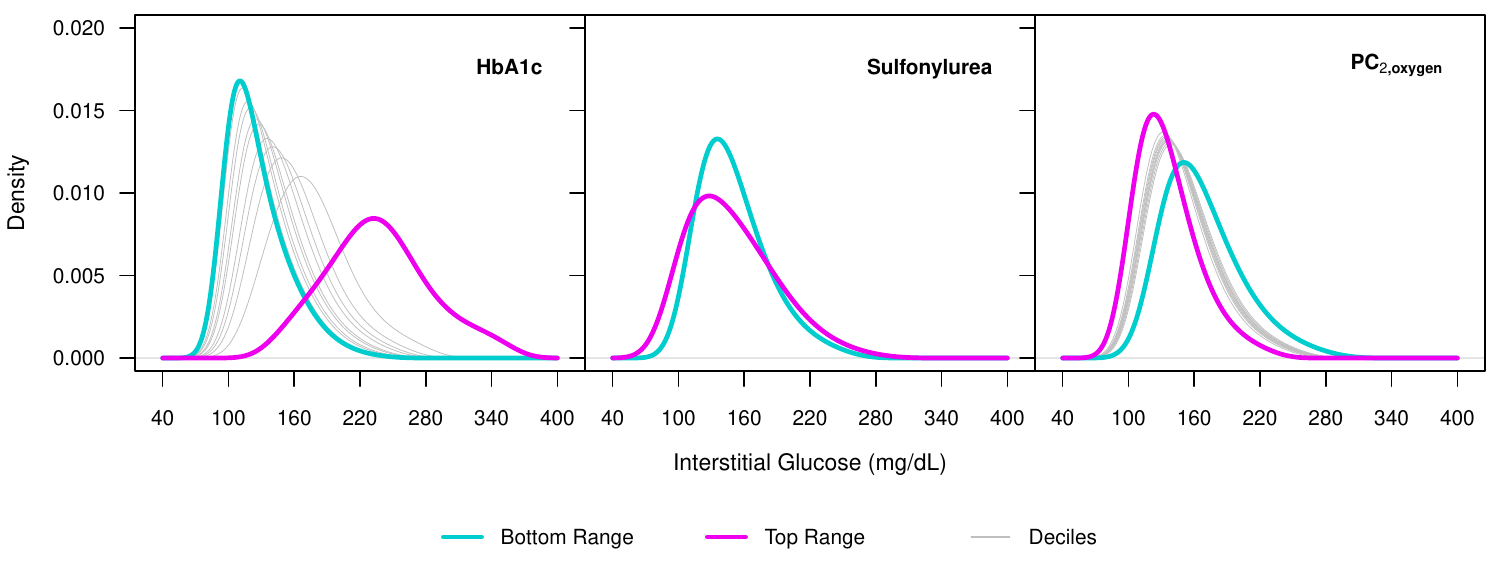}
    \caption{Predicted densities using Fréchet regression with 3 selected variables. In each panel, the levels of the corresponding covariate are varied from the lowest/``no" (bottom range, cyan) to the highest/``yes" (top range, magenta). Grey lines correspond to evaluations at 0.1 decile changes in corresponding variable. The remaining variables are kept constant, at sample mean for HbA1c, ``no" for sulfonylurea use, and zero for $\text{PC}_{2,\text{oxygen}}$.}
    \label{fig:marginal_plots}
\end{figure}

\textbf{Effects interpretation:} To elucidate how the selected variables affect the distribution of glucose values, we obtain a refitted model using~\eqref{eq:qfEstimationOne} with the selected three covariates. To investigate the sensitivity of inference to this selection, Supplement~\ref{sup:hypnos} provides the same analysis using the larger subset with NSAID, H1 antagonist, and A2RB indicators included; the interpretations of top three variables remain the same.  Figure~\ref{fig:marginal_plots} illustrates predicted densities obtained by individually varying each of the three covariates, while keeping the remaining two covariates at fixed levels (HbA1c $= 7.5\%$; sulfonylurea selected as ``no", median value; $\text{PC}_{2,\text{oxygen}} = 0$). 

For HbA1c, higher values correspond to higher glucose readings ($126.2 \rightarrow 236.7$ mg/dL mean change across the observed HbA1c range, i.e. from $6.5\%$ to $11.2\%$). This is an expected result, as HbA1c corresponds to the average glucose over the preceding 2-3 months. While this HbA1c measurement was obtained before CGM placement, there is a significant correlation in average glucose values in subsequent months in the absence of interventions \citep{huangCorrelationShortMidterm2021}, and thus the association with mean CGM values in the preceding two weeks is expected.  We found that higher HbA1c values were also predictive of higher variability ($28.9 \rightarrow 49.2$ mg/dL \textit{SD} change across the observed range).

For sulfonylurea, we find that the use of medication is associated with an increase in glucose variability ($33.2 \rightarrow 42.1$ mg/DL \textit{SD} change), but no mean change ($<1$ mg/dL). This is an unexpected result since sulfonylurea is a glucose-lowering medication that promotes insulin release from the pancreatic beta-cell; thus, we anticipated a lower mean value with medication use. However, in performing a literature search, we found that the same result was obtained when analyzing longitudinal HbA1c values of patients with type 2 diabetes in the ACCORD trial \citep{germanWiSERRobustScalable2022}. Since neither HYPNOS nor ACCORD trials were designed to evaluate the effect of sulfonylurea, a possible explanation is that the patients who were prescribed sulfonylurea had higher glucose mean values at baseline than the patients who were not, and that resulting increased insulin secretion led to lowering of the mean (matching the one of patients not on medication) while at the same time increasing variability (more common lower values due to insulin). Further studies would be required to validate this conclusion.

For $\text{PC}_{2,\text{oxygen}}$, we found that high values are associated with lower glucose levels ($167.3 \rightarrow 135.5 = -31.8$ mg/dL mean change across the observed $\text{PC}_{2,\text{oxygen}}$ range) with a decrease in variability ($37.1 \rightarrow 29.9$ mg/dL \textit{SD} change). To interpret this result in the context of original covariates, we consider the loadings associated with $\text{PC}_{2,\text{oxygen}}$, and create pairwise scatterplots of all five oxygen covariates, coloring the patients by low and high $\text{PC}_{2,\text{oxygen}}$ values (see Table~\ref{tab:loadings} and Figure~\ref{fig:oxygen_pairs} in Supplement~\ref{sup:hypnos}). We observe that low values of $\text{PC}_{2,\text{oxygen}}$ correspond to unstable oxygen saturation levels overnight, that is higher standard deviation of oxygen saturation overnight, coupled with more frequent or deep oxygen desaturation events. In contrast, patients with high values of $\text{PC}_{2,\text{oxygen}}$ (lower predicted glucose levels) tend to have lower overall saturation values and significantly lower saturation level variability overnight. Surprisingly, the $\text{PC}_{1,\text{oxygen}}$ covariate, which is reflective of the overall oxygen desaturation levels,  isn't close to the top variable set (rank $25$-$34$ of $34$ across $\tau$; $\max_{\tau}\hatPi(k; \tau) = 0.46$).

\section{Discussion}\label{sec:discussion}

In this work, we develop a new algorithm for fast distributional regression that is fast and accurate, making methodology computationally feasible on large datasets and enabling the use of subsampling-based methods (stability selection) for variable selection inference. We demonstrate the effectiveness of our approach on both synthetic data and in application to CGM data from the HYPNOS study. We anticipate that our method opens the door to applying sparse distributional regression to other large-scale datasets.

In the analysis of CGM data from the HYPNOS study, we identified a few important covariates. While the selection of HbA1c, a measure of pre-study glucose levels, is not surprising,  we found that sulfonylurea usage is associated with higher glucose variability without evidence of association with glucose mean, an association that regression methods on glucose summary statistics might fail to identify. Since the HYPNOS trial was not designed to evaluate the effect of sulfonylurea, further studies would be of interest to validate this association. We also found that overnight oxygen desaturation variability, as captured by the $2^{\text{nd}}$ principal component associated with oxygen saturation covariates, has a stronger association with glucose distributions than the overall oxygen desaturation levels, as captured by the 1st component. To the best of our knowledge, this result is new and also unexpected in light of the common practice of only using ODI$_{4}$ covariate (measuring the rate of oxygen desaturation events) to characterize OSA severity, including prior analyses on HYPNOS data \citep{auroraGlucoseProfilesObstructive2022, Sergazinov:2023}. Our findings suggest that the "stability" of saturation profiles overnight plays a crucial role in glucose regulation and that this "stability" is not fully captured by ODI$_{4}$ on its own. Subsequently, in future research, it would be of interest to better define the clinical implications of this association and validate it via replications in other studies. Finally, our analyses revealed potential associations between glycemic control and three other medication types (non-steroidal anti-inflammatory drug, H1 antagonist, and angiotensin-II receptor blocker), albeit their selection frequencies were slightly smaller than the stability selection threshold. These medications are prescribed for conditions other than diabetes and are thus not typically considered as part of the diabetes management plan. Our literature search indicated that prior research in animal models also indicated associations between these medications and glycemic control \citep{anvari:2015,chu:2006,mortazavi:2019}, providing potential corroboration for our results and warranting follow-up validation studies.

From the methodological perspective, the scalability of our method is primarily achieved due to newly obtained closed-form expressions of gradient and Hessian associated with the sparse distributional regression problem. As indicated by \citet{Tucker:2023}, characterization of the gradient is required for theoretical analysis of convergence guarantees of the selection operator~\eqref{eq:friso}, and we hope our work opens the door for such further analysis. Furthermore, while we pursued geodesic descent algorithm, other optimization methods can take advantage of closed-form gradient expressions, such as conditional gradient family methods \citep{Braun:2022}.  Finally, in developing our method, we discovered that the idea of replacing the simplex constraint with a spherical constraint has also been explored by~\citet{Li:2023} in other optimization settings, who also provide theoretical convergence guarantees when objective function is convex.

Direct application of this result to our case is not possible as objective function $f(\blambda)$ in~\eqref{eq:friso} can fail to be convex due to possibility of negative eigenvalues in Hessian in~\eqref{eq:Hessian_lambda}. While in practice our algorithm has always converged, a formal convergence analysis of the proposed algorithm and complete characterization of the stationary points of $f(\blambda)$ remain open problems for future research.

\subsection*{Funding}
This work was supported by the NSF DMS-2044823.

\subsection*{Supplementary materials} Detailed derivations, propositions proofs, and additional simulation and real data analysis results.

\bibliographystyle{agsm}
\bibliography{Bibliography_arXiv}

@article{auroraGlucoseProfilesObstructive2022,
	author = {Aurora, R. Nisha and Gaynanova, Irina and Patel, Pratik and Punjabi, Naresh M.},
	doi = {10.1016/j.sleep.2022.04.007},
	issn = {13899457},
	journal = {Sleep Medicine},
	langid = {english},
	month = jul,
	pages = {105--111},
	title = {Glucose Profiles in Obstructive Sleep Apnea and Type 2 Diabetes Mellitus},
	urldate = {2022-09-27},
	volume = {95},
	year = {2022}}

@article{battelinoContinuousGlucoseMonitoring2022,
	author = {Battelino, Tadej and Alexander, Charles M and Amiel, Stephanie A and {Arreaza-Rubin}, Guillermo and Beck, Roy W and Bergenstal, Richard M and Buckingham, Bruce A and Carroll, James and Ceriello, Antonio and Chow, Elaine and Choudhary, Pratik and Close, Kelly and Danne, Thomas and Dutta, Sanjoy and Gabbay, Robert and Garg, Satish and Heverly, Julie and Hirsch, Irl B and Kader, Tina and Kenney, Julia and Kovatchev, Boris and Laffel, Lori and Maahs, David and Mathieu, Chantal and Mauricio, D{\'\i}dac and Nimri, Revital and Nishimura, Rimei and Scharf, Mauro and Del Prato, Stefano and Renard, Eric and Rosenstock, Julio and Saboo, Banshi and Ueki, Kohjiro and Umpierrez, Guillermo E and Weinzimer, Stuart A and Phillip, Moshe},
	doi = {10.1016/S2213-8587(22)00319-9},
	issn = {22138587},
	journal = {The Lancet Diabetes \& Endocrinology},
	langid = {english},
	month = dec,
	pages = {S2213858722003199},
	shorttitle = {Continuous Glucose Monitoring and Metrics for Clinical Trials},
	title = {Continuous Glucose Monitoring and Metrics for Clinical Trials: An International Consensus Statement},
	urldate = {2022-12-09},
	year = {2022}}

@article{eddelbuettelRcppSeamlessIntegration2011,
	author = {Eddelbuettel, Dirk and Francois, Romain},
	copyright = {Copyright (c) 2010 Dirk Eddelbuettel, Romain Francois},
	doi = {10.18637/jss.v040.i08},
	issn = {1548-7660},
	journal = {Journal of Statistical Software},
	langid = {english},
	month = apr,
	pages = {1--18},
	shorttitle = {Rcpp},
	title = {Rcpp: {{Seamless R}} and {{C}}++ {{Integration}}},
	urldate = {2023-01-31},
	volume = {40},
	year = {2011}}

@article{gaynanovaDigitalBiomarkersGlucose2022,
	author = {Gaynanova, Irina},
	journal = {Biopharmaceutical Report},
	langid = {english},
	number = {1},
	pages = {21--26},
	title = {Digital Biomarkers of Glucose Control - Reproducibility Challenges and Opportunities},
	volume = {29},
	year = {2022}}

@article{germanWiSERRobustScalable2022,
	author = {German, Christopher A. and Sinsheimer, Janet S. and Zhou, Jin and Zhou, Hua},
	copyright = {{\textcopyright} 2021 The International Biometric Society.},
	doi = {10.1111/biom.13506},
	issn = {1541-0420},
	journal = {Biometrics},
	langid = {english},
	number = {4},
	pages = {1313--1327},
	shorttitle = {{{WiSER}}},
	title = {{{WiSER}}: {{Robust}} and Scalable Estimation and Inference of within-Subject Variances from Intensive Longitudinal Data},
	urldate = {2023-12-05},
	volume = {78},
	year = {2022}}

@article{ghosalDistributionalDataAnalysis2021,
	author = {Ghosal, Rahul and Varma, Vijay R and Volfson, Dmitri and Hillel, Inbar and Urbanek, Jacek and Hausdorff, Jeffrey M and Watts, Amber and Zipunnikov, Vadim},
	doi = {10.1093/biostatistics/kxab041},
	issn = {1465-4644, 1468-4357},
	journal = {Biostatistics},
	langid = {english},
	month = nov,
	pages = {kxab041},
	title = {Distributional Data Analysis via Quantile Functions and Its Application to Modeling Digital Biomarkers of Gait in {{Alzheimer}}'s {{Disease}}},
	urldate = {2023-05-11},
	year = {2021}}

@article{huangCorrelationShortMidterm2021,
	author = {Huang, Jen-Hung and Lin, Yung-Kuo and Lee, Ting-Wei and Liu, Han-Wen and Chien, Yu-Mei and Hsueh, Yu-Chun and Lee, Ting-I and Chen, Yi-Jen},
	doi = {10.1186/s13098-021-00714-8},
	issn = {1758-5996},
	journal = {Diabetology \& Metabolic Syndrome},
	month = sep,
	number = {1},
	pages = {94},
	title = {Correlation between Short- and Mid-Term Hemoglobin {{A1c}} and Glycemic Control Determined by Continuous Glucose Monitoring},
	urldate = {2023-12-05},
	volume = {13},
	year = {2021}}

@article{petersenModelingProbabilityDensity2022,
	author = {Petersen, Alexander and Zhang, Chao and Kokoszka, Piotr},
	doi = {10.1016/j.ecosta.2021.04.004},
	issn = {24523062},
	journal = {Econometrics and Statistics},
	langid = {english},
	month = jan,
	pages = {159--178},
	title = {Modeling {{Probability Density Functions}} as {{Data Objects}}},
	urldate = {2023-05-11},
	volume = {21},
	year = {2022}}

@article{wangFunctionalDataAnalysis2016a,
	author = {Wang, Jane-Ling and Chiou, Jeng-Min and M{\"u}ller, Hans-Georg},
	doi = {10.1146/annurev-statistics-041715-033624},
	issn = {2326-8298, 2326-831X},
	journal = {Annual Review of Statistics and Its Application},
	langid = {english},
	month = jun,
	number = {1},
	pages = {257--295},
	title = {Functional {{Data Analysis}}},
	urldate = {2023-05-26},
	volume = {3},
	year = {2016}}

@article{Tucker:2023,
	author = {Tucker, Danielle C. and Wu, Yichao and M\"{u}ller, Hans-Georg},
	doi = {10.1080/01621459.2021.1969240},
	issn = {1537-274X},
	journal = {Journal of the American Statistical Association},
	month = sep,
	number = {542},
	pages = {1023--1037},
	publisher = {Informa UK Limited},
	title = {Variable Selection for Global Fr{\'e}chet Regression},
	volume = {118},
	year = {2023}}

@article{Li:2023,
	author = {Li, Qiuwei and McKenzie, Daniel and Yin, Wotao},
	doi = {10.1093/imaiai/iaad017},
	eprint = {https://academic.oup.com/imaiai/article-pdf/12/3/1898/51602970/iaad017.pdf},
	issn = {2049-8772},
	journal = {Information and Inference: A Journal of the IMA},
	month = {06},
	number = {3},
	pages = {1898-1937},
	title = {{From the simplex to the sphere: faster constrained optimization using the Hadamard parametrization}},
	volume = {12},
	year = {2023}}

@misc{Turlach:2019,
	author = {Berwin A. Turlach and Andreas Weingessel and Cleve Moler},
	note = {R package version 1.5-8},
	title = {quadprog: Functions to Solve Quadratic Programming Problems},
	url = {https://CRAN.R-project.org/package=quadprog},
	year = {2019}}

@article{Sergazinov:2023,
	author = {Sergazinov, Renat and Leroux, Andrew and Cui, Erjia and Crainiceanu, Ciprian and Aurora, R Nisha and Punjabi, Naresh M and Gaynanova, Irina},
	copyright = {http://creativecommons.org/licenses/by-nc/4.0/},
	journal = {Biometrics},
	language = {en},
	month = may,
	number = {4},
	pages = {3873-3882},
	publisher = {Wiley},
	title = {A case study of glucose levels during sleep using multilevel fast function on scalar regression inference},
	volume = {79},
	year = 2023}

@article{Ahmad:2023,
	author = {Ahmad, Farida B. and Cisewski, Jodi A. and Xu, Jiaquan and Anderson, Robert N.},
	doi = {10.15585/mmwr.mm7218a3},
	issn = {1545-861X},
	journal = {MMWR. Morbidity and Mortality Weekly Report},
	month = {May},
	number = {18},
	pages = {488--492},
	publisher = {Centers for Disease Control MMWR Office},
	title = {Provisional Mortality Data --- United States, 2022},
	volume = {72},
	year = {2023}}

@article{Matabuena:2023,
	author = {Matabuena, Marcos and Petersen, Alexander},
	doi = {10.1093/jrsssc/qlad007},
	eprint = {https://academic.oup.com/jrsssc/article-pdf/72/2/294/54801470/qlad007.pdf},
	issn = {0035-9254},
	journal = {Journal of the Royal Statistical Society Series C: Applied Statistics},
	month = {02},
	number = {2},
	pages = {294-313},
	title = {{Distributional data analysis of accelerometer data from the NHANES database using nonparametric survey regression models}},
	volume = {72},
	year = {2023}}

@misc{Mersmann:2019,
	author = {Olaf Mersmann},
	note = {R package version 1.4.10},
	title = {microbenchmark: Accurate Timing Functions},
	url = {https://CRAN.R-project.org/package=microbenchmark},
	year = {2023}}

@article{Shah:2013,
	author = {Rajen D. Shah and Richard J. Samworth},
	issn = {13697412, 14679868},
	journal = {Journal of the Royal Statistical Society. Series B (Statistical Methodology)},
	number = {1},
	pages = {55--80},
	publisher = {Wiley},
	title = {Variable selection with error control: another look at stability selection},
	urldate = {2023-11-13},
	volume = {75},
	year = {2013}}

@misc{Faletto:2022,
	archiveprefix = {arXiv},
	author = {Gregory Faletto and Jacob Bien},
	eprint = {2201.00494},
	note = {preprint, arXiv:2201.00494 [stat.ME].},
	primaryclass = {stat.ME},
	title = {Cluster Stability Selection},
	year = {2022}}

@article{Goldfarb:1983,
	author = {D. Goldfarb and A. Idnani},
	doi = {10.1007/bf02591962},
	journal = {Mathematical Programming},
	month = sep,
	number = {1},
	pages = {1--33},
	publisher = {Springer Science and Business Media {LLC}},
	title = {A numerically stable dual method for solving strictly convex quadratic programs},
	volume = {27},
	year = {1983}}

@misc{Braun:2022,
	archiveprefix = {arXiv},
	author = {G{\'a}bor Braun and Alejandro Carderera and Cyrille W. Combettes and Hamed Hassani and Amin Karbasi and Aryan Mokhtari and Sebastian Pokutta},
	eprint = {2211.14103},
	note = {preprint, arXiv:2211.14103 [math.OC].},
	primaryclass = {math.OC},
	title = {Conditional Gradient Methods},
	year = {2022}}

@article{Law:2015,
	author = {Law, Graham R and Ellison, George T H and Secher, Anna L and Damm, Peter and Mathiesen, Elisabeth R and Temple, Rosemary and Murphy, Helen R and Scott, Eleanor M},
	copyright = {http://www.diabetesjournals.org/site/license},
	journal = {Diabetes Care},
	language = {en},
	month = jul,
	number = 7,
	pages = {1319--1325},
	publisher = {American Diabetes Association},
	title = {Analysis of continuous glucose monitoring in pregnant women with diabetes: Distinct temporal patterns of glucose associated with large-for-gestational-age infants},
	volume = 38,
	year = 2015}

@article{Reutrakul:2017,
	author = {Reutrakul, Sirimon and Mokhlesi, Babak},
	journal = {Chest},
	language = {en},
	month = nov,
	number = 5,
	pages = {1070--1086},
	title = {Obstructive sleep apnea and diabetes: A state of the art review},
	volume = 152,
	year = 2017}

@article{Lindberg:2012,
	author = {Lindberg, Eva and Theorell-Hagl{\"o}w, Jenny and Svensson, Malin and Gislason, Thorarinn and Berne, Christian and Janson, Christer},
	journal = {Chest},
	language = {en},
	month = oct,
	number = 4,
	pages = {935--942},
	publisher = {Elsevier BV},
	title = {Sleep apnea and glucose metabolism: a long-term follow-up in a community-based sample},
	volume = 142,
	year = 2012}

@article{Punjabi:2002,
	author = {Punjabi, Naresh M and Sorkin, John D and Katzel, Leslie I and Goldberg, Andrew P and Schwartz, Alan R and Smith, Philip L},
	journal = {Am. J. Respir. Crit. Care Med.},
	language = {en},
	month = mar,
	number = 5,
	pages = {677--682},
	publisher = {American Thoracic Society},
	title = {Sleep-disordered breathing and insulin resistance in middle-aged and overweight men},
	volume = 165,
	year = 2002}

@article{Matabuena:2021,
	author = {Matabuena, Marcos and Petersen, Alexander and Vidal, Juan C and Gude, Francisco},
	journal = {Stat. Methods Med. Res.},
	language = {en},
	month = jun,
	number = 6,
	pages = {1445--1464},
	publisher = {SAGE Publications},
	title = {Glucodensities: A new representation of glucose profiles using distributional data analysis},
	volume = 30,
	year = 2021}

@article{Scott:2020,
	author = {Scott, Eleanor M and Feig, Denice S and Murphy, Helen R and Law, Graham R and {CONCEPTT Collaborative Group}},
	copyright = {https://www.diabetesjournals.org/content/license},
	journal = {Diabetes Care},
	language = {en},
	month = jun,
	number = 6,
	pages = {1178--1184},
	publisher = {American Diabetes Association},
	title = {Continuous Glucose Monitoring in pregnancy: Importance of analyzing temporal profiles to understand clinical outcomes},
	volume = 43,
	year = 2020}

@article{Gaynanova:2022,
	author = {Gaynanova, Irina and Punjabi, Naresh and Crainiceanu, Ciprian},
	copyright = {https://academic.oup.com/journals/pages/open\_access/funder\_policies/chorus/standard\_publication\_model},
	journal = {Biostatistics},
	language = {en},
	month = jan,
	number = 1,
	pages = {223--239},
	publisher = {Oxford University Press (OUP)},
	title = {Modeling continuous glucose monitoring ({CGM}) data during sleep},
	volume = 23,
	year = 2022}

@article{Lam:2010,
	author = {Lam, David C L and Lui, Macy M S and Lam, Jamie C M and Ong, Liza H Y and Lam, Karen S L and Ip, Mary S M},
	journal = {Chest},
	language = {en},
	month = nov,
	number = 5,
	pages = {1101--1107},
	publisher = {Elsevier BV},
	title = {Prevalence and recognition of obstructive sleep apnea in Chinese patients with type 2 diabetes mellitus},
	volume = 138,
	year = 2010}

@article{Foster:2009,
	author = {Foster, Gary D and Sanders, Mark H and Millman, Richard and Zammit, Gary and Borradaile, Kelley E and Newman, Anne B and Wadden, Thomas A and Kelley, David and Wing, Rena R and Sunyer, F Xavier Pi and Darcey, Valerie and Kuna, Samuel T and {Sleep AHEAD Research Group}},
	copyright = {http://creativecommons.org/licenses/by-nc-nd/3.0/},
	journal = {Diabetes Care},
	language = {en},
	month = jun,
	number = 6,
	pages = {1017--1019},
	publisher = {American Diabetes Association},
	title = {Obstructive sleep apnea among obese patients with type 2 diabetes},
	volume = 32,
	year = 2009}

@article{Singh:2021,
	author = {Singh, Ankita and Chaudhary, Shyam Chand and Gupta, Kamlesh K and Sawlani, Kamal K and Singh, Abhishek and Singh, Abhishek B and Verma, Ajay K},
	journal = {Ann. Afr. Med.},
	language = {en},
	month = jul,
	number = 3,
	pages = {206--211},
	title = {Prevalence of obstructive sleep apnea in diabetic patients},
	volume = 20,
	year = 2021}

@article{Rooney:2021,
	author = {Rooney, Mary R and Aurora, R Nisha and Wang, Dan and Selvin, Elizabeth and Punjabi, Naresh M},
	journal = {Contemp. Clin. Trials},
	language = {en},
	month = feb,
	number = 106248,
	pages = {106248},
	publisher = {Elsevier BV},
	title = {Rationale and design of the Hyperglycemic Profiles in Obstructive Sleep Apnea ({HYPNOS}) trial},
	volume = 101,
	year = 2021}

@incollection{Rosenquist:2018,
	author = {Klara J. Rosenquist and Caroline S. Fox},
	booktitle = {Diabetes in America},
	chapter = {36},
	edition = {3},
	editor = {Catherine C. Cowie and Sarah Stark Casagrande and Andy Menke and Michelle A. Cissell and Mark S. Eberhardt and James B. Meigs and Edward W. Gregg and William C. Knowler and Elizabeth Barrett-Connor and Dorothy J. Becker and Frederick L. Brancati and Edward J. Boyko and William H. Herman and Barbara V. Howard and K. M. Venkat Narayan and Marian Rewers and Judith E. Fradkin},
	pages = {36:1-14},
	publisher = {National Institute of Diabetes and Digestive and Kidney Diseases},
	title = {Mortality Trends in Type 2 Diabetes},
	year = {2018}}

@article{Jonas:2021,
	author = {Jonas, Daniel E and Crotty, Karen and Yun, Jonathan D Y and Middleton, Jennifer Cook and Feltner, Cynthia and Taylor-Phillips, Sian and Barclay, Colleen and Dotson, Andrea and Baker, Claire and Balio, Casey P and Voisin, Christiane E and Harris, Russell P},
	journal = {JAMA},
	language = {en},
	month = aug,
	number = 8,
	pages = {744--760},
	publisher = {American Medical Association (AMA)},
	title = {Screening for prediabetes and type 2 diabetes: Updated evidence report and systematic review for the {US} Preventive Services Task Force},
	volume = 326,
	year = 2021}

@article{Callaghan:2020,
	author = {Callaghan, Timothy and Ferdinand, Alva O and Akinlotan, Marvellous A and Towne, Jr, Samuel D and Bolin, Jane},
	copyright = {http://onlinelibrary.wiley.com/termsAndConditions\#vor},
	journal = {J. Rural Health},
	language = {en},
	month = jun,
	number = 3,
	pages = {410--415},
	publisher = {Wiley},
	title = {The changing landscape of diabetes mortality in the United States across region and rurality, 1999-2016},
	volume = 36,
	year = 2020}

@article{Kodl:2008,
	author = {Kodl, Christopher T and Seaquist, Elizabeth R},
	journal = {Endocr. Rev.},
	language = {en},
	month = jun,
	number = 4,
	pages = {494--511},
	publisher = {The Endocrine Society},
	title = {Cognitive dysfunction and diabetes mellitus},
	volume = 29,
	year = 2008}

@article{Moxey:2011,
	author = {Moxey, P W and Gogalniceanu, P and Hinchliffe, R J and Loftus, I M and Jones, K J and Thompson, M M and Holt, P J},
	journal = {Diabet. Med.},
	language = {en},
	month = oct,
	number = 10,
	pages = {1144--1153},
	publisher = {Wiley},
	title = {Lower extremity amputations--a review of global variability in incidence},
	volume = 28,
	year = 2011}

@article{Resnick:2002,
	author = {Resnick, Helaine E and Howard, Barbara V},
	journal = {Annu. Rev. Med.},
	language = {en},
	number = 1,
	pages = {245--267},
	publisher = {Annual Reviews},
	title = {Diabetes and cardiovascular disease},
	volume = 53,
	year = 2002}

@article{Sobrin:2011,
	author = {Sobrin, Lucia and Green, Todd and Sim, Xueling and Jensen, Richard A and Tai, E Shyong and Tay, Wan Ting and Wang, Jie Jin and Mitchell, Paul and Sandholm, Niina and Liu, Yiyuan and Hietala, Kustaa and Iyengar, Sudha K and {Family Investigation of Nephropathy and Diabetes-Eye Research Group} and Brooks, Matthew and Buraczynska, Monika and Van Zuydam, Natalie and Smith, Albert V and Gudnason, Vilmundur and Doney, Alex S F and Morris, Andrew D and Leese, Graham P and Palmer, Colin N A and {Wellcome Trust Case Control Consortium 2} and Swaroop, Anand and Taylor, Jr, Herman A and Wilson, James G and Penman, Alan and Chen, Ching J and Groop, Per-Henrik and Saw, Seang-Mei and Aung, Tin and Klein, Barbara E and Rotter, Jerome I and Siscovick, David S and Cotch, Mary Frances and Klein, Ronald and Daly, Mark J and Wong, Tien Y},
	journal = {Invest. Ophthalmol. Vis. Sci.},
	language = {en},
	month = sep,
	number = 10,
	pages = {7593--7602},
	title = {Candidate gene association study for diabetic retinopathy in persons with type 2 diabetes: the Candidate gene Association Resource ({CARe})},
	volume = 52,
	year = 2011}

@misc{CDC:2023,
	author = {{Centers for Disease Control and Prevention}},
	lastchecked = {2023-05-10},
	note = {Accessed 2023-05-10},
	title = {{National Diabetes Statistics Report website}},
	url = {https://www.cdc.gov/diabetes/data/statistics-report/index.html},
	year = {2023}}

@article{Meinshausen:2010,
	author = {Meinshausen, Nicolai and B{\"u}hlmann, Peter},
	journal = {Journal of the Royal Statistical Society: Series B (Statistical Methodology)},
	number = {4},
	pages = {417--473},
	publisher = {Wiley Online Library},
	title = {Stability selection},
	volume = {72},
	year = {2010}}

@article{Wu:2021,
	author = {Yichao Wu},
	doi = {10.1080/00401706.2020.1791254},
	eprint = {https://doi.org/10.1080/00401706.2020.1791254},
	journal = {Technometrics},
	number = {2},
	pages = {263-271},
	publisher = {Taylor & Francis},
	title = {Can't Ridge Regression Perform Variable Selection?},
	volume = {63},
	year = {2021}}

@article{Petersen:2019,
	author = {Alexander Petersen and Hans-Georg M{\"u}ller},
	doi = {10.1214/17-AOS1624},
	journal = {The Annals of Statistics},
	number = {2},
	pages = {691 -- 719},
	publisher = {Institute of Mathematical Statistics},
	title = {{Fr{\'e}chet regression for random objects with Euclidean predictors}},
	volume = {47},
	year = {2019}}

@article{doherty:2017,
	author = {Doherty, Aiden and Jackson, Dan and Hammerla, Nils and Pl{\"o}tz, Thomas and Olivier, Patrick and Granat, Malcolm H and White, Tom and Van Hees, Vincent T and Trenell, Michael I and Owen, Christoper G and others},
	journal = {PloS one},
	number = {2},
	pages = {e0169649},
	publisher = {Public Library of Science},
	title = {Large scale population assessment of physical activity using wrist worn accelerometers: the UK biobank study},
	volume = {12},
	year = {2017}}

@article{chu:2006,
	author = {Chu, Kwan Yi and Lau, Tung and Carlsson, Per-Ola and Leung, Po Sing},
	journal = {Diabetes},
	number = {2},
	pages = {367--374},
	publisher = {Am Diabetes Assoc},
	title = {Angiotensin II type 1 receptor blockade improves $\beta$-cell function and glucose tolerance in a mouse model of type 2 diabetes},
	volume = {55},
	year = {2006}}

@article{mortazavi:2019,
	author = {Mortazavi-Jahromi, Seyed Shahabeddin and Alizadeh, Shahab and Javanbakht, Mohammad Hassan and Mirshafiey, Abbas},
	journal = {Archives of physiology and biochemistry},
	number = {5},
	pages = {435--440},
	publisher = {Taylor \& Francis},
	title = {Anti-diabetic effect of $\beta$-D-mannuronic acid (M2000) as a novel NSAID with immunosuppressive property on insulin production, blood glucose, and inflammatory markers in the experimental diabetes model},
	volume = {125},
	year = {2019}}

@article{anvari:2015,
	author = {Anvari, Ebrahim and Wang, Xuan and Sandler, Stellan and Welsh, Nils},
	journal = {Upsala Journal of Medical Sciences},
	number = {1},
	pages = {40--46},
	publisher = {Taylor \& Francis},
	title = {The H1-receptor antagonist cetirizine ameliorates high-fat diet-induced glucose intolerance in male C57BL/6 mice, but not diabetes outcome in female non-obese diabetic (NOD) mice},
	volume = {120},
	year = {2015}}

\renewcommand{\thesection}{S\arabic{section}}
\renewcommand{\thetable}{S\arabic{table}}
\renewcommand{\thefigure}{S\arabic{figure}}
\def\theequation{S\arabic{equation}}
\renewcommand{\thelemma}{S.\arabic{lemma}}
\newcommand{\thecor}{S.\arabic{cor}}

\setcounter{section}{0}
\setcounter{equation}{0}
\setcounter{figure}{0}
\setcounter{table}{0}
\clearpage
\begingroup
\centering
\Large \textbf{SUPPLEMENTARY MATERIAL TO\\ ``Fast variable selection for distributional regression with application to continuous glucose monitoring data"}\\
\endgroup

\begin{abstract}
This Supplementary Material is structured as follows: Supplement~\ref{sup:lemmas} gives supporting lemmas and proofs; Supplement~\ref{sup:proofs} gives proofs of propositions in the main text; Supplement~\ref{sup:fastEtaGradient} describes in more detail how matrix multiplication can be avoided in the embedded problem algorithm; Supplement~\ref{sup:Implementation} describes our implementation of geodesic descent to efficiently scale in $n$ and $p$, via algebraic manipulation of the gradient objects to avoid handling matrices of size $n\times n$ (in the $n \gg p$ case) or size $p \times p$ (in the $p \gg n$ case); Supplement~\ref{sup:algorithms} details the algorithms used in this paper, including modified coordinate descent (MCD) of \citet{Tucker:2023}, our geodesic second-order descent (GSD) algorithm, and our KKT multipliers method for the embedded problem; Supplement~\ref{sup:hypnos} details the HYPNOS dataset we analyzed in this paper.
\end{abstract}

\section{Supplementary Lemmas}\label{sup:lemmas}

\begin{lemma}\label{lem:lambda_gamma_derivative}
    For functions $f(\lambda) : \R \to \R$ and $\calF(\gamma) = \gamma^2 : \R \to \R$, the derivative of the composite function $g(\gamma) := (f \smallcirc \calF)(\gamma)$ with respect to $\gamma$ is
    \begin{align*}
        \frac{d}{d\gamma} g(\gamma) = 2\gamma \cdot \frac{d}{d\lambda} f(\lambda).
    \end{align*}
\end{lemma}
\proof This follows immediately from the chain rule
\begin{align*}
    \frac{d}{d\gamma} g(\gamma) = \frac{d}{d\gamma} (f \smallcirc \calF)(\gamma) = \frac{d}{d\lambda} f(\lambda)\cdot\frac{d}{d\gamma} \calF(\gamma) = \frac{d}{d\gamma} \gamma^2 \cdot \frac{d}{d\lambda} f(\lambda) = 2\gamma \cdot \frac{d}{d\lambda} f(\lambda).\enskip\qed
\end{align*}

\begin{lemma}
    Let $\underline{\mathbf{u}}$ and $\underline{\mathbf{v}}$ be orthonormal vectors in $\R^{p}$, $p \geq 2$.  Then the unique matrix which rotates $\underline{\mathbf{u}} \rightarrow \underline{\mathbf{v}}$ by angle $\theta \in \R$, while fixing the orthogonal complement of $\mathrm{span}\{\underline{\mathbf{u}}, \underline{\mathbf{v}}\}$, is
    \begin{align}\label{eq:Rtheta}
        \bR_{\theta} = \bI_p + (\sin\theta)(\underline{\mathbf{v}}\underline{\mathbf{u}}^{\top} - \underline{\mathbf{u}}\underline{\mathbf{v}}^{\top}) + (\cos\theta - 1)(\underline{\mathbf{u}}\underline{\mathbf{u}}^{\top} + \underline{\mathbf{v}}\underline{\mathbf{v}}^{\top}).
    \end{align}
\end{lemma}
\proof We prove this in parts.
\begin{enumerate}[noitemsep, nolistsep]
    \item \textbf{$\bR_{\theta}$ is a proper rotation matrix:}\\
    A matrix $\bR$ is a proper rotation matrix iff $\bR^{\top} = \bR^{-1}$ and $\abs{\bR} = 1$.  We start with $\bR_{\theta}^{\top}\bR_{\theta}$
    \begin{align*}
        \bR_{\theta}^{\top}\bR_{\theta} &= \bR_{\theta}^{\top}\bI_p + (\sin\theta) \bR_{\theta}^{\top}(\underline{\mathbf{v}}\underline{\mathbf{u}}^{\top} - \underline{\mathbf{u}}\underline{\mathbf{v}}^{\top}) + (\cos\theta - 1) \bR_{\theta}^{\top}(\underline{\mathbf{u}}\underline{\mathbf{u}}^{\top} + \underline{\mathbf{v}}\underline{\mathbf{v}}^{\top}).
    \end{align*}
    The first term is
    \begin{align*}
        \bR_{\theta}^{\top}\bI_p &= \bI_p + (\sin\theta)(\underline{\mathbf{u}}\underline{\mathbf{v}}^{\top} - \underline{\mathbf{v}}\underline{\mathbf{u}}^{\top}) + (\cos\theta - 1)(\underline{\mathbf{u}}\underline{\mathbf{u}}^{\top} + \underline{\mathbf{v}}\underline{\mathbf{v}}^{\top}).
    \end{align*}
    The second term is
    \begin{align*}
        \begin{split}
            (\sin\theta)\bR_{\theta}^{\top}(\underline{\mathbf{v}}\underline{\mathbf{u}}^{\top} - \underline{\mathbf{u}}\underline{\mathbf{v}}^{\top}) &= (\sin\theta)(\underline{\mathbf{v}}\underline{\mathbf{u}}^{\top} - \underline{\mathbf{u}}\underline{\mathbf{v}}^{\top}) + \sin^2\theta(\underline{\mathbf{u}}\underline{\mathbf{v}}^{\top} - \underline{\mathbf{v}}\underline{\mathbf{u}}^{\top})(\underline{\mathbf{v}}\underline{\mathbf{u}}^{\top} - \underline{\mathbf{u}}\underline{\mathbf{v}}^{\top}) \:+\\
            &\phe\quad (\sin\theta)(\cos\theta - 1)(\underline{\mathbf{u}}\underline{\mathbf{u}}^{\top} + \underline{\mathbf{v}}\underline{\mathbf{v}}^{\top})(\underline{\mathbf{v}}\underline{\mathbf{u}}^{\top} - \underline{\mathbf{u}}\underline{\mathbf{v}}^{\top}) \\
            &= (\sin\theta)(\underline{\mathbf{v}}\underline{\mathbf{u}}^{\top} - \underline{\mathbf{u}}\underline{\mathbf{v}}^{\top}) + \sin^2\theta(\underline{\mathbf{u}}\underline{\mathbf{u}}^{\top} + \underline{\mathbf{v}}\underline{\mathbf{v}}^{\top}) \:+\\
            &\phe\quad (\sin\theta)(\cos\theta - 1)(\underline{\mathbf{v}}\underline{\mathbf{u}}^{\top} - \underline{\mathbf{u}}\underline{\mathbf{v}}^{\top})
        \end{split}\\
        &= \sin^2\theta(\underline{\mathbf{u}}\underline{\mathbf{u}}^{\top} + \underline{\mathbf{v}}\underline{\mathbf{v}}^{\top}) + (\sin\theta\cos\theta)(\underline{\mathbf{v}}\underline{\mathbf{u}}^{\top} - \underline{\mathbf{u}}\underline{\mathbf{v}}^{\top}).
    \end{align*}
    The third term is
    \begin{align*}
        \begin{split}
            (\cos\theta - 1) \bR_{\theta}^{\top}(\underline{\mathbf{u}}\underline{\mathbf{u}}^{\top} + \underline{\mathbf{v}}\underline{\mathbf{v}}^{\top}) &= (\cos\theta - 1)(\underline{\mathbf{u}}\underline{\mathbf{u}}^{\top} + \underline{\mathbf{v}}\underline{\mathbf{v}}^{\top}) + (\cos\theta - 1)^2(\underline{\mathbf{u}}\underline{\mathbf{u}}^{\top} + \underline{\mathbf{v}}\underline{\mathbf{v}}^{\top})^2 \:+\\
            &\phe\quad (\sin\theta)(\cos\theta - 1) (\underline{\mathbf{u}}\underline{\mathbf{v}}^{\top} - \underline{\mathbf{v}}\underline{\mathbf{u}}^{\top}) (\underline{\mathbf{u}}\underline{\mathbf{u}}^{\top} + \underline{\mathbf{v}}\underline{\mathbf{v}}^{\top}) \\
            &= (\cos\theta - 1 + \cos^2\theta - 2\cos\theta + 1)(\underline{\mathbf{u}}\underline{\mathbf{u}}^{\top} + \underline{\mathbf{v}}\underline{\mathbf{v}}^{\top}) \:+\\
            &\phe\quad (\sin\theta)(\cos\theta - 1)(\underline{\mathbf{u}}\underline{\mathbf{v}}^{\top} - \underline{\mathbf{v}}\underline{\mathbf{u}}^{\top})
        \end{split}\\
        \begin{split}
            &= (\cos^2\theta - \cos\theta)(\underline{\mathbf{u}}\underline{\mathbf{u}}^{\top} + \underline{\mathbf{v}}\underline{\mathbf{v}}^{\top}) \:+\\
            &\phe\quad (\sin\theta)(\cos\theta - 1)(\underline{\mathbf{u}}\underline{\mathbf{v}}^{\top} - \underline{\mathbf{v}}\underline{\mathbf{u}}^{\top}).
        \end{split}
    \end{align*}
    Matching pieces by their trigonometric constants, and using identity $\cos^2\theta + \sin^2\theta = 1$, we see all terms except for $\bI_p$ cancel, hence $\bR_{\theta}^{\top}\bR_{\theta} = \bI_p$.  A nearly identical argument shows $\bR_{\theta}\bR_{\theta}^{\top} = \bI_p$, hence $\bR_{\theta}^{\top} = \bR_{\theta}^{-1}$.
    
    It follows $\abs{\bR_{\theta}} = \pm 1$.  Note however the eigenvalues of $(\sin\theta)(\underline{\mathbf{v}}\underline{\mathbf{u}}^{\top} - \underline{\mathbf{u}}\underline{\mathbf{v}}^{\top})$ are zero except for two complex conjugates, and the eigenvalues of $(\cos\theta - 1)(\underline{\mathbf{u}}\underline{\mathbf{u}}^{\top} + \underline{\mathbf{v}}\underline{\mathbf{v}}^{\top})$ are zero except for two identical (negative real) values.  Since the eigenspaces of these terms are both $\mathrm{span}(\underline{\mathbf{u}}, \underline{\mathbf{v}})$, the rank of $\bR_{\theta} - \bI_p$ is also 2, with eigenvalues also zero except for two complex conjugates.  Adding back $\bI_p$, the product of the eigenvalues of $\bR_{\theta}$ (i.e. its determinant) must then be positive—specifically, $+1$—hence $\bR_{\theta}$ is a proper rotation matrix.
    
    \item \textbf{$\bR_{\theta}$ rotates $\underline{\mathbf{u}} \rightarrow \underline{\mathbf{v}}$ by angle $\theta$:}\\
    For $\theta > 0$, the matrix $\bR_{\theta}$ applied to $\underline{\mathbf{u}}$ gives
    \begin{align*}
        \bR_{\theta}\underline{\mathbf{u}} &= \underline{\mathbf{u}} + (\sin\theta)(\underline{\mathbf{v}}\underline{\mathbf{u}}^{\top} - \underline{\mathbf{u}}\underline{\mathbf{v}}^{\top})\underline{\mathbf{u}} + (\cos\theta - 1)(\underline{\mathbf{u}}\underline{\mathbf{u}}^{\top} + \underline{\mathbf{v}}\underline{\mathbf{v}}^{\top})\underline{\mathbf{u}} \\
        &= \underline{\mathbf{u}} + (\sin\theta) (\underline{\mathbf{v}}) + (\cos\theta - 1)(\underline{\mathbf{u}}) \\
        &= (\cos\theta)(\underline{\mathbf{u}}) + (\sin\theta)(\underline{\mathbf{v}}).
    \end{align*}
    Clearly for any $\theta$, the angle between $\bR_{\theta}\underline{\mathbf{u}}$ and $\underline{\mathbf{u}}$ is $\theta$; in particular for $\theta = \pi/2$, $\bR_{\theta}\underline{\mathbf{u}} = \underline{\mathbf{v}}$.

    \item \textbf{$\bR_{\theta}$ fixes the orthogonal complement of $\mathrm{span}(\underline{\mathbf{u}}, \underline{\mathbf{v}})$:}\\
    Clearly for any $\bz \perp \mathrm{span}(\underline{\mathbf{u}}, \underline{\mathbf{v}})$
    \begin{align*}
        \bR_{\theta}\bz = \left\{ \bI_p + (\sin\theta)(\underline{\mathbf{v}}\underline{\mathbf{u}}^{\top} - \underline{\mathbf{u}}\underline{\mathbf{v}}^{\top}) + (\cos\theta - 1)(\underline{\mathbf{u}}\underline{\mathbf{u}}^{\top} + \underline{\mathbf{v}}\underline{\mathbf{v}}^{\top}) \right\}\bz = \bI_p\bz + \bzero + \bzero = \bz.
    \end{align*}

    \item \textbf{$\bR_{\theta}$ is unique:}\\
    Suppose $\bR_*$ performs the same rotation operation as $\bR_{\theta}$, i.e. $\bR_*\bz = \bR_{\theta}\bz$ for all $\bz \in \R^p$.  Then $(\bR_* - \bR_{\theta})\bz = \bzero_p$ for all $\bz$; but this is only true if $\bR_* - \bR_{\theta} = \bzero \: \iff \: \bR_* = \bR_{\theta}$.\qed
\end{enumerate}
\color{black}

\clearpage
\section{Proofs of Propositions} \label{sup:proofs}

\subsection*{Proof of Proposition \ref{prop:FrobeniusNormMinimization}}

Let $\bX \in \R^{n\times p}$ be a column-centered variable matrix, $\mathbf{x}_* \in \R^p$ be a vector centered the same as $\bX$, and let $\bY \in \R^{n\times m}$ be a matrix holding a set of empirical quantile functions defined on a shared, uniformly dispersed $m$-grid in $(0, 1)$.  The estimated conditional Fréchet mean $\widehat{\mathbf{q}}_* \equiv \widehat{\mathbf{q}}(\mathbf{x}_*)$ is the solution to the optimization problem
\begin{align}\label{eq:Frechet_mean_bhat_*}
    \widehat{\mathbf{q}}_* := \argmin_{\mathbf{q} \in \R^m} \sum_{i=1}^n s_i(\mathbf{x}_*) \norm{\mathbf{q} - \mathbf{y}_i}_2^2,\quad\mbox{subject to}\quad \mathbf{b}- \bA^{\top}\mathbf{q} \leq \bzero_a,
\end{align}
where $s_i(\mathbf{x}_*) = \frac{1}{n} + \mathbf{x}_*^{\top}( \bX^{\top} \bX)^- \mathbf{x}_i$.  The objective function can be rewritten
\begin{align*}
    \sum_{i=1}^n s_i(\mathbf{x}_*) \norm{\mathbf{q} - \mathbf{y}_i}_2^2 &= \sum_{i=1}^n s_i(\mathbf{x}_*) \sum_{j=1}^m (\text{y}_{i,j} - q_j)^2 \\
    &= \sum_{i=1}^n s_i(\mathbf{x}_*) \sum_{j=1}^m \left( \text{y}_{i,j}^2 - 2\text{y}_{i,j}q_j + q_j^2 \right) \\
    &= \sum_{j=1}^m \left[ \left( \sum_{i=1}^n s_i(\mathbf{x}) \text{y}_{i,j}^2 \right) - 2\left( q_j \sum_{i=1}^n s_i(\mathbf{x}_*) \text{y}_{i,j} \right) + \left( q_j^2 \sum_{i=1}^n s_i(\mathbf{x}_*) \right) \right] \\
    &= \kappa + \mathbf{q}^{\top}\mathbf{q} - 2\mathbf{q}^{\top}\widehat{\mathbf{y}}_*,
\end{align*}
where $\kappa$ is a constant free of $\mathbf{q}$, and $\widehat{\mathbf{y}}_*^{\top} = (n^{-1}\bone_n^{\top} + \mathbf{x}_*^{\top}(\bX^{\top} \bX)^- \bX^{\top}) \bY$.  Then since the $\argmin$ is invariant to $\kappa$ and to scaling, we have
\begin{align}\label{eq:qhatQuadProg}
    \widehat{\mathbf{q}}_* = \argmin_{\mathbf{q} \in \R^m} \frac{1}{2}\norm{\mathbf{q} - \widehat{\mathbf{y}}_*}_2^2, \quad\mbox{subject to}\quad \mathbf{b}- \bA^{\top}\mathbf{q} \leq \bzero_a.
\end{align}
This proves the first part of the proposition.  Let $\bX_* \in \R^{d\times p}$ be a matrix of variable vectors column-centered as $\bX$, and $\bhatQ_* \in \R^{d\times m}$ the matrix row-wise consisting of the solutions to \eqref{eq:qhatQuadProg}.  Since Frobenius norm is element-wise, the joint result follows immediately by row-adjoining
\begin{align*}
    \bhatQ_* := \argmin_{\bQ \in \R^{d\times m}} \oldnorm{\bQ - \bhatY_*}_F^2,\quad\mbox{subject to}\quad \bB - \bQ\bA \leq \bzero_{d\times (m+1)},
\end{align*}
where the rows of $\bB$ are each $\mathbf{b}$, and $\bhatY_* := (n^{-1}\bone_{d\times n} + \bX_*(\bX^{\top} \bX)^- \bX^{\top})\bY$.

\subsection*{Proof of Proposition \ref{prop:gradient_Hessian_lambda}}

The objective function of ~\eqref{eq:friso} 
is
\begin{align}\label{eq:flambda}
    f(\blambda) = \frac{1}{2} \oldnorm{\bY - \bhatQ(\blambda)}_F^2,\quad\mbox{subject to}\quad \blambda \in \bTau,
\end{align}
where
\begin{align*}
    \bhatQ(\blambda) = \argmin_{\bQ \in \R^{n\times m}} \oldnorm{\bQ - \bhatY(\blambda)}_F^2,\quad\mbox{subject to}\quad \bB - \bQ \bA \leq \bzero_{n\times (m+1)},
\end{align*}
and
\begin{align*}
    \bhatY(\blambda) &= n^{-1}\bone_{n\times n}\bY + \btilX(\btilX{}^{\top} \btilX + \bD_{\blambda}^{-1})^{-1}\btilX{}^{\top}\bY.
\end{align*}
To prevent division by zero, note we can rewrite this matrix using the push-through identity
\begin{align*}
    \bhatY(\blambda) &= n^{-1}\bone_{n\times n}\bY + \btilX \bD_{\blambda}^{1/2}(\bD_{\blambda}^{1/2}\btilX{}^{\top}\btilX \bD_{\blambda}^{1/2} + \bI_p )^{-1}\bD_{\blambda}^{1/2}\btilX{}^{\top}\bY \\
    &= n^{-1}\bone_{n\times n}\bY + \btilX \bD_{\blambda} \btilX{}^{\top}(\btilX \bD_{\blambda} \btilX{}^{\top} + \bI_n )^{-1} \bY \\
    &= n^{-1} \bone_{n\times n} \bY + \bY - (\btilX \bD_{\blambda} \btilX{}^{\top} + \bI_n)^{-1}\bY.
\end{align*}
We define $\bG \equiv \bG(\blambda) := (\btilX \bD_{\blambda} \btilX{}^{\top} + \bI_n)^{-1}$, so that
\begin{align}\label{eq:yhatG}
    \bhatY(\blambda) = n^{-1}\bone_{n\times n}\bY + \bY - \bG \bY.
\end{align}
Let $\bhatQ = \bhatQ(\blambda)$ and $\bhatY = \bhatY(\blambda)$. The partial derivative of~\eqref{eq:flambda} with respect to $\lambda_k$ is:
\begin{align}\label{eq:objDerStart}
    \frac{\partial}{\partial \lambda_k}f(\blambda) = \tr\left\{ (\bhatQ - \bY) \left(\frac{\partial}{\partial\lambda_k} \bhatQ \right)^{\top} \right\}.
\end{align}
To evaluate the partial derivative of $\bhatQ$, consider the optimality conditions associated with ~\eqref{eq:embeddedProblem}.  
The optimal $\bhatQ$, with corresponding optimal Lagrange multiplier $\bhatEta \in \R^{n\times (m + 1)}$, satisfy KKT conditions of stationarity of the associated Lagrangian
\begin{align} \label{eq:stabilityCondition}
    \frac{\partial}{\partial\bQ} \calL(\bhatQ, \bhatEta ) := \left. \frac{\partial}{\partial\bQ} \left[ \oldnorm{\bQ - \bhatY}_F^2 + \tr\left\{ (\bB - \bQ\bA)\bEta^{\top} \right\} \right]\right|_{\bhatQ, \bhatEta} = \bhatQ - \bhatY - \bhatEta  \bA^{\top} = \bzero_{n\times m}
\end{align}
and complementary slackness
\begin{align} \label{eq:complementarySlacknessCondition}
    ( \bB - \bhatQ \bA ) \circ \bhatEta  = \bzero_{n\times (m+1)}.
\end{align}
Denote $\pmb{\calQ} := \frac{\partial}{\partial\lambda_k}\bhatQ$, $\pmb{\calH} := \frac{\partial}{\partial\lambda_k} \bhatEta $, and $\pmb{\calY} := \frac{\partial}{\partial\lambda_k}\bhatY$.  Applying the partial derivative operator to each of \eqref{eq:stabilityCondition} and \eqref{eq:complementarySlacknessCondition}, we respectively obtain:
\begin{equation} \label{eq:stabilityDerivative}
    \frac{\partial}{\partial\lambda_k}( \bhatQ - \bhatY - \bhatEta  \bA^{\top} ) = \frac{\partial}{\partial\lambda_k}\bzero_{n\times m} \quad \iff \quad
    \pmb{\calQ} - \pmb{\calY} - \pmb{\calH} \bA^{\top} = \bzero_{n\times m},
\end{equation}
and:
\begin{equation} \label{eq:complementarySlacknessDerivative}
    \frac{\partial}{\partial\lambda_k} \{ ( \bB - \bhatQ \bA ) \circ \bhatEta \} = \frac{\partial}{\partial\lambda_k}\bzero_{n\times (m+1)} \quad \iff \quad
    ( \bB - \bhatQ \bA) \circ \pmb{\calH} = (\pmb{\calQ}\bA) \circ \bhatEta
\end{equation}
From~\eqref{eq:stabilityDerivative}, our desired $\pmb{\calQ}$ is a function of two other gradients $\pmb{\calY}$ and $\pmb{\calH}$ which we find next.

First, using~\eqref{eq:yhatG} and recalling $\partial_x\bM^{-1}(x) = -\bM^{-1}(x)\{\partial_x \bM(x)\}\bM^{-1}(x)$, we have
\begin{align}\label{eq:yhatGradient}
    \pmb{\calY} = \frac{\partial}{\partial\lambda_k}(n^{-1}\bone_{n\times n}\bY + \bY - \bG \bY) = -\frac{\partial}{\partial\lambda_k}(\btilX \bD_{\blambda} \btilX{}^{\top} + \bI_n)^{-1}\bY = \bG \btilX_k\btilX{}_k^{\top} \bG \bY.
\end{align}
To evaluate the gradient $\pmb{\calH}$, let $\bC := \bB - \bhatQ \bA$ be the constraint matrix associated with $\bhatQ$, where zero-valued entries indicate active constraints.  Consider two cases: element-wise, if $c_{i,\ell} < 0$, then by \eqref{eq:complementarySlacknessCondition} we have $\hath_{i,\ell} = 0$, which through \eqref{eq:complementarySlacknessDerivative} implies $\eta_{i,\ell} = 0$; on the other hand, if $c_{i,\ell} = 0$, then using \eqref{eq:stabilityDerivative} for $\pmb{\calQ}$
\begin{align*}
    [\bB - \bhatQ \bA]_{i,\ell} = 0 \quad \Rightarrow \quad
    \frac{\partial}{\partial\lambda_k} b_{i,\ell} = [\pmb{\calQ} \bA]_{i,\ell} \iff
    0 &= [\pmb{\calY} \bA + \pmb{\calH} \bA^{\top} \bA]_{i,\ell}.
\end{align*}
The constraint matrix $\bC$ then implicates the derivative matrix $\pmb{\calH} = [\eta_{i,\ell}]$ in the system of equations
\begin{align}\label{eq:system_calH}
    \begin{split}
        \begin{matrix*}[l] c_{i,\ell} = 0 & \Rightarrow & [\pmb{\calH} \bA^{\top} \bA]_{i,\ell} = -[\pmb{\calY} \bA]_{i,\ell}, \\
        c_{i,\ell} < 0 & \Rightarrow & \eta_{i,\ell} = 0. \end{matrix*}
    \end{split}
\end{align}
This system can be solved independently for rows of $\pmb{\calH}$.  Denote the \tharg{i} rows of $\pmb{\calH}$, $\pmb{\calY}$, and $\bC$ with $\pmb{\eta}_i = (\begin{matrix} \eta_{i,1} & \cdots & \eta_{i,m+1} \end{matrix})^{\top}$, $\pmb{y}_i = (\begin{matrix} y_{i,1} & \cdots & y_{i,m+1} \end{matrix})^{\top}$, and $\mathbf{c}_i = (\begin{matrix} c_{i,1} & \cdots & c_{i,m+1} \end{matrix})^{\top}$, respectively.  If $\mathbf{c}_i < \bzero$, then we take $\pmb{\eta}_i = \bzero$. The case $\mathbf{c}_i = \bzero$ is not possible, as it implies all constraints are active for $\widehat{\mathbf{q}}_i$, namely the quantile function is constant and equal to both box constraints $b_L$, $b_U$.  But this gives rise to a contradiction since $b_L < b_U$ for the support.

The remaining case is where some elements of $\mathbf{c}_i$ are zero (active constraints) and some are negative (inactive constraints). Without loss of generality, let $\bA = [\begin{matrix} \bA_{0,i} & \bA_{-,i} \end{matrix}]$ be the block decomposition of $\bA$ into those columns corresponding to the \tharg{i} active constraints (i.e. $\bA_{0,i}$) and the \tharg{i} inactive constraints (i.e. $\bA_{-,i}$), with $\pmb{\eta}_i^{\top} = [\begin{matrix} \pmb{\eta}_{0,i}^{\top} & \pmb{\eta}_{-,i}^{\top} \end{matrix}]$ split analogously.  The \tharg{i} system from~\eqref{eq:system_calH} is
\begin{align*}
    \left[ \begin{matrix} \bA_{0,i}^{\top}\bA_{0,i} & \bA_{0,i}^{\top}\bA_{-,i} \\ \bzero & \bI \end{matrix} \right] \left[ \begin{matrix} \pmb{\eta}_{0,i} \\ \pmb{\eta}_{-,i} \end{matrix} \right] &= -\left[ \begin{matrix} \bA_{0,i}^{\top}\pmb{y}_i \\ \bzero \end{matrix} \right]
\end{align*}
We immediately have $\pmb{\eta}_{-,i} = \bzero$, as desired, and since $\bA_{0,i}$ is guaranteed to be full column rank
\begin{align*}
    \bA_{0,i}^{\top}\bA_{0,i}\pmb{\eta}_{0,i} &= -\bA_{0,i}^{\top} \pmb{y}_i \\
    \pmb{\eta}_{0,i} &= -(\bA_{0,i}^{\top}\bA_{0,i})^{-1}\bA_{0,i}^{\top} \pmb{y}_i.
\end{align*}
With the solution for each \tharg{i} row of $\pmb{\calH}$, recall from~\eqref{eq:stabilityDerivative} that $\pmb{\calQ} = \pmb{\calY} + \pmb{\calH} \bA^{\top}$. Thus
\begin{align*}
    \bA\pmb{\eta}_i &= [\begin{matrix} \bA_{0,i} & \bA_{-,i} \end{matrix}] \left[ \begin{matrix} \pmb{\eta}_{0,i} \\ \pmb{\eta}_{-,i} \end{matrix} \right] 
    = \bA_{0,i} \pmb{\eta}_{0,i} + \bA_{-,i}\pmb{\eta}_{-,i} 
    = \bA_{0,i}\left\{-(\bA_{0,i}^{\top}\bA_{0,i})^{-1}\bA_{0,i}^{\top} \pmb{y}_i \right\} + \bzero 
    = -\proj{\bA_{0,i}} \pmb{y}_i,
\end{align*}
where we interpret $\proj{\bA_{0,i}} = \bzero_{m\times m}$ when $\mathbf{c}_i < \bzero$.  Plugging in we get
\begin{align*}
    \pmb{\calQ} &= \pmb{\calY} + \pmb{\calH} \bA^{\top} = \left[ \begin{matrix*}
        \pmb{y}_1^{\top}(\bI_m - \proj{\bA_{0,1}}) \\
        \vdots \\
        \pmb{y}_n^{\top}(\bI_m - \proj{\bA_{0,n}})
    \end{matrix*} \right].
\end{align*}
Plugging this back into~\eqref{eq:objDerStart}
\begin{align}\label{eq:objDerNext}
    \frac{\partial}{\partial\lambda_k} f(\blambda) &= \tr\left\{ ( \bhatQ - \bY ) \left[ \begin{matrix*}
        \pmb{y}_1^{\top}(\bI_m - \proj{\bA_{0,1}}) \\
        \vdots \\
        \pmb{y}_n^{\top}(\bI_m - \proj{\bA_{0,n}})
    \end{matrix*} \right]^{\top} \right\}.
\end{align}
From~\eqref{eq:yhatGradient} we can row-wise express $\pmb{\calY}$
\begin{align*}
    \pmb{\calY} = \left[ \begin{matrix} \pmb{y}_1^{\top} \\ \vdots \\ \pmb{y}_n^{\top} \end{matrix} \right] = \left[ \begin{matrix} \underline{\mathbf{e}}_1^{\top}\bG \btilX_k \btilX{}_k^{\top} \bG \bY \\ \vdots \\ \underline{\mathbf{e}}_n^{\top} \bG \btilX_k \btilX{}_k^{\top} \bG \bY \end{matrix} \right].
\end{align*}
Inserting these individual row vectors into~\eqref{eq:objDerNext} we obtain
\begin{align}\label{eq:flambda_derivative_k}
    \frac{\partial}{\partial\lambda_k} f(\blambda) &= \tr\left\{ ( \bhatQ - \bY )\left[ \begin{matrix*}
        \underline{\mathbf{e}}_1^{\top}\bG \btilX_k \btilX{}_k^{\top} \bG \bY(\bI_m - \proj{\bA_{0,1}}) \\
        \vdots \\
        \underline{\mathbf{e}}_n^{\top} \bG \btilX_k \btilX{}_k^{\top} \bG \bY(\bI_m - \proj{\bA_{0,n}})
    \end{matrix*} \right]^{\top} \right\} \nonumber\\
    &= \sum_{i=1}^n \left\{ (\widehat{\mathbf{q}}_i - \mathbf{y}_i)^{\top}(\bI_m - \proj{\bA_{0,i}}) \bY^{\top}\bG\btilX_k \right\} \left\{ \btilX{}_k^{\top}\bG \underline{\mathbf{e}}_i \right\} \nonumber\\
    &= \sum_{i=1}^n \left\{ \btilX{}_k^{\top}\bG \underline{\mathbf{e}}_i \right\} \left\{ (\widehat{\mathbf{q}}_i^{\top} - \mathbf{y}_i^{\top})(\bI_m - \proj{\bA_{0,i}}) \bY^{\top}\bG\btilX_k \right\} \nonumber\\
    &= \btilX{}_k^{\top} \bG \left\{ \sum_{i=1}^n \underline{\mathbf{e}}_i (\widehat{\mathbf{q}}_i^{\top} - \mathbf{y}_i^{\top})(\bI_m - \proj{\bA_{0,i}}) \right\} \bY^{\top}\bG\btilX_k.
\end{align}
Since $\bA_{0,i}$ are the columns of $\bA$ corresponding to the active constraints on $\widehat{\mathbf{q}}_i$, we can rewrite the optimization problem for $\widehat{\mathbf{q}}_i$ in terms of only those constraints via
\begin{align*}
    \widehat{\mathbf{q}}_i = \argmin_{\mathbf{q}_i \in \R^m} \frac{1}{2}\oldnorm{\mathbf{q}_i - \widehat{\mathbf{y}}_i}_2^2,\quad\mbox{subject to}\quad \bb_{0,i} - \bA_{0,i}^{\top} \mathbf{q}_i = \bzero.
\end{align*}
The Lagrangian associated with the above objective function is
\begin{align*}
    \calL(\mathbf{q}_i, \bbeta_i) = \frac{1}{2}(\mathbf{q}_i^{\top}\mathbf{q}_i - 2\mathbf{q}_i^{\top}\widehat{\mathbf{y}}_i + \widehat{\mathbf{y}}^{\top}_i\widehat{\mathbf{y}}_i) + \bbeta^{\top}_i(\bb_{0,i} - \bA_{0,i}^{\top}\mathbf{q}_i).
\end{align*}
The optimality conditions are primal feasibility and stability, given respectively by
\begin{align*}
    \bb_{0,i} - \bA_{0,i}^{\top}\widehat{\mathbf{q}}_i &= \bzero, \qquad \frac{\partial}{\partial\mathbf{q}_i}\calL(\widehat{\mathbf{q}}_i, \bhatbeta_i) = \widehat{\mathbf{q}}_i - \widehat{\mathbf{y}}_i - \bA_{0,i}\bhatbeta_i = \bzero.
\end{align*}
Combining these expressions, we obtain
\begin{align*}
    \bhatbeta_i &= (\bA_{0,i}^{\top}\bA_{0,i})^{-1}(\bb_{0,i} - \bA_{0,i}^{\top}\widehat{\mathbf{y}}_i),\qquad \widehat{\mathbf{q}}_i = (\bI_m - \proj{\bA_{0,i}})\widehat{\mathbf{y}}_i + \bA_{0,i}(\bA_{0,i}^{\top}\bA_{0,i})^{-1}\bb_{0,i}.
\end{align*}
Plugging the expression for $\widehat{\mathbf{q}}_i$ into~\eqref{eq:flambda_derivative_k}
\begin{align}\label{eq:flambda_derivative_k_final}
    \frac{\partial}{\partial\lambda_k} f(\blambda) &= \btilX_k\bG\left[ \sum_{i=1}^n \underline{\mathbf{e}}_i \left\{ \widehat{\mathbf{y}}_i^{\top}(\bI_m - \proj{\bA_{0,i}}) + \bb_{0,i}^{\top}(\bA_{0,i}^{\top}\bA_{0,i})^{-1}\bA_{0,i}^{\top} -\mathbf{y}_i^{\top} \right\} (\bI_m - \proj{\bA_{0,i}}) \right] \bY^{\top}\bG\btilX_k \nonumber\\
    &= \btilX{}_k^{\top} \bG\left\{ \sum_{i=1}^n \underline{\mathbf{e}}_i ( \widehat{\mathbf{y}}_i^{\top} - \mathbf{y}_i^{\top} ) (\bI_m - \proj{\bA_{0,i}}) \right\} \bY^{\top}\bG\btilX_k \nonumber\\
    &= \btilX{}_k^{\top} \bG\left\{ \sum_{i=1}^n \underline{\mathbf{e}}_i\underline{\mathbf{e}}_i^{\top}(\bhatY - \bY)(\bI_m - \proj{\bA_{0,i}})\right\} \bY^{\top}\bG\btilX_k.
\end{align}
Finally, combining across $k$'s, the full gradient of~\eqref{eq:flambda} is
\begin{align*}
    \nabla f(\blambda) &= \diag (\bN),
\end{align*}
where
\begin{align}\label{eq:N}
    \bN := \btilX{}^{\top} \bG \left\{ \sum_{i=1}^n \underline{\mathbf{e}}_i \underline{\mathbf{e}}_i^{\top} (\bhatY - \bY)(\bI_m - \proj{\bA_{0,i}}) \right\} \bY^{\top}\bG\btilX.
\end{align}
Before deriving the Hessian, we make two key observations. The first is that the matrices $\{ \proj{\bA_{0,i}} \}_{i=1}^n$ are implicitly functions of $\blambda$, since changing $\blambda$ can change which constraints become active or inactive on $\bhatQ$.  However, as a function on the simplex $\bTau$, $\proj{\bA_{0,i}}(\blambda)$ is piece-wise constant, since any change in the projection matrix occurs by inclusion or exclusion of new columns of $\bA$.  There is only a finite number of ways to select columns of $\bA$ to create $\bA_{0,i}$, and so with respect to the Lebesgue measure $\mu$, $\mu\left\{ \frac{\partial}{\partial\lambda_k} \proj{\bA_{0,i}} \neq \bzero \right\} = 0$.

The second is that these points of discontinuity in $\proj{\bA_{0,i}}$ over $\bTau$ correspond to instances where the unconstrained solution $\bhatY$ \textit{is exactly} on the boundary specified by $\bB - \bhatY\bA \leq \bzero$. From \eqref{eq:yhatG}, each column in $\bhatY(\blambda)$ is a weighted mean of the corresponding column of $\bY$, so there are two scenarios under which this can happen.
\begin{itemize}[noitemsep, nolistsep]
    \item All observations in one or more columns $\bY_j$ are equal to each other. Then no matter what $\blambda$ is, the weighted mean takes the same value, and so the unconstrained solution can rest exactly on a box constraint (if all observed values lie on that box constraint) or exhibit constancy over sequential columns (if all observations are the same across sequential columns). In such a case, the corresponding optimal Lagrange multipliers are always zero, and we arrive at the same conclusion $\eta = 0$ from \eqref{eq:system_calH} whether we include or exclude the corresponding columns in $\bA$ from $\bA_{0,i}$.
    \item We do not have constant values in a column $\bY_j$, but the value of $\blambda$ is such that the weighted mean(s) of the observations makes some constraints exactly met. As stated above, when tracing out a countably long sequence $\{ \blambda^{(0)}, \blambda^{(1)},  \cdots \}$ during gradient descent, it is almost sure that $\frac{\partial}{\partial\lambda_k^{(t)}} \proj{\bA_{0,i}}(\blambda^{(t)}) = 0$ for all $i$, $k$, and $t$.
\end{itemize}
Given these observations, we treat $\proj{\bA_{0,i}}(\blambda)$ as locally constant with respect to $\blambda$ in deriving the Hessian $\nabla^2 f(\blambda)$. The Hessian can be found element-wise
\begin{align*}
    \frac{\partial^2}{\partial\lambda_h \partial \lambda_k} f(\blambda) = \frac{\partial}{\partial \lambda_h} \bN_{k,k} &= \btilX{}_k^{\top} \left[ \left( \frac{\partial}{\partial \lambda_h} \bG \right) \left\{ \sum_{i=1}^n \underline{\mathbf{e}}_i \underline{\mathbf{e}}_i^{\top} (\bhatY - \bY)(\bI_m - \proj{\bA_{0,i}}) \right\} \bY^{\top}\bG \: + \right. \\
    &\phe\qquad\: \left. \bG\left\{ \sum_{i=1}^n \underline{\mathbf{e}}_i \underline{\mathbf{e}}_i^{\top} \left( \frac{\partial}{\partial\lambda_h}\bhatY \right)(\bI_m - \proj{\bA_{0,i}}) \right\} \bY^{\top}\bG \: + \right. \\
    &\phe\qquad\: \left. \bG\left\{ \sum_{i=1}^n \underline{\mathbf{e}}_i \underline{\mathbf{e}}_i^{\top} (\bhatY - \bY)(\bI_m - \proj{\bA_{0,i}}) \right\} \bY^{\top}\left( \frac{\partial}{\partial\lambda_h} \bG\right) \right] \btilX_k \\
    &= \btilX{}_k^{\top} \left[ -\bG\btilX_h\btilX_h^{\top}\bG \left\{ \sum_{i=1}^n \underline{\mathbf{e}}_i \underline{\mathbf{e}}_i^{\top} (\bhatY - \bY)(\bI_m - \proj{\bA_{0,i}}) \right\} \bY^{\top}\bG \: + \right. \\
    &\phe\qquad\: \left. \bG\left\{ \sum_{i=1}^n \underline{\mathbf{e}}_i \underline{\mathbf{e}}_i^{\top}\bG\btilX_h \btilX_h^{\top} \bG \bY(\bI_m - \proj{\bA_{0,i}}) \right\} \bY^{\top}\bG \: - \right. \\
    &\phe\qquad\: \left. \bG\left\{ \sum_{i=1}^n \underline{\mathbf{e}}_i \underline{\mathbf{e}}_i^{\top} (\bhatY - \bY)(\bI_m - \proj{\bA_{0,i}}) \right\} \bY^{\top}\bG\btilX_h\btilX_h^{\top}\bG \right] \btilX_k,
\end{align*}
where we used~\eqref{eq:yhatGradient} for the gradient of $\bG$ and $\bhatY$.

The first and third terms are transposes and contain $\bN$ (or $\bN^{\top}$), and can be combined
\begin{align}\label{eq:flambda_hessian_kh_final}
    \begin{split}
        \frac{\partial^2}{\partial\lambda_h \partial \lambda_k} f(\blambda) &=  \left[ \sum_{i=1}^n (\btilX{}^{\top} \bG \underline{\mathbf{e}}_i \underline{\mathbf{e}}_i^{\top} \bG \btilX) \circ \left\{ \btilX{}^{\top}\bG\bY(\bI_m - \proj{\bA_{0,i}})\bY^{\top}\bG \btilX\right\} \right]_{k,h} \: -\\
        &\phe\quad \left[ (\btilX{}^{\top} \bG \btilX) \circ (\bN + \bN^{\top}) \right]_{k,h}.
    \end{split}
\end{align}
Finally, the Hessian of~\eqref{eq:flambda} is
\begin{align*}
    \begin{split}
        \nabla^2 f(\blambda) &= (\btilX{}^{\top}\bG^2\btilX) \circ (\btilX{}^{\top}\bG\bY\bY^{\top}\bG\btilX) - (\btilX{}^{\top}\bG\btilX) \circ ( \bN + \bN^{\top} ) \:-\\
        &\phe\qquad \sum_{i=1}^n (\btilX{}^{\top} \bG \underline{\mathbf{e}}_i\underline{\mathbf{e}}_i^{\top} \bG \btilX) \circ (\btilX{}^{\top}\bG\bY\proj{\bA_{0,i}}\bY^{\top}\bG\btilX).
    \end{split}\tag*{\qed}
\end{align*}

\subsection*{Proof of Proposition \ref{prop:gradient_Hessian_gamma}}

For $\blambda = \calF(\bgamma) := \bgamma \circ \bgamma$, we have $g(\bgamma) = (f \smallcirc \calF)(\bgamma) = f(\blambda)$.  From Lemma~\ref{lem:lambda_gamma_derivative}
\begin{align*}
    \frac{\partial}{\partial\gamma_k} g(\bgamma) = \frac{\partial}{\partial\gamma_k} (f \smallcirc \calF)(\bgamma) = 2\gamma_k \cdot \frac{\partial}{\partial\lambda_k}f(\blambda),
\end{align*}
where the right term is given by Proposition~\ref{prop:gradient_Hessian_lambda}. 
Then the full gradient is
\begin{align*}
    \nabla g(\bgamma) &= 2\bgamma \circ \nabla f(\blambda),
\end{align*}
as desired. The Hessian $\nabla^2 g(\bgamma)$ is element-wise
\begin{align*}
    \frac{\partial^2}{\partial\gamma_h \partial\gamma_k} g(\bgamma) &= \frac{\partial}{\partial\gamma_h}\left\{ 2\gamma_k \cdot \frac{\partial}{\partial\lambda_k}f(\blambda) \right\} \\
    &= 2\frac{\partial}{\partial\lambda_k} f(\blambda)\ind(k = h) + 2\gamma_k \cdot \frac{\partial}{\partial\gamma_h} \left\{ \frac{\partial}{\partial\lambda_k}f(\blambda) \right\} \\
    &= 2\frac{\partial}{\partial\lambda_k} f(\blambda)\ind(k = h) + 4\gamma_k\gamma_h \cdot \frac{\partial^2}{\partial\lambda_h\partial\lambda_k} f(\blambda),
\end{align*}
where the right term is given by Proposition~\ref{prop:gradient_Hessian_lambda}.  
Then the full Hessian is
\begin{align*}
    \nabla^2 g(\bgamma) = 2 \bD_{\nabla f(\blambda)} + 4 (\bgamma \bgamma^{\top}) \circ \nabla^2 f(\blambda).\tag*{\qed}
\end{align*}

\subsection*{Proof of Proposition \ref{prop:theta_derivatives}}

Fix $\bgamma \in \calS_{\sqrt{\tau}}$.  The gradient $\nabla g(\bgamma)$ has an orthogonal decomposition into a component in the tangent space of $\calS_{\bgamma}$ (i.e. perpendicular to $\bgamma$) and a component normal to it (i.e. parallel to $\bgamma$)
\begin{align*}
    \nabla g(\bgamma) &= (\bI_p - \proj{\bgamma}) \nabla g(\bgamma) + \proj{\bgamma}\nabla g(\bgamma)
\end{align*}
Suppose $\nabla_t g(\bgamma) \neq \bzero_p$, and let $\mathbf{v} := -(\bI_p - \proj{\bgamma})g(\bgamma)$.  Let $\{(\sqrt{\tau}, \theta) : \theta \in [0, 2\pi)\}$ be the radius-angle polar coordinate system which parameterizes the unique great circle in $\calS_{\sqrt{\tau}}$ that contains $\bgamma$ and is coplanar with $\nabla_t g(\bgamma)$.  Without loss of generality, for $\theta \geq 0$ we use $\bR_{\theta}\bgamma$ to represent rotations of $\bgamma$ around this circle in the direction of $\mathbf{v}$, with $\bR_{\theta}$ given by~\eqref{eq:Rtheta}.  Observe
\begin{align}\label{eq:Rgamma}
    \bR_{\theta}\bgamma &= \norm{\bgamma}_2 \bR_{\theta}\underline{\bgamma} = \norm{\bgamma}_2 \left\{ (\cos\theta)\underline{\bgamma} + (\sin\theta)\underline{\mathbf{v}} \right\}.
\end{align}
Define the composite function $g_*(\theta) := (g \smallcirc \bR_{\theta})(\bgamma)$.  The first total derivative of $g_*(\theta)$ with respect to $\theta$ is given by
\begin{align*}
    \frac{d}{d\theta} g_*(\theta) &= \frac{d}{d\theta} (g \smallcirc \bR_{\theta})(\bgamma) \\
    &= \inner{\nabla g(\bgamma), \: \frac{\partial}{\partial\theta} \bR_{\theta} \bgamma} \\
    &= \inner{\proj{\bgamma} \nabla g(\bgamma) - \norm{\mathbf{v}}_2 \underline{\mathbf{v}},\: \norm{\bgamma}_2 \left\{ (\cos\theta)\underline{\mathbf{v}} - (\sin\theta)\underline{\bgamma} \right\}} \\
    &= \norm{\bgamma}_2 \left[ (\cos\theta) \{ \nabla g(\bgamma) \}^{\top}\proj{\bgamma}\underline{\mathbf{v}} - (\sin\theta)\{ \nabla g(\bgamma) \}^{\top}\proj{\bgamma}\underline{\bgamma} \:- \right.\\
    &\phe\qquad\qquad \left. (\cos\theta)\norm{\mathbf{v}}_2 \underline{\mathbf{v}}^{\top}\underline{\mathbf{v}} + (\sin\theta)\norm{\mathbf{v}}_2 \underline{\mathbf{v}}^{\top}\underline{\bgamma}\right] \\
    &= \norm{\bgamma}_2 \left[ 0 - (\sin\theta)\{ \nabla g(\bgamma)\}^{\top}\underline{\bgamma} - (\cos\theta)\norm{\mathbf{v}}_2 + 0 \right] \\
    &= -(\sin\theta)\norm{\bgamma}_2 \{ \nabla g(\bgamma)\}^{\top}\underline{\bgamma} - (\cos\theta)\norm{\bgamma}_2\norm{\mathbf{v}}_2.
\end{align*}
Evaluated at $\theta = 0$, we obtain
\begin{align*}
    \frac{d}{d\theta}g_*(0) &= -\norm{\bgamma}_2 \norm{\mathbf{v}}_2,
\end{align*}
as desired.  The second total derivative of $g_*(\theta)$ with respect to $\theta$ is given by
\begin{align}\label{eq:der2_gstar_1}
    \frac{d^2}{d\theta^2}g_*(\theta) &= \frac{d^2}{d\theta^2} (g \smallcirc \bR_{\theta})(\bgamma) \nonumber\\ 
    &= \inner{\nabla\frac{d}{d\theta}g_*(\theta), \: \norm{\bgamma}_2 \{ (\cos\theta)\underline{\mathbf{v}} - (\sin\theta)\underline{\bgamma} \}} \nonumber\\
    &= -\norm{\bgamma}_2\inner{\nabla \left[ (\sin\theta)\{\nabla g(\bgamma) \}^{\top}\bgamma + (\cos\theta)\norm{\bgamma}_2\norm{\mathbf{v}}_2 \right], \: (\cos\theta)\underline{\mathbf{v}} - (\sin\theta)\underline{\bgamma}} \nonumber\\
    &= -(\cos^2\theta)\norm{\bgamma}_2 \underline{\mathbf{v}}^{\top}\nabla( \norm{\bgamma}_2\norm{\mathbf{v}}_2 ) + (\text{$\sin\theta$ terms}).
\end{align}
Our goal is to evaluate this expression at $\theta = 0$, so each term involving $\sin\theta$ will be zero.  As such, we focus on only the first term, especially the gradient (which is with respect to $\bgamma$).  As a preliminary matter, we evaluate $\nabla \underline{\bgamma}$
\begin{align*}
    \frac{\partial}{\partial\gamma_h} \frac{\gamma_k}{\norm{\bgamma}_2} &= \frac{\ind(k=h)\norm{\bgamma}_2 - \gamma_k \frac{1}{2\norm{\bgamma}_2} 2\gamma_h }{\norm{\bgamma}_2^2}, \\
    \Rightarrow \qquad \nabla \underline{\bgamma} &= \frac{1}{\norm{\bgamma}_2} (\bI_p - \underline{\bgamma}\underline{\bgamma}^{\top}).
\end{align*}
The gradient in~\eqref{eq:der2_gstar_1} involves solving both $\nabla \norm{\bgamma}_2$ and $\nabla \norm{\mathbf{v}}_2$.  The first is
\begin{align}\label{eq:grad_unorm}
    \nabla \norm{\bgamma}_2 &= \nabla (\bgamma^{\top}\bgamma)^{1/2} = \frac{1}{2\norm{\bgamma}_2} \nabla \bgamma^{\top}\bgamma = \underline{\bgamma}.
\end{align}
Next, since $\norm{\mathbf{v}}_2 = \left[ \{\nabla g(\bgamma)\}^{\top}(\bI_p - \proj{\bgamma})\{\nabla g(\bgamma)\} \right]^{1/2} = \left[ \{\nabla g(\bgamma)\}^{\top}\nabla g(\bgamma) - \{\underline{\bgamma}^{\top}\nabla g(\bgamma)\}^2 \right]^{1/2}$
\begin{align}\label{eq:grad_vnorm}
    \nabla \norm{\mathbf{v}}_2 &= \nabla \left[ \{\nabla g(\bgamma)\}^{\top}\nabla g(\bgamma) - \{\underline{\bgamma}^{\top}\nabla g(\bgamma)\}^2 \right]^{1/2} \nonumber\\
    &= \frac{1}{2\norm{\mathbf{v}}_2}\left[ 2\{\nabla^2 g(\bgamma)\}\nabla g(\bgamma) - 2\underline{\bgamma}^{\top}\nabla g(\bgamma) \left\{ \frac{1}{\norm{\bgamma}_2}(\bI_p - \underline{\bgamma}\underline{\bgamma}^{\top})\nabla g(\bgamma) + \{\nabla^2 g(\bgamma)\}\underline{\bgamma}\right\}\right] \nonumber\\
    &= \frac{1}{\norm{\mathbf{v}}_2}\left[ \{\nabla^2 g(\bgamma)\}\nabla g(\bgamma) + \underline{\bgamma}^{\top}\nabla g(\bgamma)\left\{ \frac{\mathbf{v}}{\norm{\bgamma}_2} - \{ \nabla^2 g(\bgamma) \}\underline{\bgamma}\right\}\right] \nonumber\\
    &= \frac{1}{\norm{\mathbf{v}}_2}\left[ \mathbf{v} \cdot \frac{\underline{\bgamma}^{\top}\nabla g(\bgamma)}{\norm{\bgamma}_2} + \{\nabla^2 g(\bgamma)\}( \bI_p - \underline{\bgamma}\underline{\bgamma}^{\top})\nabla g(\bgamma) \right] \nonumber\\
    &= \underline{\mathbf{v}} \cdot \frac{\underline{\bgamma}^{\top}\nabla g(\bgamma)}{\norm{\bgamma}_2} - \{\nabla^2 g(\bgamma)\}\underline{\mathbf{v}}.
\end{align}
Combining~\eqref{eq:grad_unorm} and~\eqref{eq:grad_vnorm}, we complete~\eqref{eq:der2_gstar_1}
\begin{align*}
    \frac{d^2}{d\theta^2}g_*(\theta) &= -(\cos^2\theta)\norm{\bgamma}_2 \underline{\mathbf{v}}^{\top}\left\{ \norm{\bgamma}_2 \nabla \norm{\mathbf{v}}_2 + \norm{\mathbf{v}}_2 \nabla \norm{\bgamma}_2 \right\} + (\text{$\sin\theta$ terms})\\
    &= -(\cos^2\theta)\norm{\bgamma}_2 \left[ \norm{\bgamma}_2 \frac{\underline{\bgamma}^{\top}\nabla g(\bgamma)}{\norm{\bgamma}_2} - \norm{\bgamma}_2 \underline{\mathbf{v}}^{\top}\{ \nabla^2 g(\bgamma) \}\underline{\mathbf{v}} + \norm{\mathbf{v}}_2 \cdot 0 \right]  + (\text{$\sin\theta$ terms})\\
    &= (\cos^2\theta) \left[ \norm{\bgamma}_2^2 \underline{\mathbf{v}}^{\top}\{ \nabla^2 g(\bgamma) \}\underline{\mathbf{v}} - \bgamma^{\top}\nabla g(\bgamma) \right] + (\text{$\sin\theta$ terms}).
\end{align*}
Finally, evaluated at $\theta = 0$, the second total derivative of $g_*(\theta)$ is
\begin{align*}
    \frac{d^2}{d\theta^2} g_*(0) &= \tau \cdot \underline{\mathbf{v}}^{\top}\{ \nabla^2 g(\bgamma) \}\underline{\mathbf{v}} - \bgamma^{\top}\nabla g(\bgamma),
\end{align*}
as desired.\qed

\clearpage\section{Fast Embedded Problem Solving}\label{sup:fastEtaGradient}

Recall the embedded quantile estimation problem consists of solving
\begin{align} \label{eq:quantileEstimation}
    \bhatQ := \argmin_{\bQ \in \R^{n\times m}} \oldnorm{\bQ - \bhatY}_F^2,\quad\mbox{subject to}\quad \bB - \bQ\bA \leq \bzero_{n\times m},
\end{align}
where $\bhatY$ is given by~\eqref{eq:hatY}.  
The matrices $\bA \in \R^{m\times (m+1)}$ and $\bB \in \R^{n\times (m+1)}$ impose monotonicity constraints and box constraints $b_L \leq q_{i,j} \leq b_U$ for all $i,j$.  We require $b_L < b_U$, but we permit both $b_L = -\infty$ and $b_U = \infty$.  $\bA$ and $\bB$ respectively take form
\begin{align}
    \bA &= \left[ \begin{matrix*}[r] +1 & -1 &  0 &  0 & \cdots &  0 &  0 \\
                                     0 & +1 & -1 &  0 & \cdots &  0 &  0 \\
                                     0 &  0 & +1 & -1 & \cdots &  0 &  0 \\
                                     0 &  0 &  0 & +1 & \cdots &  0 &  0 \\
                                     \vdots & \vdots & \vdots & \vdots & \ddots & \vdots & \vdots \\
                                     0 &  0 &  0 &  0 & \cdots & +1 & -1 \end{matrix*}\right],\label{eq:A}\\
    \bB &= [\begin{matrix} b_L\bone_n & \bzero_{n\times(m-1)} & -b_U\bone_n \end{matrix}].\label{eq:B}
\end{align}
The Lagrangian of the embedded optimization problem is
\begin{align}
    \calL(\bQ, \bEta) = \frac{1}{2}\oldnorm{\bQ - \bhatY}_F^2 + \tr\left\{ (\bB - \bQ\bA)^{\top}\bEta\right\},
\end{align}
where $\bEta \in \R^{n\times(m+1)} \geq \bzero_{n\times(m+1)}$ is the associated Lagrange multiplier.  Optimality is given by the stability and primal feasibility KKT conditions
\begin{align*}
    \frac{\partial}{\partial\bQ} \calL(\bhatQ, \bhatEta) &= \bzero_{n\times m}, \\
    \bB - \bhatQ \bA &= \bzero_{n\times (m+1)},
\end{align*}
where the second condition is equivalent to $\frac{\partial}{\partial\bEta} \calL(\bhatQ, \bhatEta) = \bzero_{n\times(m+1)}$; the solution pair $\bhatQ, \bhatEta$ form a saddle point in $(\bQ, \bEta)$ space.  The derivative of $\calL(\bQ, \bEta)$ with respect to $\bQ$ is $\frac{\partial}{\partial \bQ} \calL(\bQ, \bEta) = \bQ - \bhatY - \bEta\bA^{\top}$, and hence $\bhatQ = \bhatY + \bhatEta\bA^{\top}$.  This allows us to perform projected gradient ascent on $\bEta$ alone, where the projection step is onto the positive part of $\bEta$, to obey $\bEta \geq \bzero$.  Given \tharg{t} iterate $\bEta^{(t)}$ and dampening parameter $\alpha > 0$, we substitute $\bQ^{(t)} = \bhatY + \bEta^{(t)}\bA^{\top}$ and write the ascent step
\begin{align}
    \bEta^{(t+1)} &= \left\{ \bEta^{(t)} + \alpha \cdot \frac{\partial}{\partial\bEta} \calL(\bQ^{(t)}, \bEta^{(t)}) \right\}_+ \nonumber\\
    &= \left\{ \bEta^{(t)} + \alpha (\bB - \bQ^{(t)}\bA) \right\}_+ \nonumber\\
    &= \left\{ \bEta^{(t)} + \alpha (\bB - \bhatY\bA - \bEta^{(t)}\bA^{\top}\bA) \right\}_+ \nonumber\\
    &= \left\{ \alpha(\bB - \bhatY\bA) + \alpha \bEta^{(t)}(\alpha^{-1}\bI_{m+1} - \bA^{\top}\bA)\right\}_+.\label{eq:Eta_step}
\end{align}
We take $\alpha = 1/2$, so that each gradient iteration involves evaluating $\bEta^{(t)}(2 \bI_{m+1} - \bA^{\top}\bA)$.  The form of $\bA$ from~\eqref{eq:A} admits
\begin{align*}
    2\bI_{m+1} - \bA^{\top}\bA &= \left[ \begin{matrix} 1 & 1 & 0 & 0 & \cdots & 0 & 0 \\
                                                        1 & 0 & 1 & 0 & \cdots & 0 & 0 \\
                                                        0 & 1 & 0 & 1 & \cdots & 0 & 0 \\
                                                        0 & 0 & 1 & 0 & \cdots & 0 & 0 \\
                                                        \vdots & \vdots & \vdots & \vdots & \ddots & \vdots & \vdots \\
                                                        0 & 0 & 0 & 0 & \cdots & 0 & 1 \\
                                                        0 & 0 & 0 & 0 & \cdots & 1 & 1 \end{matrix} \right],
\end{align*}
so that right-multiplication of $\bEta^{(t)}$ by $2\bI_{m+1} - \bA^{\top}\bA$ is simply a summation operation over columns of $\bEta^{(t)}$.  Specifically, this is
\begin{align}
    \bEta^{(t)}(2\bI_{m+1} - \bA^{\top}\bA) &= [\begin{matrix} \bEta_{2:(m+1)}^{(t)} & \bEta_{m+1}^{(t)} \end{matrix}] + [\begin{matrix} \bEta_1^{(t)} & \bEta_{1:m}^{(t)} \end{matrix}], \label{eq:Eta_column_sum}
\end{align}
where $\bEta_{i:j} := [\begin{matrix} \bEta_i & \bEta_{i+1} & \cdots & \bEta_j \end{matrix}]$.  Defining $\bC_{\bhatY} := \bB - \bhatY\bA$ and plugging~\eqref{eq:Eta_column_sum} into~\eqref{eq:Eta_step}, we have a final gradient step of
\begin{align}
    \bEta^{(t+1)} &= \left\{ \frac{1}{2} \left( \bC_{\bhatY} + [\begin{matrix} \bEta_{2:(m+1)}^{(t)} & \bEta_{m+1}^{(t)} \end{matrix}] + [\begin{matrix} \bEta_1^{(t)} & \bEta_{1:m}^{(t)} \end{matrix}] \right)\right\}_+,\label{eq:Eta_step_final}
\end{align}
which is simultaneous across all $i$ and which does not involve vector or matrix multiplication.  Importantly too, equation~\eqref{eq:Eta_step_final} updates rows of $\bEta^{(t)}$ independently of each other, and if a row $\widehat{\mathbf{y}}_i$ of $\bhatY$ does not violate any constraints in $\bB - \bhatY\bA \leq \bzero$, the corresponding row $\boldeta_i$ of $\bhatEta$ will be $\bzero_{m+1}$.  As such, gradient ascent~\ref{eq:Eta_step_final} may be focused on those rows $\boldeta_i$ where $\mathbf{b}- \bA^{\top}\widehat{\mathbf{y}}_i \leq \bzero_{m+1}$ is violated.

\clearpage
\section{Specific Implementation of Geodesic Descent}\label{sup:Implementation}

In the $p \gg n$ case or the $n \gg p$ case, efficient implementation should avoid handling matrices of size $p \times p$ or $n \times n$, respectively.  Here we algebraically manipulate the matrix operations to ``fold in" the larger dimension, leaving most matrix operations to occur between objects of the smaller dimension.

To begin, note the matrix product $\bG \btilX$ can be rearranged with the push-through identity:
\begin{align}
    \bG \btilX &= (\btilX \bD_{\bgamma}^2 \btilX{}^{\top} + \bI_n)^{-1}\btilX \nonumber\\
    &= \btilX(\bD_{\bgamma}^2\btilX{}^{\top} \btilX + \bI_p)^{-1} \nonumber\\
    &=: \btilX\bF,\label{eq:GF}
\end{align}
where $\bF := (\bD_{\bgamma}^2 \btilX{}^{\top}\btilX + \bI_p)^{-1}$ has been defined for convenience.  Since $\bG \in \R^{n\times n}$ and $\bF \in \R^{p\times p}$, the identity $\bG \btilX = \btilX \bF$ will be helpful in reducing object sizes when $n$ and $p$ are different in scale.

\subsection{Gradient \texorpdfstring{$\nabla f(\blambda)$}{v f(l)}}

The gradient $\nabla f(\blambda)$ information is contained in the diagonal entries of $\bN$, which again are
\begin{align}\label{eq:partial_grad}
    \frac{\partial}{\partial\lambda_k} f(\blambda) &= \btilX{}_k^{\top} \bG\left\{ \sum_{i=1}^n \underline{\mathbf{e}}_i\underline{\mathbf{e}}_i^{\top}(\bhatY - \bY)(\bI_m - \proj{\bA_{0,i}})\right\} \bY^{\top}\bG\btilX_k
\end{align}
Let $\btilE \in \R^{n\times m}$ be the matrix inside the braces in \eqref{eq:partial_grad}, formed by row-wise modifying $\bhatE = \bhatY - \bY$ through right-multiplication with $\bI_m - \proj{\bA_{0,i}}$. Note $\btilE = \sum_{i=1}^n \underline{\mathbf{e}}_i\underline{\mathbf{e}}_i^{\top}\btilE$.  Equation \eqref{eq:partial_grad} can be rewritten:
\begin{align*}
    \frac{\partial}{\partial\lambda_k} f(\blambda) &= \sum_{i=1}^n \btilX{}_k^{\top}\bG\bY\btilE{}^{\top}\underline{\mathbf{e}}_i \cdot \btilX{}_k^{\top}\bG\underline{\mathbf{e}}_i, \qquad\text{hence}\\
    \nabla f(\blambda) &= \sum_{i=1}^n (\btilX{}^{\top}\bG\bY\btilE{}^{\top} \underline{\mathbf{e}}_i ) \circ ( \btilX{}^{\top}\bG\underline{\mathbf{e}}_i ) \\
    &= \left\{ (\btilX{}^{\top}\bG\bY\btilE{}^{\top}) \circ (\btilX{}^{\top}\bG) \right\} \bone_n.
\end{align*}
Since $\bG \in \R^{n \times n}$, this form is suitable when $p \gg n$, otherwise we use the identity in \eqref{eq:GF}
\begin{align}\label{eq:gradient_lambda_new}
    \nabla f(\blambda) &= \left\{(\bF^{\top}\btilX{}^{\top}\bY\btilE{}^{\top}) \circ (\bF^{\top}\btilX{}^{\top}) \right\} \bone_n,
\end{align}
where $\bF \in \R^{p \times p}$. The matrix $\btilX{}^{\top}\bY \in \R^{p \times m}$ can be calculated once and stored in memory.

\subsection{\texorpdfstring{$\mathbf{2}^{\mathrm{nd}}$}{2nd} Derivative \texorpdfstring{$g_*''(0)$}{g*''(0)}}

The geodesic step involves evaluating second derivative, which can be written
\begin{align*}
    g_*''(0) &= \tau \cdot \underline{\mathbf{v}}^{\top}\left\{ \nabla^2 g(\bgamma) \right\} \underline{\mathbf{v}} - \bgamma^{\top}\nabla g(\bgamma).
\end{align*}
Letting $\blambda = \calF(\bgamma) = \bgamma \circ \bgamma$, the Hessian $\nabla^2 g(\bgamma)$ is
\begin{align*}
    \nabla^2 g(\bgamma) &= 2\bD_{\nabla f(\blambda)} + 4 (\bgamma \bgamma^{\top}) \circ \nabla^2 f(\blambda),
\end{align*}
which we do not want to directly calculate when $p \gg n$.  Instead, we can express the corresponding component of the second derivative as a sum of terms
\begin{align}\label{eq:der2_quadratic}
    \tau \cdot \underline{\mathbf{v}}^{\top}\left\{ \nabla^2 g(\bgamma) \right\} \underline{\mathbf{v}} &= 2\tau \cdot \{\underline{\mathbf{v}} \circ \underline{\mathbf{v}} \circ \nabla f(\blambda)\}\bone_p + 4\tau^2 \cdot (\underline{\mathbf{v}} \circ \underline{\bgamma})^{\top}\{ \nabla^2 f(\blambda) \}(\underline{\mathbf{v}} \circ \underline{\bgamma}).
\end{align}
For the second term of \eqref{eq:der2_quadratic}, we combine the Hessian terms from Proposition 2 into a single summation over $i \in [n]$
\begin{align*}
    \nabla^2 f(\blambda) &= \sum_{i=1}^n \left[ \left\{ \btilX{}^{\top}\bG \underline{\mathbf{e}}_i\underline{\mathbf{e}}_i^{\top}\bG\btilX\right\} \circ \left\{ \btilX{}^{\top}\bG\bY(\bI_m - \proj{\bA_{0,i}})\bY^{\top}\bG\btilX\right\} \: - \right.\\
    &\phe\qquad \left. \left\{ \btilX{}^{\top}\bG\btilX \right\} \circ \left\{ \btilX{}^{\top}\bG( \bY\btilE{}^{\top}\underline{\mathbf{e}}_i\underline{\mathbf{e}}_i^{\top} + \underline{\mathbf{e}}_i\underline{\mathbf{e}}_i^{\top}\btilE\bY^{\top} )\bG\btilX\right\} \right],
\end{align*}
so that now
\begin{align*}
    (\underline{\mathbf{v}} \circ \underline{\bgamma})^{\top}\{ \nabla^2 f(\blambda) \}(\underline{\mathbf{v}} \circ \underline{\bgamma}) &= \sum_{k=1}^p \sum_{h=1}^p \underline{\gamma}_k\underline{\gamma}_h\underline{\text{v}}_k\underline{\text{v}}_h\sum_{i=1}^n \left[ \btilX{}_k^{\top}\bG\underline{\mathbf{e}}_i\underline{\mathbf{e}}_i^{\top}\bG\btilX_h \btilX{}_k^{\top}\bG\bY(\bI_m - \proj{\bA_{0,i}})\bY^{\top}\bG\btilX_{h} \right. \:-\\
    &\phe \qquad\qquad\qquad\qquad\qquad \left. \btilX{}_k^{\top}\bG\btilX_h \btilX{}_k^{\top}\bG(\bY\btilE{}^{\top}\underline{\mathbf{e}}_i\underline{\mathbf{e}}_i^{\top} + \underline{\mathbf{e}}_i\underline{\mathbf{e}}_i^{\top}\btilE\bY^{\top})\bG\btilX_h \right]\\
    &= \sum_{i=1}^n \sum_{k=1}^p \underline{\gamma}_k\underline{\text{v}}_k \sum_{h=1}^p \btilX{}_k^{\top}\bG\underline{\mathbf{e}}_i\underline{\mathbf{e}}_i^{\top}\bG\btilX(\underline{\gamma}_h\underline{\text{v}}_h\underline{\mathbf{e}}_h\underline{\mathbf{e}}_h^{\top} )\btilX{}^{\top}\bG\bY(\bI_m - \proj{\bA_{0,i}})\bY^{\top}\bG\btilX_k\:-\\
    &\phe \sum_{i=1}^n \sum_{k=1}^p \underline{\gamma}_k\underline{\text{v}}_k \sum_{h=1}^p \btilX{}_k^{\top}\bG\btilX(\underline{\gamma}_h\underline{\text{v}}_h\underline{\mathbf{e}}_h\underline{\mathbf{e}}_h^{\top})\btilX{}^{\top}\bG(\bY\btilE{}^{\top}\underline{\mathbf{e}}_i\underline{\mathbf{e}}_i^{\top} + \underline{\mathbf{e}}_i\underline{\mathbf{e}}_i^{\top}\btilE\bY^{\top})\bG\btilX_k \\
    &= \sum_{i=1}^n \sum_{k=1}^p \sum_{h=1}^p \underline{\mathbf{e}}_i^{\top}\bG\btilX(\underline{\gamma}_k\underline{\text{v}}_k\underline{\mathbf{e}}_k\underline{\mathbf{e}}_k^{\top})\btilX{}^{\top}\bG\bY(\bI_m - \proj{\bA_{0,i}})\bY^{\top}\bG\btilX(\underline{\gamma}_h\underline{\text{v}}_h\underline{\mathbf{e}}_h\underline{\mathbf{e}}_h^{\top})\btilX{}^{\top}\bG\underline{\mathbf{e}}_i \:-\\
    &\phe\sum_{i=1}^n \sum_{p=1}^p \sum_{h=1}^p \underline{\mathbf{e}}_i^{\top}\btilE\bY^{\top}\bG\btilX(\underline{\gamma}_h\underline{\text{v}}_h\underline{\mathbf{e}}_h\underline{\mathbf{e}}_h^{\top})\btilX{}^{\top}\bG\btilX(\underline{\gamma}_k\underline{\text{v}}_k\underline{\mathbf{e}}_k\underline{\mathbf{e}}_k^{\top})\btilX{}^{\top}\bG\underline{\mathbf{e}}_i \:-\\
    &\phe\sum_{i=1}^n \sum_{p=1}^p \sum_{h=1}^p \underline{\mathbf{e}}_i^{\top}\bG\btilX(\underline{\gamma}_h\underline{\text{v}}_h\underline{\mathbf{e}}_h\underline{\mathbf{e}}_h^{\top})\btilX{}^{\top}\bG\btilX(\underline{\gamma}_k\underline{\text{v}}_k\underline{\mathbf{e}}_k\underline{\mathbf{e}}_k^{\top})\btilX{}^{\top}\bG\bY\btilE{}^{\top}\underline{\mathbf{e}}_i \\
    &= \sum_{i=1}^n \underline{\mathbf{e}}_i^{\top}\bG\btilX\bD_{\underline{\bgamma} \circ \underline{\mathbf{v}}}\btilX{}^{\top}\bG\bY(\bI_m - \proj{\bA_{0,i}})\bY^{\top}\bG\btilX\bD_{\underline{\bgamma} \circ \underline{\mathbf{v}}}\btilX{}^{\top}\bG\underline{\mathbf{e}}_i \:-\\
    &\phe 2 \sum_{i=1}^n \underline{\mathbf{e}}_i^{\top} \btilE \bY^{\top}\bG\btilX\bD_{\underline{\bgamma} \circ \underline{\mathbf{v}}}\btilX{}^{\top}\bG\btilX \bD_{\underline{\bgamma} \circ \underline{\mathbf{v}}} \btilX{}^{\top}\bG\underline{\mathbf{e}}_i.
\end{align*}
Then defining:
\begin{align*}
    \begin{aligned}
    \bV &:= \bY^{\top} \bG \btilX \bD_{\underline{\bgamma} \circ \underline{\mathbf{v}}}\btilX{}^{\top}\bG &&= \bY^{\top} \btilX \bF \bD_{\underline{\bgamma} \circ \underline{\mathbf{v}}} \bF^{\top} \btilX{}^{\top}, \\
    \bW &:= \bY^{\top} \bG \btilX \bD_{\underline{\bgamma} \circ \underline{\mathbf{v}}}\btilX{}^{\top}\bG\btilX \bD_{\underline{\bgamma} \circ \underline{\mathbf{v}}} \btilX{}^{\top} \bG &&= \bY^{\top} \btilX \bF \bD_{\underline{\bgamma} \circ \underline{\mathbf{v}}} \bF^{\top} \btilX{}^{\top}\btilX \bD_{\underline{\bgamma} \circ \underline{\mathbf{v}}} \bF^{\top} \btilX{}^{\top},
    \end{aligned}
\end{align*}
the second derivative is:
\begin{align}\label{eq:theta_derivative_2_new}
    g_*''(0) &= 2\tau \cdot \{ \underline{\mathbf{v}} \circ \underline{\mathbf{v}} \circ \nabla f(\blambda) \} \bone_p + 4\tau^2 \sum_{i=1}^n \bV_i^{\top}(\bI_m - \proj{\bA_{0,i}})\bV_i - 16\tau^2 \inner{\bW, \btilE{}^{\top}}_F - \bgamma^{\top}\nabla g(\bgamma).
\end{align}
Assuming $n, p > m$, the largest object which needs to be calculated and stored is $p \times n$ in size.

\clearpage\section{Algorithms}\label{sup:algorithms}

Algorithm~\ref{alg:mcd} recounts the modified coordinate descent algorithm as proposed by \citet{Tucker:2023}. Algorithm~\ref{alg:embedded} describes our procedure for solving the embedded problem using the gradient methods discussed in Section~\ref{sup:fastEtaGradient}. Algorithm~\ref{alg:GSD} describes our proposed geodesic algorithm.

\begin{algorithm}
\DontPrintSemicolon
  
  \KwInput{Total allowance $\tau > 0$; error tolerance parameters $\varepsilon_1 > 0$ and $\varepsilon_2 > 0$; constraint matrices $\bB \in \R^{n\times (m+1)}$ and $\bA \in \R^{m\times (m+1)}$; column-centered regressor matrix $\bX \in \R^{n \times p}$; empirical quantile functions $\bY \in \R^{n \times m}$; initial allowance vector $\blambda^{(0)}$; maximum iteration count $T \in \Narg{1}$.}
  
  \KwOutput{Allowance vector $\bhatlambda$ solving the FRiSO sparsity problem \eqref{eq:friso}.}
  
  Initialize $t \mapsfrom 0$. \\
  
  Initialize $\mathbf{error} \mapsfrom \varepsilon_1 + 1$.
  
  \While{$\mathbf{error} > \varepsilon_1 \enskip\mathrm{and}\enskip t < T$}{

    Update $t \mapsfrom t + 1$.
    
    Set $\blambda^{(t)} \mapsfrom \blambda^{(t - 1)}$.
    
    \For{$k = 1, \hdots, p$}{
        \If{$\sum_{k' \neq k}\lambda_{k'}^{(t)} > \varepsilon_2$}{
            Define $\bL^{(t)}_k(\alpha) := \{\btillambda(\alpha) : 0 \leq \alpha \leq 1\} \subset \bTau$ as the line segment spanning $\bTau$, which connects $\blambda^{(t)}$ and the \tharg{k} corner of $\bTau$ (given by $\tau \cdot \underline{\mathbf{e}}_k$), where $\tillambda_k(1) = \tau$ and $\tillambda_k(0) = 0$. \\
            
            Calculate:
            \begin{align*}
                \hatalpha &= \argmin_{0 \leq \alpha \leq 1} \sum_{i=1}^n d^2_W\left( \mathbf{y}_i, \: \widehat{\mathbf{q}}(\mathbf{x}_i; \btillambda(\alpha))\right).
            \end{align*}
            That is, perform a line search to minimize the FRiSO objective function along $\bL^{(t)}_k(\alpha)$.\\
            
            Update $\blambda^{(t)} \mapsfrom \btillambda(\hatalpha)$.
        }
        
    }
    Update $\mathbf{error} \mapsfrom \norm{\blambda^{(t)} - \blambda^{(t - 1)}}_{\infty}$.
  }
  Set $\bhatlambda \mapsfrom \blambda^{(t)}$.
\caption{Modified Coordinate Descent (MCD)}
\label{alg:mcd}
\end{algorithm}

\clearpage\begin{algorithm}

\DontPrintSemicolon
  
  \KwInput{Error tolerance parameter $\varepsilon > 0$; constraint matrices $\bA \in \R^{m\times (m+1)}$ and $\bB \in \R^{m\times (m+1)}$ from \eqref{eq:A} and \eqref{eq:B}, respectively; unconstrained predictor matrix $\bhatY$ from (6); 
  maximum iteration count $T \in \Narg{1}$.}
  
  \KwOutput{Estimated quantile matrix $\bhatQ$ solving the embedded problem \eqref{eq:embeddedProblem}.}

  Define $\bC := \bB - \bhatY \bA$, and identify the index set $\calI = \{ i : \mathbf{c}_i \nleq 0\}$ (i.e. the rows of $\bC$ which have at least one positive entry, indicating a violated constraint).

  \If{$\calI = \emptyset$}{

    Set $\bhatEta = \bzero_{n\times (m + 1)}$.
  
  }\Else{

    Without loss of generality, let $\bC = \left[ \begin{matrix} \bC_{\calI} \\ \bC_{\calI^c} \end{matrix} \right]$ be the block decomposition of $\bC$ into those rows indexed by $\calI$ and the relative complement set of rows, $\calI^c = [n] \setminus \calI$.
    
    Initialize $t \mapsfrom 0$.

    Initialize $\mathbf{error} \mapsfrom \varepsilon + 1$.

    Initialize $\bEta_{\calI}^{(0)} \mapsfrom (\bC_{\calI})_+$.
    
    \While{$\mathbf{error} > \varepsilon\enskip\mathrm{and}\enskip t < T$}{

        Update $t \mapsfrom t + 1$.
        
        Update $\bEta_{\calI}^{(t)}$ using $\bEta_{\calI}^{(t-1)}$ and $\bC_{\calI}$ in equation \eqref{eq:Eta_step_final}.

        Update $\mathbf{error} \mapsfrom \norm{\bEta_{\calI}^{(t+1)} - \bEta_{\calI}^{(t)}}_{\infty}$
      
    }

    Again without loss of generality, set $\bhatEta = \left[ \begin{matrix} \bhatEta_{\calI} \\ \bhatEta_{\calI^c} \end{matrix} \right] = \left[ \begin{matrix} \bhatEta_{\calI}^{(t-1)} \\ \bzero \end{matrix} \right]$
  
  }

  Set $\bhatQ \mapsfrom \bhatY + \bhatEta \bA^{\top}$.
  
\caption{Quantile Estimation (``Embedded") Problem}
\label{alg:embedded}
\end{algorithm}

\clearpage\begin{algorithm}

\DontPrintSemicolon
  
  \KwInput{Total allowance $\tau > 0$; error tolerance parameter $\varepsilon > 0$; dampening parameter $\alpha > 0$; maximum rotation angle $\theta_{\mathrm{max}} > 0$; ``nudge" parameter $\beta > 0$; constraint matrices $\bB \in \R^{n\times (m+1)}$ and $\bA \in \R^{m\times (m+1)}$; column-centered and column-scaled regressor matrix $\bX \in \R^{n\times p}$; empirical quantile functions $\bY \in \R^{n\times m}$; initial allowance vector $\blambda^{(0)} \in \R^p$; maximum iteration count $T \in \Narg{1}$.}
  
  \KwOutput{Allowance vector $\bhatlambda$ solving the sparsity problem \eqref{eq:friso}.}
  
  Initialize $\bgamma^{(0)} : \gamma^{(0)}_k = \sqrt{\lambda_k^{(0)}} + \beta$ for each $k = 1, \hdots, p$, and normalize $\bgamma^{(0)} \mapsfrom \sqrt{\tau} \cdot \underline{\bgamma}^{(0)}$.
  
  Define $\bbarY := n^{-1}\bone_{n\times n}\bY$, and define $\btilX := \bX / \sqrt{n}$.

  Initialize $t \mapsfrom 0$.

  Initialize $\mathbf{error} = \varepsilon + 1$.

  \While{$\mathbf{error} > \varepsilon \enskip\mathrm{and}\enskip t < T$}{

    Update $t \mapsfrom t + 1$.

    Calculate $\bG \mapsfrom (\btilX \bD_{\bgamma^{(t-1)}}^2 \btilX{}^{\top} + \bI_n)^{-1}$.
    
    Calculate $\bhatY \mapsfrom \bbarY + \bY - \bG\bY$.
    
    Obtain $\bhatEta$ using Algorithm \ref{alg:embedded}, and active constraint matrix $\bC := \text{sign}(\bhatEta)$.

    Calculate gradient $\nabla g(\bgamma)$
    
    Calculate $\nabla g(\bgamma^{(t-1)})$ using equations \eqref{eq:gradient_lambda_new} and \eqref{eq:gradient_gamma}.
    
    Calculate $\mathbf{v}$ from \eqref{eq:v}, 
    and calculate $g'_*(0)$ using equation \eqref{eq:theta_first_derivative}.
    
    Calculate $g''_*(0)$ using equation \eqref{eq:theta_derivative_2_new}.
    
    Calculate $\theta_* = \min\left\{ \abs{ \alpha \cdot \frac{g'_*(0)}{g''_*(0)} }, \: \theta_{\mathrm{max}}\right\}$.
    
    Set $\bgamma^{(t)} = \abs{\cos(\theta_*) \bgamma^{(t-1)} + \sqrt{\tau} \sin(\theta_*) \underline{\mathbf{v}}}$.
    
    Update $\textbf{error} = \oldnorm{\bgamma^{(t)} - \bgamma^{(t-1)}}_{\infty}$.
    
  }
  
  Set $\bhatlambda \mapsfrom \bgamma^{(t)} \circ \bgamma^{(t)}$.

\caption{Geodesic Second-Order Descent (GSD) Algorithm}
\label{alg:GSD}
\end{algorithm}

\clearpage\section{HYPNOS Data Set}\label{sup:hypnos}

\subsection{List of variables}\label{sup:hypnos_data_sets}

The dataset used for analysis consisted of $n=207$ patients which had complete information on $p=34$ variables collected during the pre-randomization period. These variables comprised:

\textbf{Demographic variables ($4$).} These included ``Age", ``Gender", ``Race\_White", and ``Race\_Other" (i.e. non-White and non-African American; see Section~\ref{sup:recoding}).

\textbf{Baseline health-related variables ($5$).} These included body mass index (``BMI"), point of care ``HbA1c", self-reported average hours of sleep during the work week (``Sleep\_Workdays"), Epworth Sleepiness Scale score (``EpworthSS"), and Berlin Questionnaire High/Low OSA risk level (``Berlin").

\textbf{Binary medication variables ($20$).} These were coded Yes/No (resp. 1/0), and excluded medications for which $< 5$ subjects indicated use. The included variables were ACE inhibitor (``ACEI"), alpha adrenergic blocker (``AAB"), angiotensin 2 receptor blocker (``A2RB"), ``Anti\_Epileptic", ``Aspirin", ``Beta\_Blocker", ``Biguanide", calcium channel blocker (``CCB"), dihydro-calcium channel blocker (``Dihydro\_CCB"), glucagon-like peptide-1 receptor agonists (``GLP1RA"), H1 antagonist (``H1\_Ant"), ``NSAID", peroxisome proliferator-activated receptor gamma (``PPAR"), proton pump inhibitor (``PPI"), serotonin-norepinephrine reuptake Inhibitor (``SNRI"), selective serotonin reuptake Inhibitor (``SRI"), ``Statin", ``Sulfonylurea", ``Thiazide", and ``Thyroxine".

\textbf{Oxygen saturation variables ($5$).} These were variables measured during the at-home overnight sleep study, and included mean oxygen saturation (``Mean\_Sat"), minimum oxygen saturation (``Min\_Sat"), standard deviation of oxygen saturation levels (``Std\_Sat"), $4\%$ oxygen desaturation index (``$\text{ODI}_4$"), and proportion of time spent below 90\% saturation (``TST$90\%$").

\subsection{Variable recoding}\label{sup:recoding}

We coded binary variables as 0/1, with binary Sex label coded as Female = 0, Male = 1. We recoded two other variables as described below.

\textbf{Oxygen saturation time.} One sleep study variable, TST$90\%$, was originally given as the amount of time (in minutes) during the sleep study during which blood oxygen saturation levels were $<90\%$.  Since this time is confounded by the length of the sleep study, which differed from subject to subject, we re-calculated TST$90\%$ instead as a proportion of total sleep study time.

\textbf{Race category.} Subject race category was originally coded with values $1$ through $6$, where $1$ corresponded to African American, $2$ corresponded to American Indian/Native American, $3$ corresponded to Asian or Pacific Islander, $4$ corresponded to Caucasian/White, $5$ corresponded to Hispanic, and $6$ corresponded to Other.  Due to small sample sizes in non-White and non-African American categories, we re-coded race into three variables: ``Race\_White" ($1$ if race $= 4$, and $0$ otherwise); ``Race\_Other" ($1$ if race $= 2, 3, 5, 6$, and $0$ otherwise); and the remaining subjects were identified as African American (race $= 1$), implicitly defined as the default category, i.e. Race\_White $= 0$ and Race\_Other $= 0$.  In total, there were $72$ subjects identified as African American, $111$ subjects identified as Race\_White, and $24$ subjects identified as Race\_Other.

\subsection{Correlation Analysis}\label{sup:correlation}

Strong positive or negative correlations between $\bX_k$ variables can lead to unstable selection between such correlated variables. During repeated subsampling for stability selection, this instability leads to ``vote splitting" between correlated variables, such that no variable's empirical selection probability is high, i.e. exceeds threshold $\pi_{\mathrm{thr}}$. Correlation analysis of HYPNOS data reveals the sleep study oxygen saturation variables have strong negative and positive correlations among themselves, forming a correlated cluster of $5$ variables.

\begin{figure}[!t]
    \centering
    \includegraphics[width=0.5\textwidth, keepaspectratio]{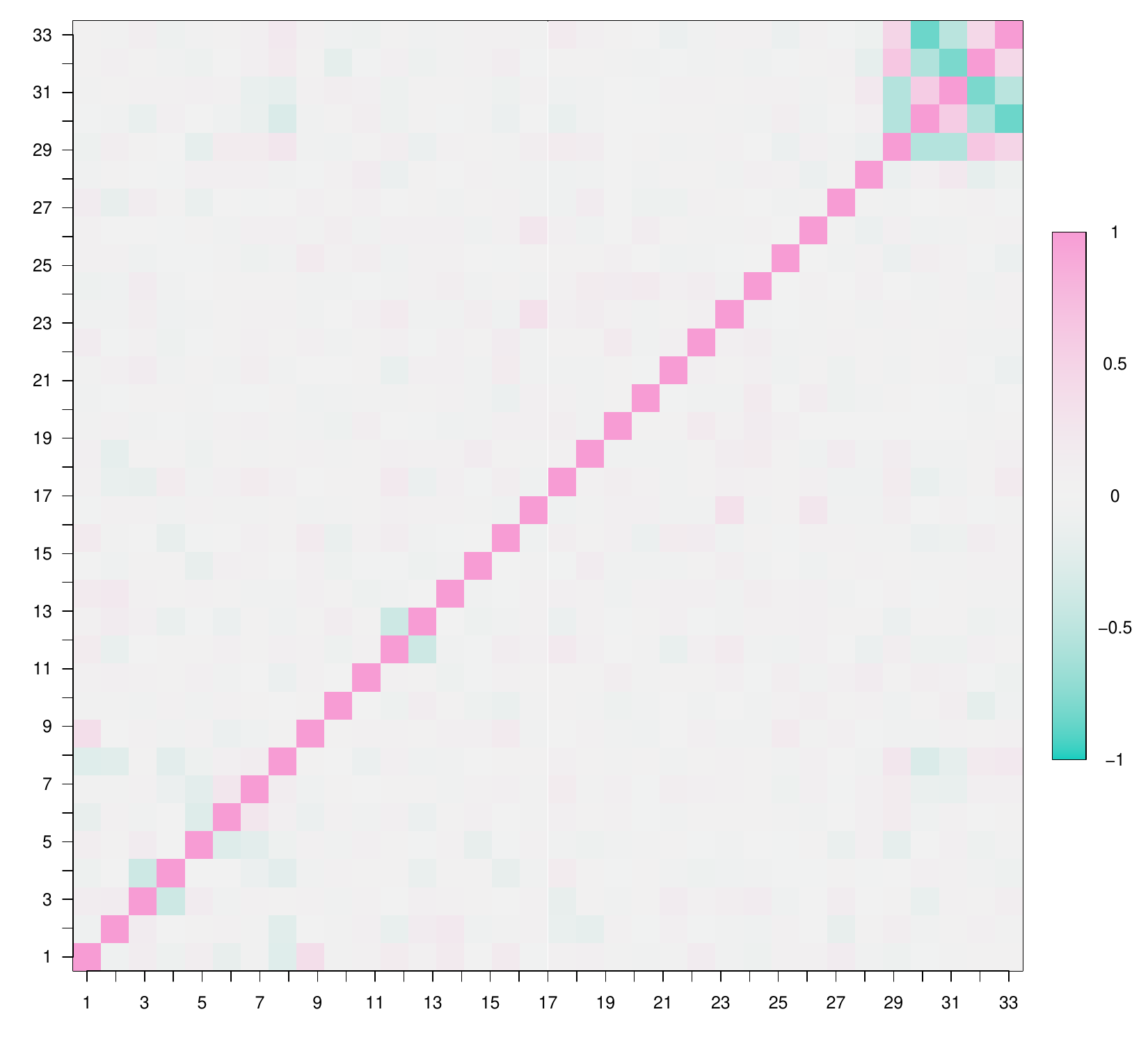}
    \caption{Correlation heatmap of covariates from HYPNOS data.  The correlated cluster of five oxygen saturation variables is viewed in the top right of the plot.  Small negative correlations at variables $3$ and $4$ correspond to indicators for self-reported race category (Race\_White, and Race\_Other, respectively), and small negative correlations at $11$ and $12$ correspond to two hypertension drugs (A2RB and ACEI, respectively), which are not co-prescribed.}
    \label{fig:correlation}
\end{figure}

\begin{figure}[!t]
    \centering
    \includegraphics[width=0.6\textwidth, keepaspectratio]{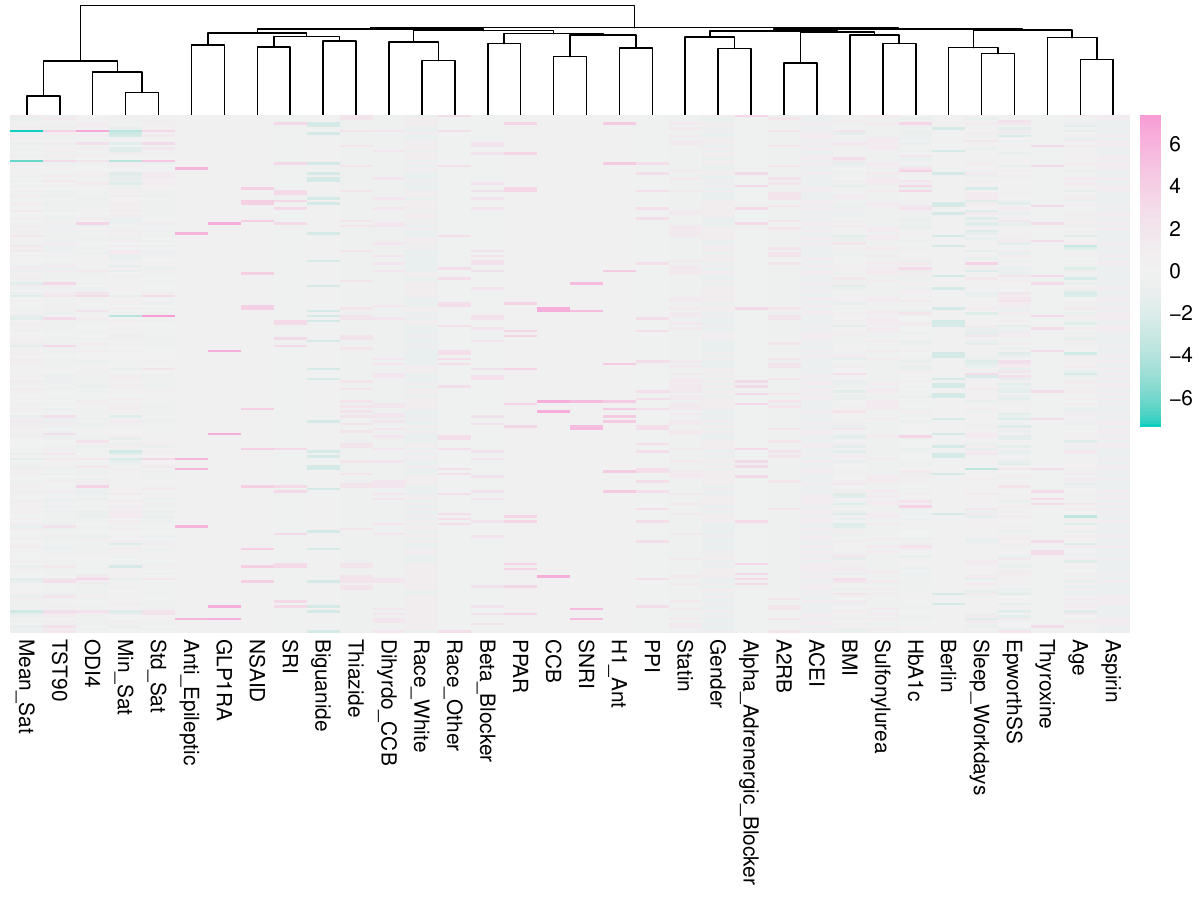}
    \caption{Dendrogram and correlation plot of $\bX_k$ variables from HYPNOS data set, using $1 - |\text{cor}(\bX_j, \bX_k)|$.  The five oxygen variables are grouped together, on the left of the graphic.}
    \label{fig:dendrogram}
\end{figure}

We provide visualizations of this cluster in two ways. First, Figure~\ref{fig:correlation} shows (signed) Pearson correlations between all variables $\bX_k$, with the oxygen variables taking the last five places in the variable ordering. The correlation plot shows strong positive and negative correlations between all oxygen saturation variables, with all pairwise correlations $\geq 0.44$ and each variable having at least one pairwise correlation $\geq 0.6$. By contrast, the largest magnitude pairwise correlation outside of this cluster is $<0.4$. Second, Figure~\ref{fig:dendrogram} illustrates the oxygen saturation cluster by column-clustered dendrogram, using similarity metric $1-\abs{\text{cor}(\bX_j, \bX_k)}$. The first split of the full variable set from the dendrogram ``root" is between the oxygen saturation variables (left-most group) and the other variables.

\begin{figure}[!t]
    \centering
    \includegraphics[width=0.7\textwidth, keepaspectratio]{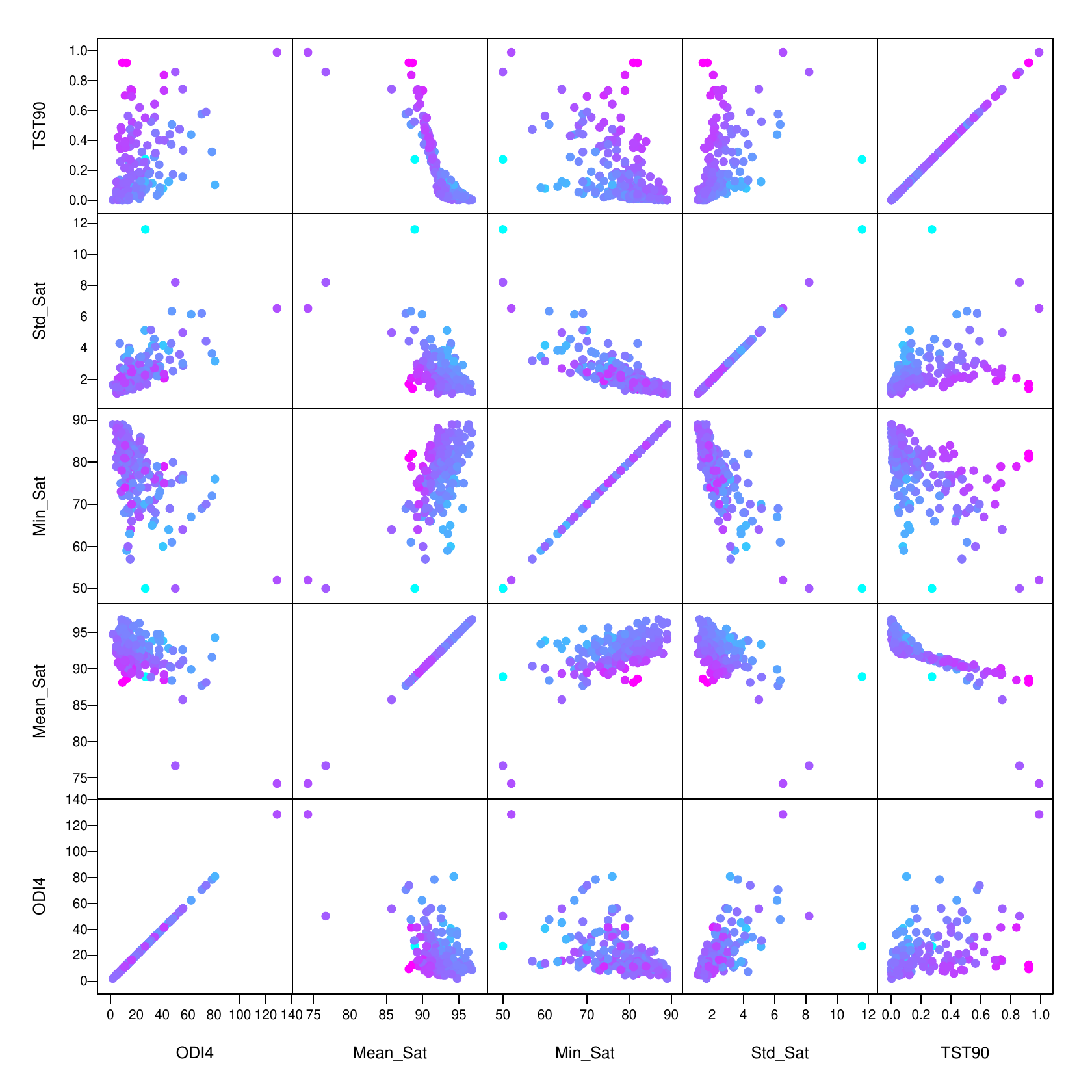}
    \caption{Pairwise scatterplots of oxygen saturation variables, with color scale according to $\text{PC}_{2,\text{oxygen}}$ value.}
    \label{fig:oxygen_pairs}
\end{figure}

To circumvent the correlation problem, we perform principal component analysis on the $5$ oxygen saturation variables, obtaining orthogonal principal components (PCs) which we use in place of the original $5$ variables. Table~\ref{tab:loadings} gives the loadings associated with the $5$ PCs. Since our data analysis shows the $2^{\mathrm{nd}}$ PC is selected, Figure~\ref{fig:oxygen_pairs} illustrates pairwise scatterplots of the original oxygen variables, colored by value of this PC.

\begin{table}[!t]
    \renewcommand{\arraystretch}{1.2}
    \centering
    \begin{tabular}{l|r|r|r|r|r}
        Variable & $PC_1$ & $PC_2$ & $PC_3$ & $PC_4$ & $PC_5$ \\\hline
        $\text{ODI}_4$ & $-0.419$ & $-0.242$ & $-0.850$ & $-0.208$ & $0.001$ \\
        Mean\_Sat & $0.470$ & $-0.462$ & $-0.024$ & $-0.314$ & $-0.683$ \\
        Min\_Sat & $0.455$ & $0.360$ & $-0.480$ & $0.626$ & $-0.202$ \\
        Std\_Sat & $-0.455$ & $-0.477$ & $0.198$ & $0.660$ & $-0.301$ \\
        TST90 & $-0.436$ & $0.609$ & $0.083$ & $-0.174$ & $-0.634$
    \end{tabular}
    \caption{PCA loadings on oxygen saturation variables.}
    \label{tab:loadings}
\end{table}

\subsection{Effects Interpretation}\label{sup:effects}

\begin{figure}[!t]
    \centering
    \includegraphics[width=\textwidth, keepaspectratio]{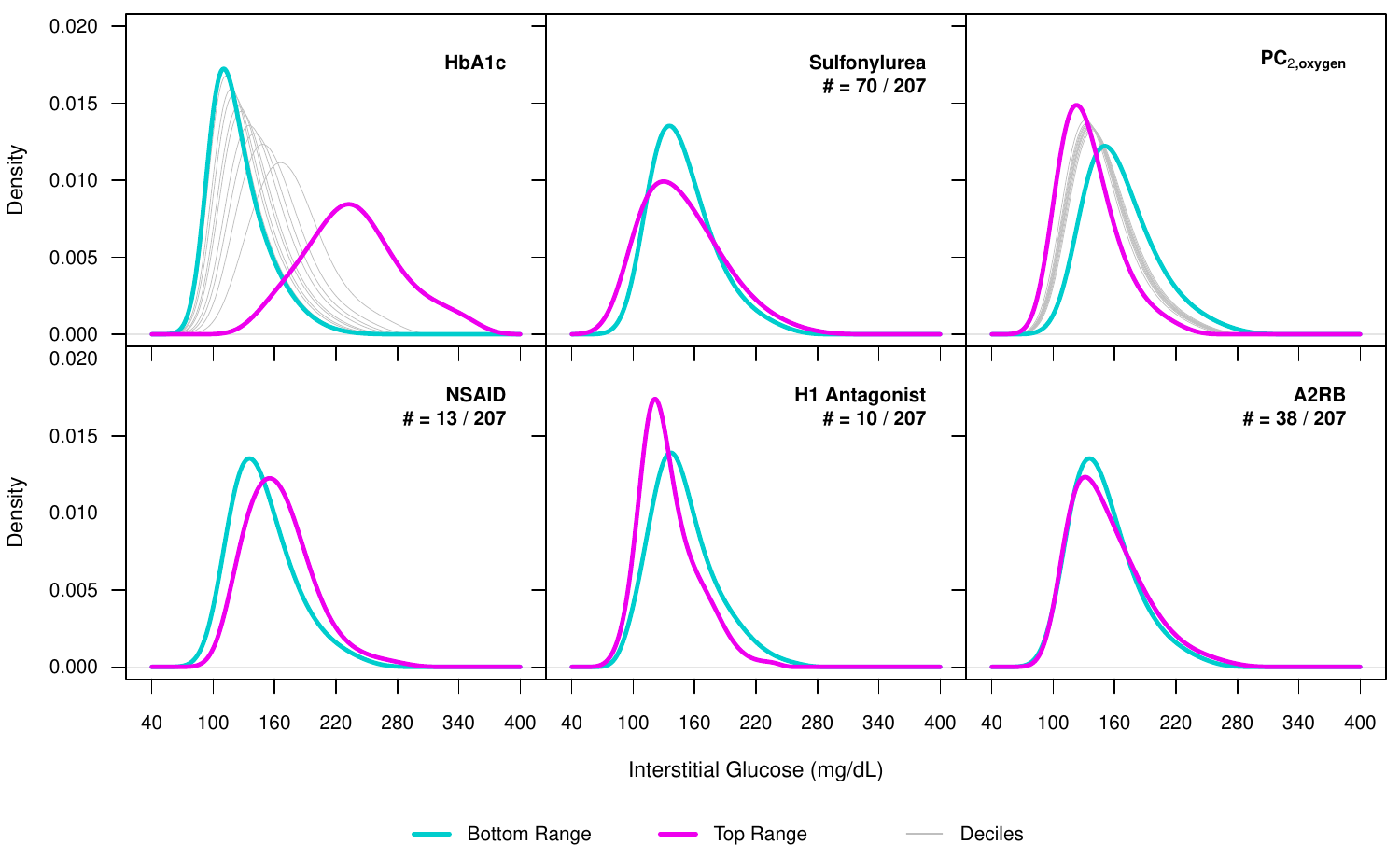}
    \caption{Predicted densities using Fréchet regression with 6 variables, including 3 selected variables (\textit{top row}) and 3 nearest non-selected variables (\textit{bottom row}). In each panel, the levels of the corresponding covariate are varied from the lowest/``no" (bottom range, cyan) to the highest/``yes" (top range, magenta). Grey lines correspond to evaluations at 0.1 decile changes in corresponding variable. The remaining variables are kept constant, at sample mean for HbA1c, ``no" for medication use variables, and zero for $\text{PC}_{2,\text{oxygen}}$.}
    \label{fig:sup_marginal_plots_6}
\end{figure}

\begin{figure}[!t]
    \centering
    \includegraphics[width=\textwidth, keepaspectratio]{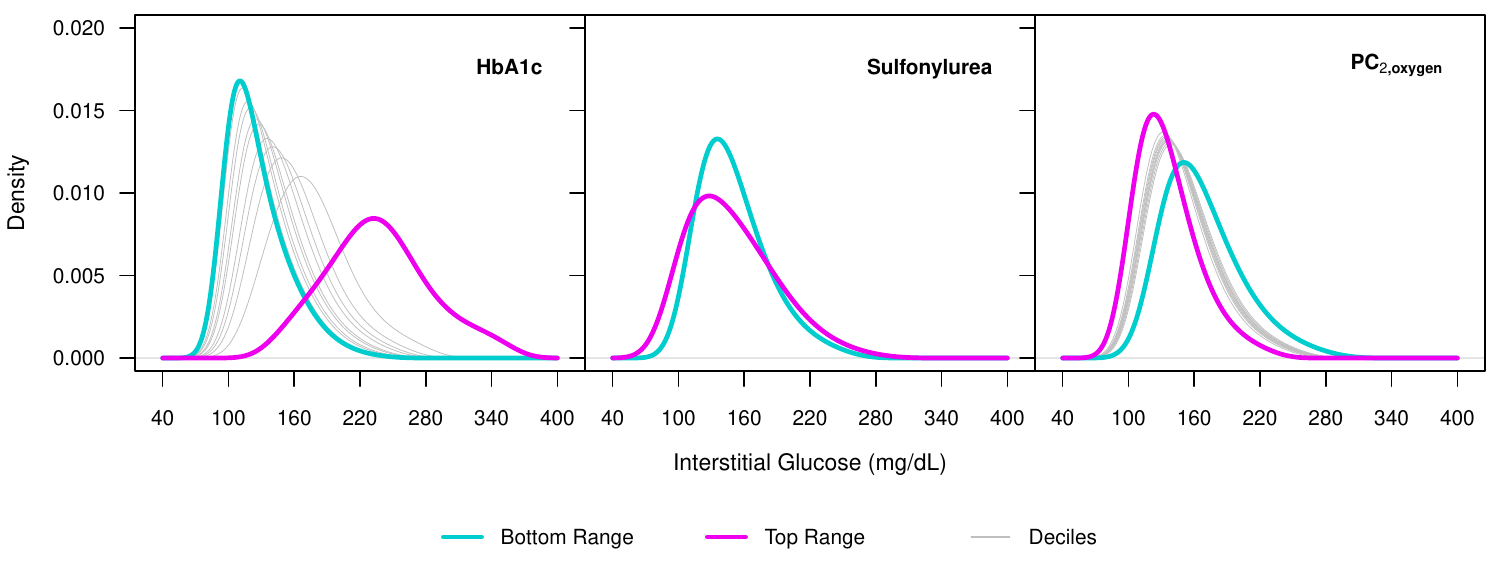}
    \caption{Predicted densities using Fréchet regression with 3 selected variables. In each panel, the levels of the corresponding covariate are varied from the lowest/``no" (bottom range, cyan) to the highest/``yes" (top range, magenta). Grey lines correspond to evaluations at 0.1 decile changes in corresponding variable. The remaining variables are kept constant, at sample mean for HbA1c, ``no" for sulfonylurea use, and zero for $\text{PC}_{2,\text{oxygen}}$.}
    \label{fig:sup_marginal_plots_3}
\end{figure}

Three variables of the $p = 34$ HYPNOS data set were selected by applying an ``any vote" procedure on the complementary pairs stability selection ``stability paths" from \citet{Shah:2013}: HbA1c, a binary indicator for sulfonylurea use, and the $2^{\text{nd}}$ oxygen principal component (Section~\ref{sup:correlation}). Three variables' stability paths were marginally close to the selection threshold (i.e. had $\hatPi_B(k; \tau)$ within $0.1$ of $\pi_{\mathrm{thr}}(\tau)$ for some $\tau$), namely, indicators for use of non-steroidal anti-inflammatory drugs (NSAIDs), use of H1 antagonists (an antihistamine family), and use of angiotensin 2 receptor blockers (A2RBs, a family of hypertension medications). To ascertain whether inclusion of these variables in the selected variable set affects our effects interpretations of the three selected variables, we replicate the marginal effects analysis while including the three new medication use variables; the results are illustrated in Figure~\ref{fig:sup_marginal_plots_6}. For ease of comparison, we repeat the original marginal effects plot in Figure~\ref{fig:sup_marginal_plots_3}.

The marginal changes in glucose mean and standard deviation from high to low input variable values, comparing the 3-variable and 6-variable models, are given in Table~\ref{tab:marginal_changes}. Including the extra three medication use variables results in only a small change in the estimated effect across the empirical range of input values. We do note the range of $\text{PC}_{2,\text{oxygen}}$ values in the HYPNOS data set might not be a representative description of input range for this principal component, e.g. there is larger separation between the top/bottom range marginal curves in the $\text{PC}_{2,\text{oxygen}}$ panels of Figures~\ref{fig:sup_marginal_plots_3} and~\ref{fig:sup_marginal_plots_6} and the decile curves. For completeness, we also present marginal effects between the empirical $2.5\%$ and $97.5\%$ $\text{PC}_{2,\text{oxygen}}$ quantiles, i.e. across the ``central" $95\%$ range, in Table~\ref{tab:marginal_changes_oxygen_95}. In contrast to a mean change of $-31.8$ mg/dL and SD change of $-7.2$ mg/dL over the full range, the mean and SD change over the central $95\%$ range are $-14.8$ and $-3.4$ mg/dL, respectively, in the 3-variable model.

\begin{table}[!t]
    \renewcommand{\arraystretch}{1.2}
    \centering
    \begin{tabular}{llrr}
         & & \multicolumn{2}{c}{Change in glucose over empirical range} \\\cline{3-4}
        Model & Variable & \multicolumn{1}{r}{Mean} & \multicolumn{1}{r}{SD} \\\hline
        3-Variable & \textbf{HbA1c} & $\mathbf{126.2 \rightarrow 236.7 \: (+110.5)}$ & $\mathbf{28.9 \rightarrow 49.2 \: (+20.3)}$ \\
         & \textbf{Sulfonylurea} & $\mathbf{150.0 \rightarrow 149.3 \: (-0.7)}$ & $\mathbf{33.2 \rightarrow 42.1 \: (+8.9)}$ \\
         & \textbf{$\text{PC}_{2,\text{oxygen}}$} & $\mathbf{167.3 \rightarrow 135.5 \: (-31.8)}$ & $\mathbf{37.1 \rightarrow 29.9 \: (-7.2)}$ \\\hline
        6-Variable & \textbf{HbA1c} & $\mathbf{124.7 \rightarrow 236.8 \: (+112.1)}$ & $\mathbf{28.1 \rightarrow 49.4 \:(+21.2)}$ \\
         & \textbf{Sulfonylurea} & $\mathbf{148.8 \rightarrow 148.7 (-0.1)}$ & $\mathbf{32.6 \rightarrow 41.5 \:(+8.9)}$ \\
         & \textbf{$\text{PC}_{2,\text{oxygen}}$} & $\mathbf{165.6 \rightarrow 134.7 \:(-30.8)}$ & $\mathbf{36.1 \rightarrow 29.7 \:(-6.3)}$ \\
         & NSAID & $148.8 \rightarrow 164.5 \:(+15.7)$ & $32.6 \rightarrow 34.3 \:(+1.7)$ \\
         & H1 Antagonist & $148.8 \rightarrow 134.4 \:(-14.4)$ & $32.6 \rightarrow 28.2 \:(-4.4)$ \\
         & A2RB & $148.8 \rightarrow 152.6 \:(+3.8)$ & $32.6 \rightarrow 36.5 \: (+3.9)$
    \end{tabular}
    \caption{Marginal effects on glucose levels across full empirical range of given input variables (i.e. minimum $\rightarrow$ maximum for numeric variables, and ``no" $\rightarrow$ ``yes" for binary variables). Top three rows show marginal changes for the model consisting of selected variable set only (Section~\ref{sec:dataAnalysis}); 
    bottom six rows show marginal changes for the model also including next-three closest variables to selection threshold. Selected variable rows are bolded. For min-max comparison of a single variable, all other inputs are set to constant levels: empirical mean value for numeric variables, and ``no" for binary variables.}
    \label{tab:marginal_changes}
\end{table}

\begin{table}[!t]
    \renewcommand{\arraystretch}{1.2}
    \centering
    \begin{tabular}{llrr}
         & & \multicolumn{2}{c}{Change in glucose over $2.5 \rightarrow 97.5\%$ range} \\\cline{3-4}
        Model & Variable & \multicolumn{1}{r}{Mean} & \multicolumn{1}{r}{SD} \\\hline
        3-Variable & $\text{PC}_{2,\text{oxygen}}$ & $157.8 \rightarrow 143.0 \: (-14.8)$ & $35.0 \rightarrow 31.6 \: (-3.4)$ \\
        6-Variable &  $\text{PC}_{2,\text{oxygen}}$ & $156.3 \rightarrow 142.0 \:(-14.3)$ & $34.2 \rightarrow 31.2 \:(-2.9)$
    \end{tabular}
    \caption{Marginal effects on glucose levels across central empirical $95\%$ range of $\text{PC}_{2,\text{oxygen}}$ values, using same models as in Table~\ref{tab:marginal_changes}. All other inputs are set to constant levels: empirical mean value for numeric variables, and ``no" for binary variables.}
    \label{tab:marginal_changes_oxygen_95}
\end{table}

\end{document}